\documentclass{aa}  

\usepackage{graphicx}
\usepackage{color, colortbl}
\usepackage{xcolor}
\usepackage{subcaption}
\usepackage{txfonts}
\usepackage{hyperref}
\usepackage{cleveref}

\crefname{equation}{Eq.}{Eqs.}
\Crefname{equation}{Equation}{Equations}
\crefname{figure}{Fig.}{Figs.}
\Crefname{figure}{Figure}{Figures}
\crefname{table}{Table}{Tables}
\Crefname{table}{Table}{Tables}
\crefname{chapter}{Chapter}{Chapters}
\Crefname{chapter}{Chapter}{Chapters}
\crefname{section}{Section}{Sections}
\Crefname{section}{Section}{Sections}
\crefname{appendix}{Appendix}{Appendices}
\Crefname{appendix}{Appendix}{Appendices}

\newcommand{\msol}{\,M$_{\odot}$}
\newcommand{\fcbm}{f_{\rm CBM}}
\newcommand{\acbm}{\alpha_{\rm CBM}}
\newcommand{\rcc}{r_{\rm cc}}
\newcommand{\rcbm}{r_{\rm CBM}}
\newcommand{\lb}{\left(}
\newcommand{\rb}{\right)}
\newcommand{\n}{\nabla_{\text{T}}}
\newcommand{\nad}{\nabla_{\text{ad}}}
\newcommand{\nrad}{\nabla_{\text{rad}}}
\newcommand{\D}[1][]{D_{\text{#1}}}
\newcommand{\Hp}[1][]{H_{\rm p {#1}}}
\newcommand{\Pe}{\text{Pe}}
\newcommand{\Y}[1]{\textbf{Y}^{\text{#1}}}
\begin{document} 
	
	\title{Probing the temperature gradient in the core boundary layer of stars with gravito-inertial modes}
	\subtitle{The case of KIC\,7760680}
	
	\author{M. Michielsen
		\inst{1}
		\and C. Aerts
		\inst{1,2,3} 
		\and D. M. Bowman
		\inst{1}   
	}
	
	\institute{\inst{1}Institute of Astronomy, KU Leuven, Celestijnenlaan 200D, B-3001 Leuven, Belgium\\
		\inst{2}Department of Astrophysics, IMAPP, Radboud University Nijmegen, P.O. Box 9010, 6500 GL Nijmegen, The Netherlands \\
		\inst{3}Max Planck Institute for Astronomy, Koenigstuhl 17, 69117 Heidelberg, Germany \\
		\email{mathias.michielsen@kuleuven.be}}
	
	\date{Received 17 November 2020; accepted 8 April 2021}

	\titlerunning{Probing the temperature gradient in near-core layers}
	\authorrunning{M.\ Michielsen et al.}
	
	% \abstract{}{}{}{}{} 
	% 5 {} token are mandatory
	
	\abstract
	{}
	{We investigated the thermal and chemical structure in the near-core region of stars with a convective core by means of gravito-inertial modes. 
		We determined the probing power of different asteroseismic observables and fitting methodologies.
		We focus on the case of the B-type star KIC\,7760680, rotating at a quarter of its critical rotation velocity.}
	{We computed grids of 1D stellar structure and evolution models for two different prescriptions of the temperature gradient and mixing profile in the near-core region. We determined which of these prescriptions is preferred according to the prograde dipole modes detected in 4 yr {\it Kepler\/} photometry of KIC\,7760680.
		We considered different sets of asteroseismic observables and compared the outcomes of the regression problem for a $\chi^2$ and a Mahalanobis distance merit function,
		where the latter takes into account realistic uncertainties for the theoretical predictions and the former does not.}
	{Period spacings of modes with consecutive radial order offer a better diagnostic than mode periods or mode frequencies for asteroseismic modelling of stars revealing only high-order gravito-inertial modes. We find KIC\,7760680 to reveal a radiative temperature gradient in models with convective boundary mixing, but less complex models without such mixing are statistically preferred for this rotating star, revealing extremely low vertical envelope mixing.}
	{Our results strongly suggest the use of measured individual period spacing values for modes of consecutive radial order as an asteroseismic diagnostic for stellar modelling of B-type pulsators with gravito-inertial modes. }
	
	
	\maketitle
	%
	%-------------------------------------------------------------------
	%%%%%%%%%%%%%%%%%%%%%%%%%%%%%%%%%%%%%%%%%%%%%%%%%%%%%%%%%%%%%%%%%%%%%%%%%%%%%%%%%%%%%%%%%%%%%%
	%%%%%%%%%%%%%%%%%%%%%%%%%%%%%%%%%%%%%%%%%%%%%%%%%%%%%%%%%%%%%%%%%%%%%%%%%%%%%%%%%%%%%%%%%%%%%%
	\section{Introduction}
	
	Asteroseismology, the study of stellar interiors by means of detected identified oscillation modes \citep{2010aste.book.....A}, has two major aims. The first goal is to provide masses, radii, and ages of stars relying on current knowledge of stellar structure, providing these three quantities 
	with far better precision than is available from single-epoch data of surface properties (such as spectroscopy or interferometry). The second goal is to improve the input physics of stellar interiors so as to achieve better models for stellar evolution. The need for better models is obvious when it comes to angular momentum transport across stellar evolution \citep{2018A&A...616A..24G,2019ARA&A..57...35A} and mixing that guides the change in the mass fractions of the chemical species \citep{2017RSOS....470192S}. Here our aim is to probe the temperature and chemical structure in the deep interior of stars and to improve our models of those quantities, since these are aspects of the stellar structure that propagate throughout and influence the star's evolution.
	
	Although angular momentum and element transport are two different physical phenomena, they are intimately related, and as such they are treated by similar diffusion coefficients in stellar evolution codes \citep{2009pfer.book.....M}. To date, asteroseismology has mainly been used to evaluate and improve the transport of angular momentum \citep[e.g.][]{2019MNRAS.485.3661F,2019A&A...621A..66E,2019A&A...631L...6E,2020A&A...634L..16D,2020A&A...641A.117D} and to assess the level of chemical mixing \citep[e.g.][]{2016A&A...589A..93D,2019MNRAS.482.2305B,2019ApJ...885..143B,2020MNRAS.493.4987A}, with a specific focus on red giants and on stars of low mass. These studies are all based on acoustic or mixed modes excited by envelope convection. In the current work we rely on gravito-inertial modes in a rotating star of intermediate mass with a well-developed convective core and a radiative envelope. These modes are excited by a heat mechanism \citep{1999AcA....49..119P} and have frequencies that are
	similar to the rotation frequency of the star, such that the modes' spin parameters $s=2\Omega/\omega > 1$, with $\Omega$ and $\omega$ the rotation and oscillation frequencies. In this case the Coriolis and buoyancy forces act together as restoring forces of the oscillations. These oscillations correspond to gravito-inertial modes \citep[see][Fig.\,5]{2019ARA&A..57...35A}.
	In particular, we focus on a B-type pulsator. The extent of and mass in the rotating core-boundary layers are critical unknowns for this type of stars in the current theory of stellar structure, yet these properties drive their evolution.
	For such stars, the asteroseismic modelling cannot rely on a perturbative treatment of the Coriolis force, in contrast to the case of acoustic or mixed modes in slowly rotating stars \citep[see e.g.][for the limits of the perturbative method]{2010A&A...518A..30B}.
	
	Despite their interest from a stellar evolution and chemical yield point of view, few B-type stars with gravito-inertial modes have been modelled asteroseismically, and the focus has been on exceptionally slow rotators \citep{2015A&A...580A..27M,TaoWu2020}. On the other hand, fast rotators offer the more interesting case to probe stellar interiors with the aim of improving stellar evolution theory
	\citep{2016ApJ...823..130M,2018A&A...616A.148B,2018MNRAS.478.2243S}. 
	One of the reasons for the limited number of modelled cases is that only the {\it Kepler\/} space telescope \citep{2010Sci...327..977B} offered sufficiently long time-series photometry to do so, and massive stars were avoided by it \citep[see][for a recent review]{2020FrASS...7...70B}. While time series covering a few months allow us to detect high-order gravito-inertial modes in main-sequence stars and to identify their mode wavenumbers $(l,m)$ from period spacing patterns \citep{2010A&A...519A..38D,2012A&A...542A..55P,2017A&A...603A..13K}, only the 4 yr {\it Kepler\/} light curves of B-type pulsators lead to sufficient frequency precision to discriminate among various profiles representing convective boundary mixing \citep[hereafter CBM;][]{2018A&A...614A.128P}. Another reason for the scarcity of asteroseismically modelled B-type pulsators is that few of those observed by {\it Kepler\/} have delivered suitable modes to perform modelling of their interiors. Finally, the methodology used to perform the modelling in the case of gravito-inertial modes is more complex than the framework based on modes in the perturbative regime of rotation \citep[see][for a review of the various modelling regimes]{Aerts2021}.

	Several prescriptions to describe the temperature gradient, $\nabla_T(r)$, and the diffusion coefficient defining the local chemical mixing profile in the convective core boundary layers, $D_{\rm mix}(r)$, exist. We consider two main options: firstly convective penetration with an adiabatic temperature gradient, $\nabla_{\rm ad}$, and a constant mixing profile in the core boundary layers, following \citet{1991A&A...252..179Z}, and secondly
	diffusive overshooting with a radiative temperature gradient, $\nabla_{\rm rad}$, 
	and an exponentially decaying CBM profile, following \citet{1996A&A...313..497F} and \citet{2000A&A...360..952H}. 
	Here we are concerned with convective boundary layers adjacent to a convective core in a main-sequence star. We take a data-driven approach based on observed gravito-inertial modes, which are excellent probes of the near-core regions deep inside stars, rather than relying on (uncalibrated) numerical simulations of these regions. 
	
	In a proof-of-concept study, \citet{2019A&A...628A..76M} showed that observed gravito-inertial modes hold the potential to discriminate between a radiative and an adiabatic temperature gradient, along with assessment of the mixing profile in the core boundary layers, provided that the mass, evolutionary stage, internal rotation, and metallicity are known. In this work we improve this initial theoretical study and apply it to one of the rotating B-type pulsators with a long series of prograde dipole gravito-inertial modes of consecutive radial order, KIC\,7760680. Our aim 
	is to assess the temperature and chemical gradients of the core boundary layers in this star from observations, but also to investigate the impact of various choices of asteroseismic observables in the modelling process on the inferred outcome. We do not focus on the observational aspects of the detected gravito-inertial modes because the frequency uncertainties are typically two orders of magnitude below those stemming from unknown input physics of stellar models \citep[see][]{2018ApJS..237...15A}. We thus consider the case of KIC\,7760680 by relying on its observational properties derived previously by \citet{2015ApJ...803L..25P}, and briefly recall the observational input of the modelling in the next section.

	\section{The gravito-inertial B-type pulsator KIC\texorpdfstring{\,}{TEXT}7760680} \label{sec:case_study}
	We have opted to use the slowly pulsating B star KIC\,7760680 as a case study for this paper. Our reasons for this choice are twofold. First, this is the only star rotating at a considerable fraction of its critical rotation rate \citep[about 26\%,][]{2016ApJ...823..130M} with an extensive identified mode period pattern (36 prograde dipole modes of consecutive radial order). Second, this star has a remarkably low level of inferred mixing in its envelope compared to slower rotating B-type pulsators, while one would expect its rotation to induce strong mixing beyond the convective core. The radial shear-induced mixing in differentially rotating massive stars is about three to ten orders of magnitude higher according to theoretical predictions than the envelope mixing found by \citet{2016ApJ...823..130M}. We refer to Fig.\,6 in \citet{2000A&A...361..101M}, Fig.\,3 in \citet{2004A&A...425..243M}, and Figs.\,15 and 16 in \citet{2009A&A...495..271D} for various examples. An explanation for this low envelope mixing proposed by \citet{2016ApJ...823..130M} is that the star is nearly a rigid-body rotator.
	
	Following up on the best model parameters for the star found by \citet{2016ApJ...823..130M} listed in \cref{tab:Moravveji2016},
	we assess whether a more structured CBM profile is meaningful and leads to a better fit to the star's observed oscillation modes. \citet{2016ApJ...823..130M} considered $\nabla_{\rm rad}$ in the CBM region, limited the parameter estimation to the use of a $\chi^2$ merit function, and performed mode frequency fitting. Given the mathematical challenges connected to it, no precision estimation for the parameters in \cref{tab:Moravveji2016} was done in this previous study. In the case of strong correlations among the parameters to estimate, as is the case here, the very small grid step size listed in \cref{tab:Moravveji2016} tends to be an underestimation of the true uncertainties because this neglects the parameter correlation structure.
	We aim to improve upon this omission of error estimation, while searching for a more accurate interior model of the star. This improvement concerns several aspects, namely the use of more optimized asteroseismic observables, a more appropriate merit function taking the correlation structure of the fitting problem into account, and a more structured $D_{\rm mix}(r)$ profile compared to the initial study by \citet{2016ApJ...823..130M}. Another reason to choose this star as a test case is that it has an excellent asteroseismic calibration of its near-core rotation frequency. \citet{2016ApJ...823..130M} derived a value of $0.4805$\,d$^{-1}$ under the assumption of rigid rotation, which corresponds to $26$\% of its critical Roche rotation frequency. To reduce dimensionality, we thus fixed this value of the rotation frequency. 
	
	\begin{table}
		\caption{Stellar parameters of the best asteroseismic model 
			of KIC\texorpdfstring{\,}{TEXT}7760680
			derived by \citet{2016ApJ...823..130M}.} 
		\label{tab:Moravveji2016}
		\centering
		\begin{tabular}{l c c}     
			\hline               
			Parameter & value & grid step size \\  
			\hline                        
			$M_{\rm ini}$ [\msol] & 3.25 & 0.05 \\   
			$Z_{\rm ini}$ & 0.020 & 0.001 \\
			$\fcbm$  & 0.024 &0.001 \\
			log($\D[env]$) & 0.75 & 0.25\\ 
			$X_c$ & 0.503 & 0.001 \\ 
			\hline
		\end{tabular}
		\begin{flushleft}
			\textbf{Notes.} From top to bottom, the parameters are  initial mass, metallicity, CBM parameter, envelope mixing, and central hydrogen content.
		\end{flushleft}
	\end{table}
	
	The {\it Kepler\/} light curve of this pulsator and its Fourier transform have already been discussed in great detail in \citet{2015ApJ...803L..25P}, so we do not repeat this information here for brevity. In practice, we use the period values of the 36 prograde dipole modes listed in Table\,1 of \citet{2015ApJ...803L..25P} as asteroseismic input for our modelling, $\Y{obs}$ composed of $\text{Y}_i^{\text{Obs}}$ with $i=1, \ldots, 36$. 
	\citet{2015ApJ...803L..25P} list the formal errors from a least-squares fit to the light curve, after  pre-whitening all the modes with amplitudes above four times the local noise level. \citet{1991MNRAS.253..198S} showed that the formal least-squares errors are statistically equivalent 
	to the errors derived in the Fourier domain if one takes into account the variance in the data, the signal-to-noise ratio, and the total time base of the data set. However, the author noted that a correction factor to the formal errors is required to take into account the correlated nature of the data, following their covariance structure. In line with the method in 
	\citet{1991MNRAS.253..198S}, we multiply the errors listed in 
	Table\,1 of \citet{2015ApJ...803L..25P} 
	by a correction factor of four, as was  done for the modelling performed by \citet{2016ApJ...823..130M}. This leads to errors on the mode periods lower than $\sim10\,$s, while the time resolution restriction stemming from the light curve that may have affected the choice of frequencies during the pre-whitening process amounts to $\sim60\,$s.
	
	As additional input to guide the modelling, we rely on T$_{\text{eff}}=11650 \pm 210 $K, log $g$=3.97 $\pm$ 0.08 dex, and [M/H]=0.14 $\pm$ 0.09 dex, again taken from \citet{2015ApJ...803L..25P}. We also rely on $\log\,(L/L_\odot)= 2.19\pm0.06$ based on Gaia DR2 astrometry deduced by \citet{2020MNRAS.495.2738P}.
	
	%%%%%%%%%%%%%%%%%%%%%%%%%%%%%%%%%%%%%%%%%%%%%%%%%%%%%%%%%%%%%%%%%%%%%%%%%%%%%%%%%%%%%%%%
	%%%%%%%%%%%%%%%%%%%%%%%%%%%%%%%%%%%%%%%%%%%%%%%%%%%%%%%%%%%%%%%%%%%%%%%%%%%%%%%%%%%%%%%%
	\section{Chemical mixing profiles}
	
	\begin{figure*}[htp]
		\centering 
		\begin{subfigure}{0.87\hsize}
			\includegraphics[width=\hsize]{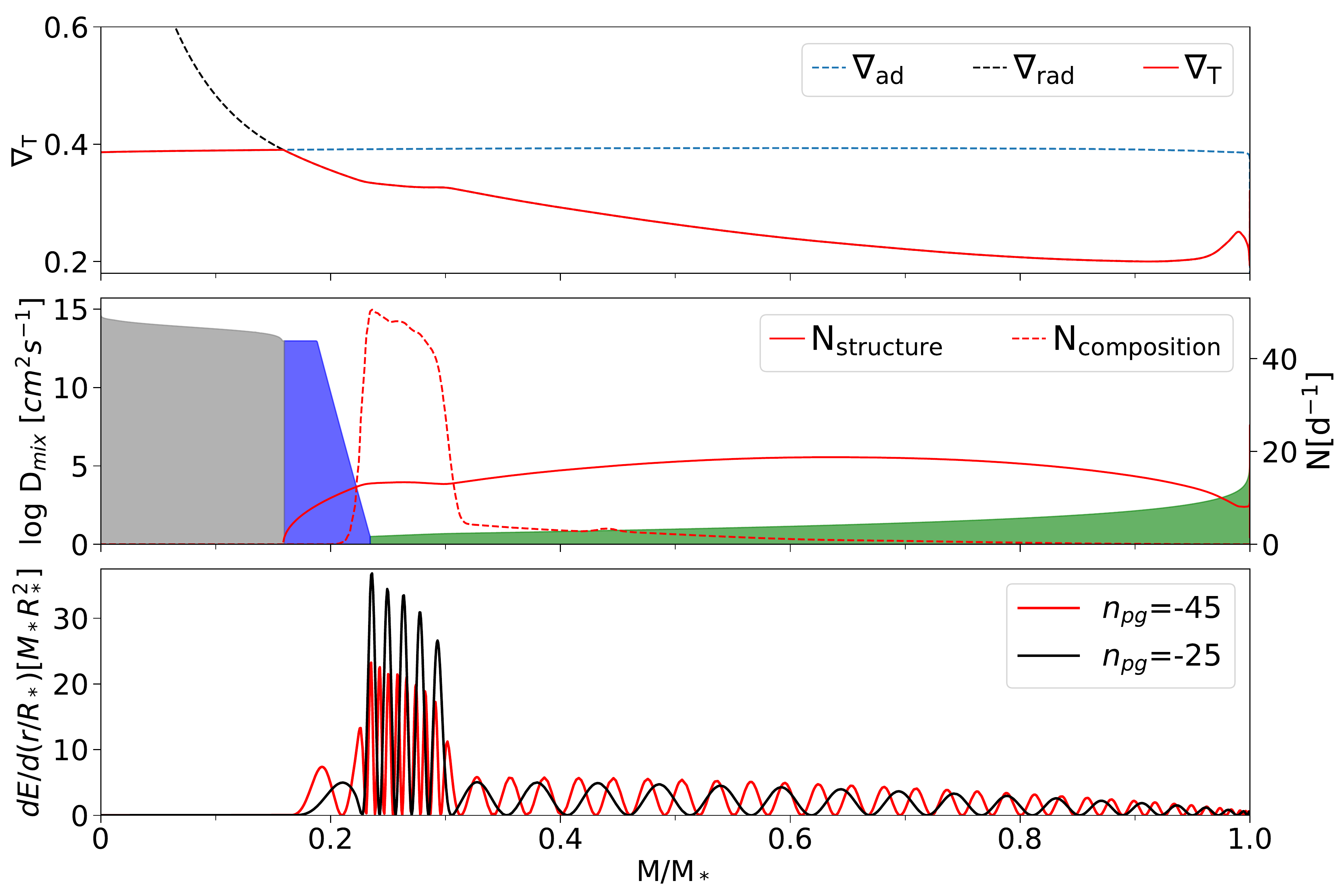}
			\caption{Near-core region with radiative temperature gradient}
			\label{fig:DO_profile}
		\end{subfigure}
		\begin{subfigure}{0.87\hsize}
			\includegraphics[width=\hsize]{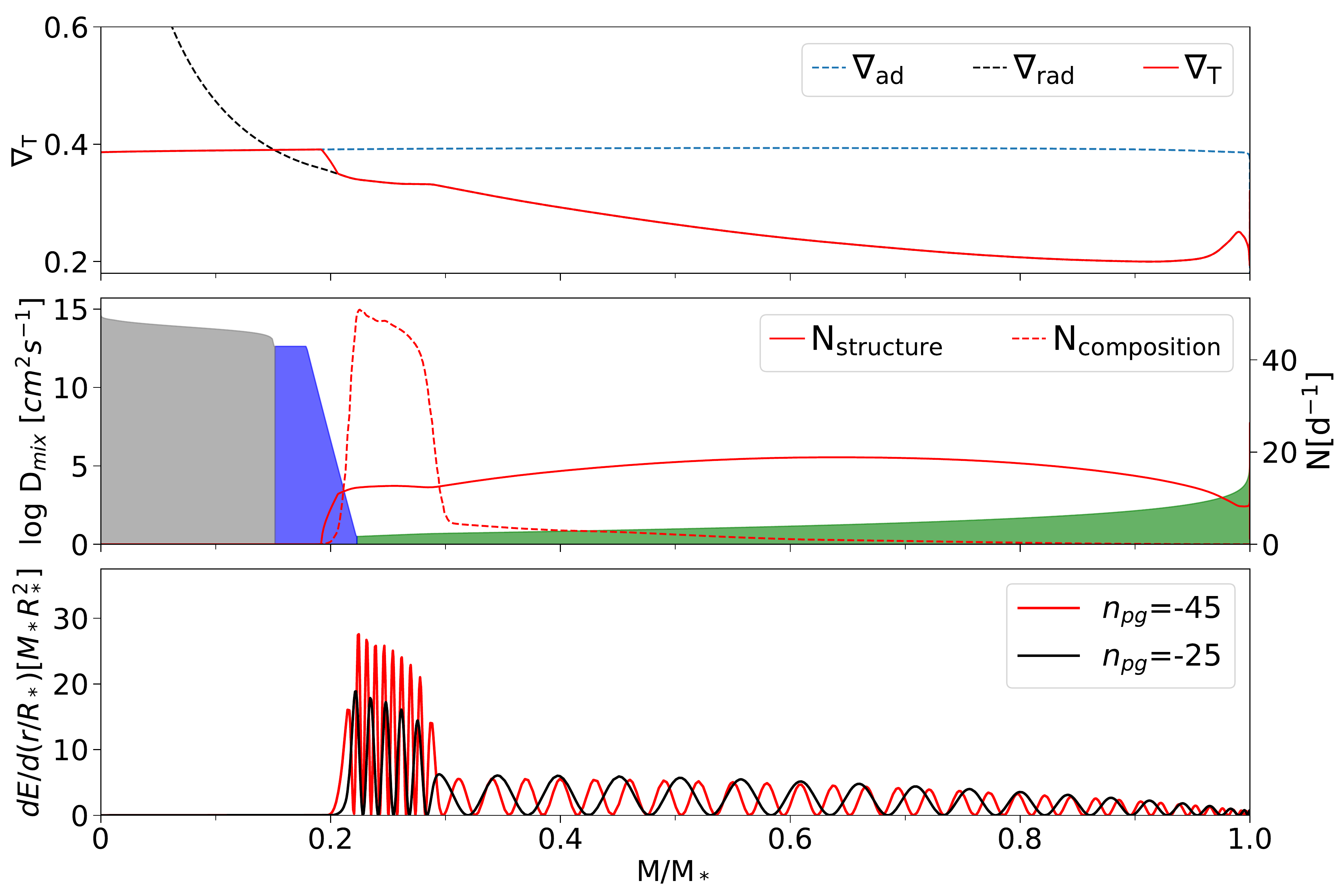}
			\caption{Near-core region with temperature gradient according to P\'eclet number}
			\label{fig:ECP_profile}
		\end{subfigure}
		\caption{Radial profiles of a 3\msol \:star halfway through the main sequence with a central hydrogen content $X_c=0.4$. The top panels show the temperature gradients. The middle panels show the structure (solid) and composition (dashed) components of the Brunt-V\"ais\"al\"a profile, as well as the shape of the mixing profiles, divided in convective core (grey), near-core mixing (blue), and diffusive mixing in the outer radiative envelope (green). The bottom panels show the mode inertia of two g-modes with different radial orders.}
		\label{fig:profiles}
	\end{figure*}
	
	In the diffusion approximation, the transport
	equation describing the rate of change in the mass fraction
	of each individual chemical element $i$ in the considered chemical mixture, denoted here as $X_i$, reads
	\begin{equation}
	\label{transport-eq}
	\displaystyle{\frac{\partial X_i(r)}{\partial t}} = \displaystyle{{\cal R}_i (r)}
	+ \displaystyle{\frac{1}{\rho (r) r^2} \frac{\partial}{\partial r} \left[\rho (r) r^2 
		D_{\rm mix} (r)
		\frac{\partial X_i (r)}{\partial r} \right],}
	\end{equation}
	with ${\cal R}_i (r)$ the change due to nuclear reactions and $D_{\rm mix} (r)$ the local mixing coefficient. In this work this mixing coefficient in the convective boundary layers is implemented as extended convective penetration described by \citet{2019A&A...628A..76M}, with an extra diffusive mixing profile in the radiative envelope due to internal gravity waves as deduced by \citet{2017ApJ...848L...1R}, following \citet{2018A&A...614A.128P}. The mixing profile outside of the regions determined to be convective by the Ledoux criterion can hence be described by
	\begin{align}
		\D[CBM](r) &= D_0 && \text{for} \quad  r_0<r<r_{\rm cp} ,\\
		\D[CBM](r) &= D_0 \exp{\lb \frac{-2 (r-r_{\rm cp})}{\fcbm \Hp }\rb}  &&\text{for} \quad  r_{\rm cp} < r < \rcbm, \label{eq:ext_exp_mix} \\
		\D[env](r)  &= \D[env] \frac{\rho (\rcbm)}{\rho (r)} && \text{for} \quad  \rcbm < r,
	\end{align}
	with $r_0 = \rcc - f_0 \Hp$ being the radius where the transition between core mixing and core boundary mixing is made, $\rcc$ being the radius of the convective core as obtained from the Ledoux criterion applied for mixing length theory \citep[MLT;][]{1958ZA.....46..108B}, and $\Hp$ the local pressure scale height; $D_0$ is the mixing coefficient at $r_0$, and is hence influenced by the choice of $f_0$.
	The edge of the step-like CBM in the transition region is given by $r_{\rm cp} = r_0+\acbm \Hp$. Beyond this value the mixing profile takes on an exponentially decaying function with free parameter $\fcbm$ until the core boundary transition region reaches the envelope of the star, whose bottom is situated at $\rcbm$ and where the envelope mixing becomes active rather than CBM. This corresponds to the radius where the convective boundary mixing coefficient becomes smaller than $\D[env]$, which stands for the mixing at the bottom of the radiative envelope. The latter mixing coefficient takes over as of this position towards the surface of the star. 
	The free parameters $\acbm$ and $\fcbm$ determine the extent of the step and exponentially decaying profile part of $D_{\rm mix}(r)$ in the core boundary region, respectively.
	
	Employing this prescription for the mixing profile, two cases are considered regarding the temperature gradient in the core boundary transition region. The first one (seen in \cref{fig:DO_profile}) adopts the radiative temperature gradient, $\n = \nrad$, outside the convective core ($r>r_0$). The second one (seen in \cref{fig:ECP_profile}) assumes that the
	overshooting material influences the entropy stratification in the transition layers, as proposed by \citet{1991A&A...252..179Z}.
	Instead of a transition in temperature gradient based on the mass coordinate in the convective boundary mixing region used in \citet{2019A&A...628A..76M}, we opt for a transition based on the P\'eclet number, as this is more physically motivated \citep[see e.g.][for an overview of the different convective regimes]{2015A&A...580A..61V}.
	The P\'eclet number is defined as the ratio of the advective transport rate to the diffusive transport rate, such that the flow can be considered diffusive when Pe $\ll$1 and adiabatic when Pe$\gg$1. Bridging these two regimes was implemented by making a gradual transition based on the P\'eclet number, in terms of a factor $h = (\log(\Pe)+2)/4$. This factor was constructed to create a small transition region between the regimes instead of introducing a discontinuity at the interface. This implementation causes Pe=1 to correspond to the temperature structure being halfway between $\nad$ and $\nrad$:
	\renewcommand{\arraystretch}{1.2}
	\begin{equation}
	\begin{array}{ll}
	\label{eq:peclet}
	\n = \nad\ & \text{for}\ \  \Pe>100 , \\
	\n = h\nad + (1-h)\nrad & \text{for}\ \  0.01<\Pe<100, \\
	\n = \nrad\ &\text{for}\ \  \Pe<0.01 .
	\end{array}
	\end{equation}
	One of our goals is to answer whether the 36 detected dipole gravito-inertial modes of KIC\,7760680 allow us to distinguish between these two options of temperature gradients in the transition layer between its convective core and radiative envelope, adopting the observational diagnostics of the star from
	\citet{2015ApJ...803L..25P}.
	
	%%%%%%%%%%%%%%%%%%%%%%%%%%%%%%%%%%%%%%%%%%%%%%%%%%%%%%%%%%%%%%%%%%%%%%%%%%%%%%%%%%%%%%%%
	\section{Computation of theoretical mode frequencies}
	\subsection{Stellar equilibrium models}
	
	Two grids of stellar models were computed as input for pulsation computations. These two grids provide equilibrium models with the same input physics, except for the adopted temperature gradient in the core boundary transition region. One grid adopts the radiative temperature gradient in that transition zone and is termed the radiative grid. For the other grid we adopt the temperature gradient according to \cref{eq:peclet} and we call it the P\'eclet grid.
	Both grids are identical in terms of the free parameter ranges listed in \cref{tab:parameters}. The ranges of the mass and central hydrogen fraction were guided by the earlier study of 
	\citet{2016ApJ...823..130M}, while the metallicity range relied on the spectroscopic results taken from \citet{2015ApJ...803L..25P}.
	
	The grids were computed using the stellar evolution code \texttt{MESA} \citep{2011ApJS..192....3P,2013ApJS..208....4P,2015ApJS..220...15P,2018ApJS..234...34P,2019ApJS..243...10P} version r12115.
	The \texttt{MESA} models contain the standard chemical mixture of OB stars in the solar neighbourhood deduced by \citet{2012A&A...539A.143N} and \citet{2013EAS....63...13P} and make use of the OP opacity tables \citep{2005MNRAS.362L...1S}. An Eddington grey atmosphere is used as atmospheric boundary condition. To set the initial mass fractions ($X_{\rm ini}, Y_{\rm ini}, Z_{\rm ini}$) we rely on the initial metallicity $Z_{\rm ini}$ as derived from the spectroscopic measurements derived by \citet{2015ApJ...803L..25P}, and vary it along its uncertainty range. The initial helium fraction is then determined by adopting an enrichment law $Y_{\rm ini} = Y_{p} + (\Delta Y/\Delta Z) Z_{\rm ini}$, where the primordial helium abundance, $Y_{p}$, and the galactic enrichment ratio, $\frac{\Delta Y}{\Delta Z}$, come into play. There is currently no consensus on the value of $\frac{\Delta Y}{\Delta Z}$ \citep[see e.g.][and references therein]{2019MNRAS.483.4678V}. Therefore, we have taken $Y_p = 0.2465$, as determined by \citet{2013JCAP...11..017A}, and we required that the enrichment ratio was able to reproduce the mass fractions of the adopted chemical mixture ($X$=0.71, $Y$=0.276, $Z$=0.014) derived by \citet{2012A&A...539A.143N}. This led us to adopt $\Delta Y/\Delta Z$=2.1. After $Y_{\rm ini}$ is determined according to this enrichment law, $X_{\rm ini}$ is set following $X_{\rm ini}=1-Y_{\rm ini}-Z_{\rm ini}$. We refer to \citet{2020MNRAS.500...54N} for an in-depth comparison between this approach of using an enrichment ratio, and including the initial helium abundance as a free variable.
	
	To determine where the transition from core to near-core mixing is made, we fixed $f_0=0.005$, with the exception of setting $f_0=0$ for the models where both $\acbm$ and $\fcbm$ are equal to zero as there is no CBM region for this case.
	The mixing length parameter was set to $\alpha_{\text{mlt}}=2.0$ in the mixing length theory as developed by \citet{1968pss..book.....C}. We used the Ledoux criterion for convection without allowing for semi-convection, since this form of slow mixing does not occur in the presence of convective boundary mixing \citep[see e.g.][]{2020MNRAS.496.1967K}, which is included in the vast majority of models in our grids. \cref{appendix:inlist} provides the link to the detailed \texttt{MESA} setup.
	
	\begin{table}
		\caption{Parameter ranges of each of the two grids of equilibrium models constituting of 117600 grid points used for the asteroseismic modelling of KIC\,7760680.} 
		\label{tab:parameters}
		\centering
		\begin{tabular}{l l l l}     
			\hline               
			Parameter & lower boundary & upper boundary & step size \\  
			\hline                        
			$M_{\rm ini}$ [\msol] & 2.8 & 3.7 & 0.1 \\      
			$Z_{\rm ini}$ & 0.015 &  0.023 & 0.004 \\
			$\acbm$  & 0 & 0.3 &  0.05 \\
			$\fcbm$   & 0 & 0.03 & 0.005 \\
			log($\D[env]$) & 0 & 2 &  0.5 \\ 
			$X_c$ & 0.3 & 0.6 &  0.02 \\ 
			\hline
		\end{tabular}
	\end{table}

	%%%%%%%%%%%%%%%%%%%%%%%%%%%%%%%%%%%%%%%%%%%%%%%%%%%%%%%%%%%%%%%%%%%%%%%%%%%%%%%%%%%%%%%%
	\subsection{Pulsation computations}
	To model the 36 consecutive dipole g~modes detected in KIC\,7760680, the pulsation mode properties of the MESA equilibrium models were computed employing the stellar oscillation code \texttt{GYRE} \citep{2013MNRAS.435.3406T,2018MNRAS.475..879T}, version 5.2. 
	We computed the dipole g-modes for a fixed rotation frequency of 0.4805\,d$^{-1}$, assuming rigid rotation as discussed previously. We relied on the traditional approximation of rotation \citep[TAR, following the implementation by][]{2018MNRAS.475..879T} and adopted the adiabatic approximation to compute the dipole gravito-inertial mode frequencies. The
	mode inertias of the considered g-modes are dominant near the core of the star, as illustrated in \cref{fig:profiles}. Hence the adiabatic approximation is sufficient for our modelling work, whilst non-adiabatic effects mainly become important in the outer stellar envelope. 
	
	\cref{fig:profiles} reveals the difference in temperature gradient between two models from the two MESA grids. Although it is relatively small and leaves the composition term of the Brunt-V\"ais\"al\"a frequency (N$_{\text{composition}}=\frac{g^2 \rho}{P} \nabla_{\mu}$) unchanged, it does change the thermal structure term (N$_{\text{structure}}=\frac{g^2 \rho}{P} (\nad-\n$)), which causes a different mode trapping. This is illustrated for the gravito-inertial modes with radial orders 25 and 45. 
	
	It is this type of difference in the mode kernels that brings about the potential to assess the temperature gradient in an observed star, provided that it delivers suitable modes to make this assessment. It was shown by \citet[][see their Fig.\,10]{2016ApJ...823..130M} that KIC\,7760680 is a suitable candidate to offer this opportunity.

	The final \texttt{GYRE} inlist is provided through the link in \cref{appendix:inlist} and provides us with a list of theoretically predicted prograde dipole mode frequencies, $\Y{Theo}$ composed of $\text{Y}_j^{\text{Theo}}$, where $j$ stands for the radial order, for each of the equilibrium models in the two grids. 
	
	%%%%%%%%%%%%%%%%%%%%%%%%%%%%%%%%%%%%%%%%%%%%%%%%%%%%%%%%%%%%%%%%%%%%%%%%%%%%%%%%%%%%%%%%
	%%%%%%%%%%%%%%%%%%%%%%%%%%%%%%%%%%%%%%%%%%%%%%%%%%%%%%%%%%%%%%%%%%%%%%%%%%%%%%%%%%%%%%%%
	\section{Modelling approaches}
	
	The asteroseismic modelling of a star involves various aspects, some of statistical nature and others connected with astrophysical considerations. The problem comes down to an estimation effort to assess the free parameters covered in the two model grids listed in \cref{tab:parameters}. We achieve this by relying on the list of observed gravito-inertial mode frequencies, or any other observables derived from them and denoted as $\Y{Obs}$ with $\text{Y}_i^{\text{Obs}}$ labelled as $i=1, \ldots, 36$, where the radial order of the observed modes is unknown a priori.
	These observables are matched to a list of 
	theoretically predicted analogues $\Y{Theo}$, with $\text{Y}_j^{\text{Theo}}$ to be computed from a list of model frequencies of radial order 
	$j=1, \ldots, 100$. There are various aspects of this matching process that have to be considered. Ultimately it comes down to solving regression problems, where various choices can be made for the observables to match and the merit functions to minimize. 
	We discuss each of these aspects in the following subsections, noting here that we do not use the spectroscopic and astrometric information as observables in the fitting procedure, but use the measured spectroscopic abundances to determine the range of initial metal and helium mass fractions.
	
	In the case of gravito-inertial modes of B-type pulsators with 4 yr {\it Kepler\/} light curves, we are in the situation where non-seismic observables from spectroscopy or astrometry, such as $T_{\rm eff}$, $\log\,g$, or 
	$\log\,(L/L_\odot)$ have relative observational uncertainties that are typically two orders of magnitude or more above those of the measured frequencies \citep[see Table\,1 in][]{2019ARA&A..57...35A}. For this reason, these classical observables are not included in the merit functions, but rather serve to eliminate stellar models in the two grids having values of these non-seismic observables outside the observed regime \citep[see e.g.][]{2018A&A...616A.148B}.
	
	\subsection{Model selection criterion}
	
	The asteroseismic modelling involves maximum likelihood estimation based on the chosen observables and their theoretically predicted counterparts computed from astrophysical models. The modelling may involve nested or non-nested statistical models \citep[see][]{Claeskens2008}. 
	We are dealing with both cases. Firstly a comparison between the performance of models within one grid of equilibrium models where one or both of the CBM parameters are fixed at zero ($\acbm, \fcbm$, or both) versus a free parameter to estimate. Such a comparison constitutes a nested model comparison allowing the most appropriate statistical model to be selected. We are doing this model selection as a test to evaluate whether the added complexity in the core boundary transition region of the models improves the fit to the data more than the entailed punishment by the selection criterion for having additional free parameters. Secondly a comparison between the performance of the radiative versus P\'eclet grids while estimating the same free parameters for both grids deals with a model selection based on non-nested models as only the input physics changes, but not the degrees of freedom.
	
	In the case of model selection for nested statistical models, we use the Akaike Information Criterion corrected for small sample size \citep[AICc,][Chapter\,2]{Claeskens2008}. 
	This type of model selection rewards fit quality, but penalizes complexity. Moreover, the
	AICc is an appropriate criterion to evaluate nested models with a limited number of observables yet considerable dimensionality of the free parameter vector, as we have here (36 modes to estimate six free parameters, see  \cref{tab:parameters}).
	The AICc is defined as
	\begin{equation}\label{eq:aicc}
	\text{AICc} = -2\ln{\cal L} + \frac{2kN}{N-k-1}
	,\end{equation}
	with $N$ and $k$ being the number of observables and free parameters, respectively, and ${\cal L}$ being the likelihood of a stellar model.
	
	In our framework of asteroseismic modelling, $N=36$ or 35 depending on whether we choose to fit mode frequencies,  periods, or period spacings.
	The number of free parameters $k$ is according to the parameters varied in the grids and is 4, 5, or 6 in our case, which means that  
	($M_{\rm ini}, Z_{\rm ini}, \acbm, \fcbm, \log(\D[env]), X_c$) are used to assess the general performance of the two grids, and 
	the same list but with only one or neither of ($\acbm, \fcbm$) included to test if increased CBM complexity performs better or worse within one grid. In this latter application, we make use of the property that two nested models A and B can be compared in performance through their difference in AICc; if $\Delta \text{AICc} = \text{AICc}_\text{A} - \text{AICc}_\text{B} >2$, model B is favoured over model A, with the evidence being (very) strong if $\Delta \text{AICc} > 6$ (10) \citep{1955...Kass,Claeskens2008}.
	These regimes are derived from the value of $2\ln[P(D|A)/P(D|B)]$, with $P(D|X)$ the probability of observing $D$ given model $X$. A difference in $\Delta \text{AICc}$ of 6 or 10 implies that the probability of reproducing the data better with model A than with model B is roughly 95\% or 99.3\%, respectively.
	%%%%%%%%%%%%%%%%%%%%%%%%%%%%%%%%%%%%%%%%%%%%%%%%%%%%%%%%%%%%%%%%%%%%%%%%%%%%%%
	
	\subsection{Merit functions: Mahalanobis distance versus \texorpdfstring{$\chi^2$}{TEXT}}
	
	Uncalibrated physical ingredients required as input physics of stellar evolution models are usually dealt with via the inclusion of free parameters. Often these ingredients are correlated. Thus,
	the systematic uncertainties originating from these limitations in the model input physics cause a variance--covariance structure in the theoretically predicted observables. 
	This, as well as non-linear correlations among the observables and hence in the overall correlated nature of the estimation problem to solve, has to be addressed in the regression problem.
	
	Since the beginnings of space asteroseismology, Bayesian inferences have been developed and applied in the modelling of solar-like oscillations of low-mass stars \citep[e.g.][]{2009A&A...506....1A,2013MNRAS.429.3645S,2013ApJ...769..141S,2014ApJS..214...27M,2015MNRAS.452.2127S,2017ApJ...835..173S}. In such applications the Coriolis acceleration is ignored in the prediction of the oscillation frequencies.
	The asteroseismic observables of rotating stars with a convective core and gravito-inertial modes 
	demand the inclusion of the Coriolis acceleration in a non-perturbative way, implying that the g-mode frequencies are strongly dependent on the rotational frequency. Moreover, the observables are also correlated in a non-linear way, for example 
	the modes are often strongly trapped in the near-core region \citep{Papics2017}.
	In addition, in the mass regime under study here,
	several of the free parameters of the equilibrium models to be estimated are also strongly correlated, following tight 
	and sometimes non-linear relationships, such as between mass and metallicity, convective-core-mass and mass, and convective-core-mass and age  \citep[e.g.][]{2009pfer.book.....M}.

	The non-linear correlations among the free parameters and observables, along with the inclusion of the uncertainties for the theoretical predictions in the estimation problem lend itself to use a more appropriate merit function 
	instead of the traditional $\chi^2$ (whose results we  include in \cref{app:chi2} for comparison). Here we take the full variance--covariance structure among the observables and the theoretical predictions into account, thereby mitigating the aforementioned shortcomings. This is achieved by building a regression model via the use of the so-called Mahalanobis distance  \citep[see][for its application to asteroseismic modelling]{2018ApJS..237...15A},
	\begin{align}
		\text{MD}_j = \lb \Y{theo}_{j} - \Y{obs} \rb ^T \lb V+\Sigma \rb ^{-1} \lb \Y{theo}_{j} - \Y{obs} \rb,
	\end{align}
	where $\Y{obs}$ is the vector of observables, $\Y{theo}_j$ is the corresponding vector of predicted values in gridpoint $j$ of the grids of equilibrium models, $V$ is the variance--covariance matrix of $\Y{theo}$ due to the unknown physical ingredients in the input physics of the equilibrium models, and $\Sigma$ is the variance due to the measurement errors of $\Y{obs}$. A major advantage of this approach is that it allows   the inclusion of theoretical uncertainties for the predictions $\Y{theo}$ used in the modelling.
	
	The likelihood function corresponding to the Mahalanobis distance to be used in the statistical model selection is assumed to adhere to normality \citep[][for arguments why this is a good approach]{2018ApJS..237...15A}. 
	Using the likelihood function of the multivariate normal distribution, we find the likelihood of parameters $\boldsymbol{\theta} = (\theta^1, \theta^2 ... \theta^k)$ given the observed data $\mathbf{D}$,
	
	\begin{align}
		\cal L (\boldsymbol{\theta} | \mathbf{D} ) = & \exp \left( \vphantom -\frac{1}{2} \left( \vphantom{\lb \sum \rb} \ln(|V+\Sigma|)  + k\ln(2\pi)  \right.\right. \nonumber \\ %vphantom is to make the brackets equal accross the line break
		&+ \left.\left. \lb \Y{theo} - \Y{obs} \rb ^T \lb V+\Sigma \rb ^{-1} \lb \Y{theo} - \Y{obs} \rb   \right) \right) \nonumber \\
		= &\exp \lb - \frac{1}{2} \lb \ln(|V+\Sigma|) + k \ln(2\pi) +\text{MD} \rb \rb,
	\end{align}
	which is proportional to the probability density function 
	\begin{align}
		\cal L (\boldsymbol{\theta} | \mathbf{D} ) \propto P( \mathbf{D} | \boldsymbol{\theta} ).
	\end{align}
	We can write the AICc in this case as 
	\begin{equation}
	\text{AICc}  = \ln(|V+\Sigma|) + k\ln(2\pi) + \text{MD}  + \frac{2kN}{N-k-1}.
	\end{equation}
	
	To determine the uncertainty region of the best solution, we use the likelihood in combination with Bayes' theorem, stating that the probability of a parameter $\theta^m$ occurring in the interval |$\theta^m_a$, $\theta^m_b$| is given by
	\begin{align}
		P(\theta^m_a < \theta^m < \theta^m_b | \mathbf{D}) =& \frac{\sum_i^q P(\mathbf{D}|\boldsymbol{\theta}_i) P(\boldsymbol{\theta}_i) }{\sum_j^Q P(\mathbf{D}|\boldsymbol{\theta}_j) P(\boldsymbol{\theta}_j) } \nonumber \\
		=& \frac{\sum_i^q P(\mathbf{D}|\boldsymbol{\theta}_i) \prod_l^k P({\theta}_i^l) }{\sum_j^Q P(\mathbf{D}|\boldsymbol{\theta}_j) \prod_l^k P({\theta}_j^l) }.
	\end{align}
	Here the index $j$ is summed over all $Q$ equilibrium models in the grid that are consistent within 3$\sigma$ of the spectroscopic observables, and index $i$ is summed over the $q$ models with the highest likelihood so that $P(\theta^m_a < \theta^m < \theta^m_b | \mathbf{D}) = 0.95$. Sometimes only one equilibrium model in the grid fulfils the requirement to reach this probability of 0.95. In that case we consider the uncertainty region to span a smaller area than the one defined by the grid step. This leaves us with the step sizes of the grid in each of its dimensions as an upper limit of the uncertainty region in that particular case.
	
	%%%%%%%%%%%%%%%%%%%%%%%%%%%%%%%%%%%%%%%%%%%%%%%%%%%%%%%%%%%%%%%%%%%%%%%%%%%%%%%%%%%%%%%%%%%%%%
	\subsection{Observables to fit}
	We investigate which is the optimal set of observables to be used in the stellar modelling procedure.
	In addition to  just the mode periods,  we also investigate the period spacing values (i.e. the differences in period between two pulsation modes of consecutive radial order, $\Delta P_n\equiv P_{n+1} - P_n$).
	The best of these sets of observables is determined from the condition numbers of the variance--covariance matrices $V+\Sigma$ following the Mahalanobis distance definition.
	
	In general, the condition number $\kappa$ of a matrix $A$
	is defined as the ratio of its maximum to minimum eigenvalue,
	\begin{equation} \label{eq:cond_nr}
	\kappa(A) = \frac{|\lambda_{max}(A)|}{|\lambda_{min}(A)|},
	\end{equation}
	and gives a handle on how well- or ill-conditioned the matrix is with respect to the inversion to be computed. 
	A rule of thumb is that $\log(\kappa)$ gives an estimate of the number of digits of accuracy lost in addition to  the loss due to arithmetic methods.
	Given this, we search for the vector of asteroseismic observables that gives the lowest condition numbers.
	
	%%%%%%%%%%%%%%%%%%%%%%%%%%%%%%%%%%%%%%%%%%%%%%%%%%%%%%%%%%%%%%%%%%%%%%%%%%%%%%%%%%%%%%%%%%%%%%
	\subsection{Constructing the theoretical period spacing pattern}
	When modelling an observed pattern of gravito-inertial modes of consecutive radial order, the theoretical mode periods are typically matched to the observed ones, as inspired by asymptotic relations for gravito-inertial modes \citep{2010aste.book.....A}. However, different choices can be made to determine the starting point to build the mode pattern. Three such choices are compared in modelling performance in this paper. In the first we begin matching mode periods starting from the theoretical period that is closest to the highest-frequency mode detected in the observed pattern. A second option is to build the pattern starting from the mode with the highest observed amplitude.
	The third method is to match each observed mode period to its best matching theoretical counterpart, and adopt the longest sequence of consecutive modes that we get in this way. In the case of multiple mode series with the same length, a final pattern selection is made based on the best match between theory and observations. We henceforth refer to these three options of pattern construction as highest frequency, highest amplitude, and longest sequence, respectively.
	
	%%%%%%%%%%%%%%%%%%%%%%%%%%%%%%%%%%%%%%%%%%%%%%%%%%%%%%%%%%%%%%%%%%%%%%%%%%%%%%%%%%%%%%%%%%%%%%
	%%%%%%%%%%%%%%%%%%%%%%%%%%%%%%%%%%%%%%%%%%%%%%%%%%%%%%%%%%%%%%%%%%%%%%%%%%%%%%%%%%%%%%%%%%%%%%
	\section{Modelling results}
	
	%%%%%%%%%%%%%%%%%%%%%%%%%%%%%%%%%%%%%%%%%%%%%%%%%%%%%%%%%%%%%%%%%%%%%%%%%%%%%%%%%%%%%%%%%%%%%%
	\subsection{Selected observables to fit}
	
	When calculating the condition number of the variance--covariance matrices according to \cref{eq:cond_nr}, we find values on the order of $10^6$ when we solely consider the observed mode periods to fit. The variance--covariance matrices for this case, which express the coverage of the mode periods across the model grids, are shown for the two grids, and for the three ways of constructing the period spacing pattern in \cref{fig:Vmatrix_DO_chisq_longest_sequence_MD_P,fig:Vmatrix_DO_highest_amplitude_MD_P,fig:Vmatrix_DO_highest_frequency_MD_P,fig:Vmatrix_ECP_chisq_longest_sequence_MD_P,fig:Vmatrix_ECP_highest_amplitude_MD_P,fig:Vmatrix_ECP_highest_frequency_MD_P}. Variances in the theoretical mode period predictions across the entire grids reach up to $\sim$3400\,s in this case.
	This is about 50 times larger than the observational uncertainties for the mode periods due to the time resolution and noise properties added together.
	
	On the other hand, the variance--covariance matrices for the case of theoretically predicted  period spacing values  shown in \cref{fig:Vmatrix_DO_chisq_longest_sequence_MD_dP,fig:Vmatrix_DO_highest_amplitude_MD_dP,fig:Vmatrix_DO_highest_frequency_MD_dP,fig:Vmatrix_ECP_chisq_longest_sequence_MD_dP,fig:Vmatrix_ECP_highest_amplitude_MD_dP,fig:Vmatrix_ECP_highest_frequency_MD_dP}
	cover up to $\sim$400\,s in variance for the theoretical predictions of the spacings, which is about 20 times larger than the formal uncertainties on the observed period spacing values quoted by \citet{2015ApJ...803L..25P} and \citet{2016ApJ...823..130M}.
	When we fit the period spacing values in the detected pattern instead of the periods, the condition numbers significantly improve, yielding $\kappa(A) \sim 10^3$.
	We therefore conclude that the period spacings are the optimal set of observables to fit for our case study of KIC\,7760680. 
	
\begin{figure}[!htp]
		\centering
		\includegraphics[width=\hsize]{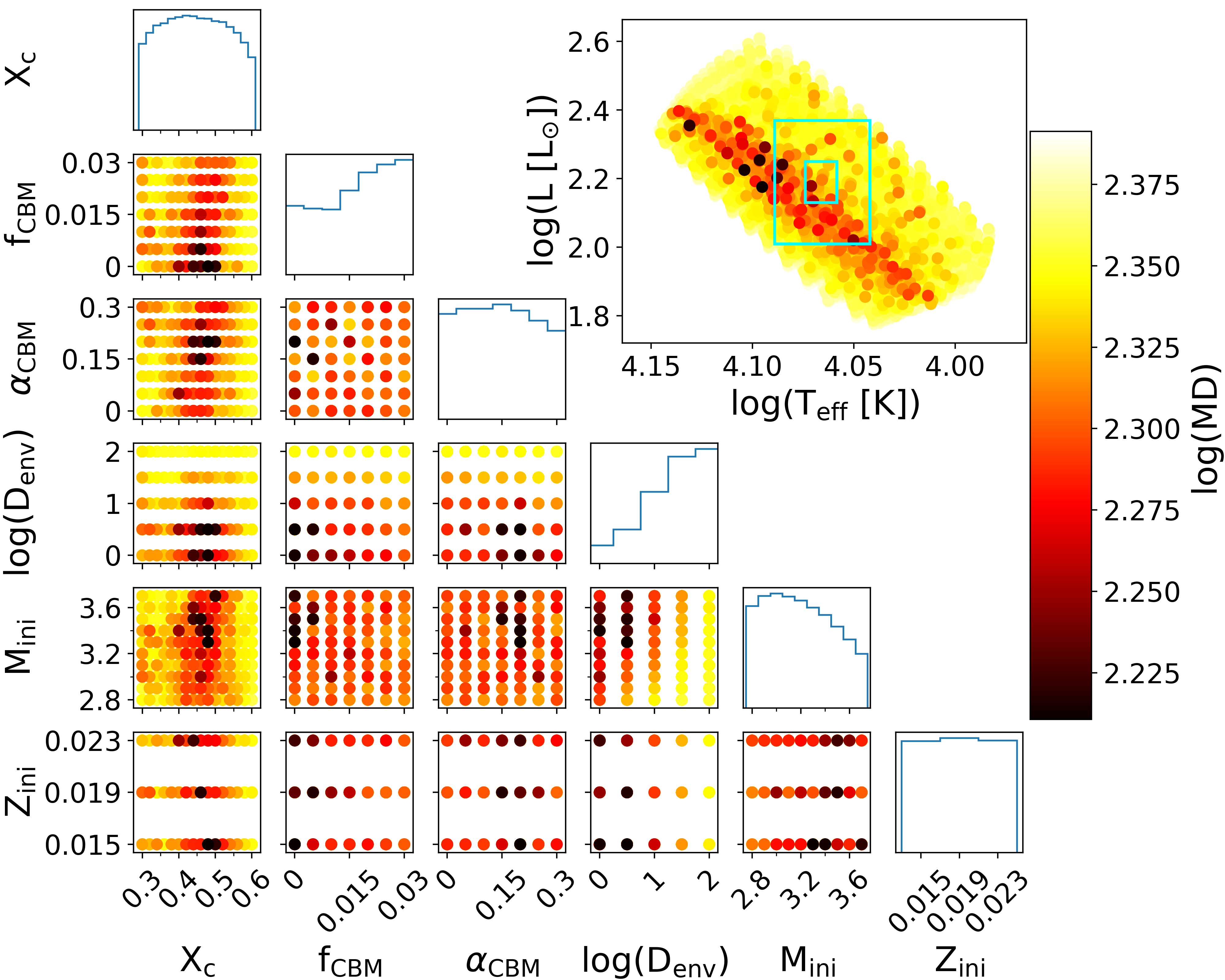}
		\caption{Correlation plot for the radiative grid, using periods in a Mahalanobis distance merit function. The theoretical pulsation pattern was constructed according to the longest sequence method. The 50\% best models are shown, colour-coded according to the log of their merit function value (at right). The figures on the diagonal show binned parameter distributions, and the larger panel  at the top right shows a Hertzsprung–Russell diagram with the 1 and 3$\sigma$ spectroscopic $T_{\rm eff}$ and astrometric luminosity error boxes.}
		\label{fig:DO_chisq_longest_sequence_MD_P}
	\end{figure}

\begin{figure}[!htp]
		\centering
		\includegraphics[width=\hsize]{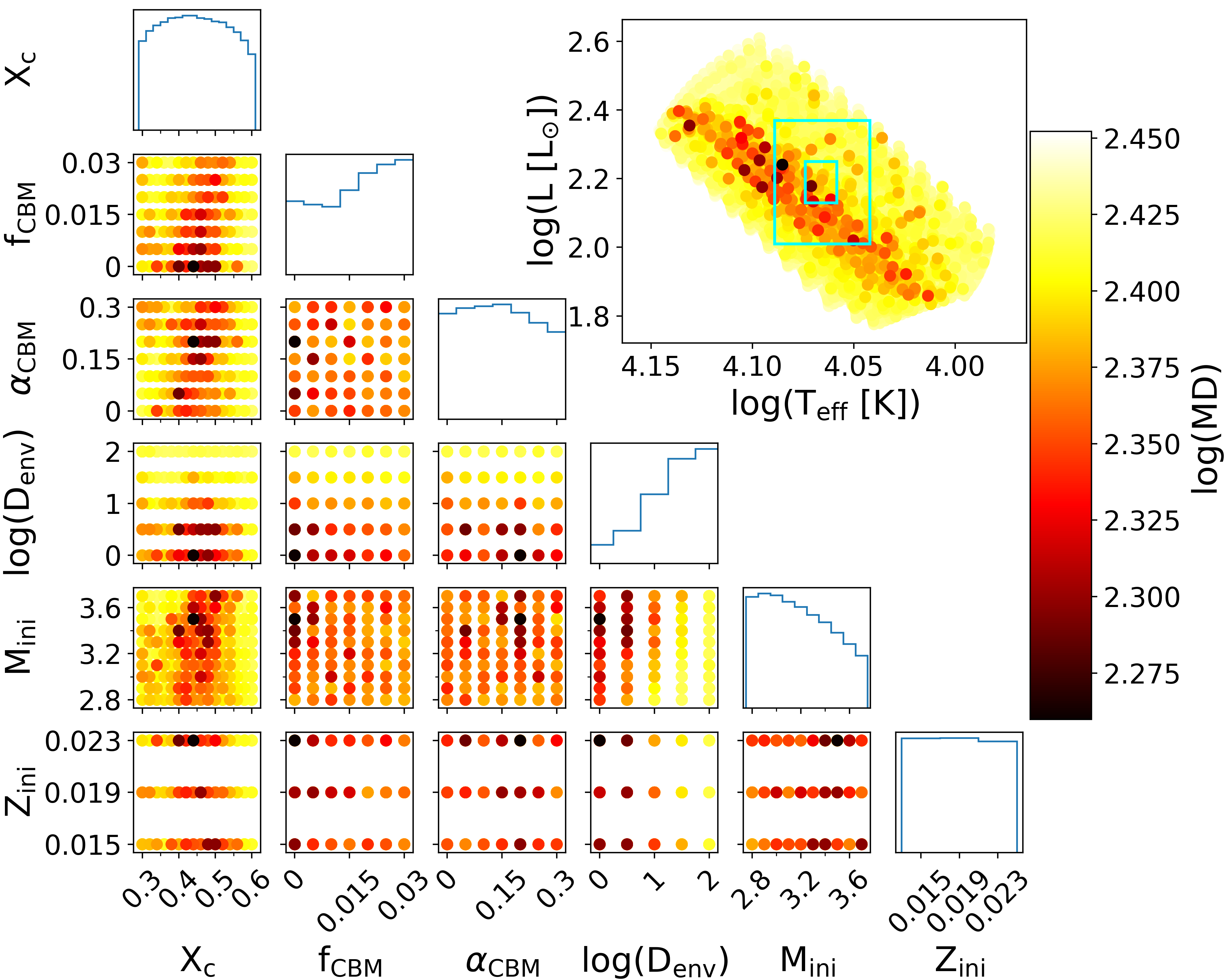}
		\caption{ Correlation plot (see \cref{fig:DO_chisq_longest_sequence_MD_P}) for the radiative grid, using period spacings in a Mahalanobis distance merit function. The theoretical pulsation pattern was constructed according to the longest sequence method.}
		\label{fig:DO_chisq_longest_sequence_MD_dP}
	\end{figure}

	The results for both sets of observables are very similar for this star in terms of the correlations and constraints on the optimal solutions. The correlation structures of the radiative grid are shown in \cref{fig:DO_chisq_longest_sequence_MD_P,fig:DO_chisq_longest_sequence_MD_dP} for the two sets of observables.
	While the correlation structures in these figures are similar, the condition numbers for the cases of the period spacing values as observables are far lower than those for the periods as observables, and hence constitute the best modelling setup in terms of numerical stability. 
	
	The better suitability of the period spacing values to fit as observables compared to the mode periods themselves also has advantages from the input physics point of view.  The mode periods are subject to systematic uncertainty connected to various choices for the chemical mixture and opacities in B-type pulsators, as illustrated in  \citet[][their Fig.\,3]{2015A&A...580A..27M}. This systematic uncertainty decreases drastically when considering mode period or mode frequency differences. 
	Similarly, dipole gravito-inertial modes of $\gamma\,$Doradus pulsators computed from equilibrium models with atomic diffusion are globally shifted compared to those without taking atomic diffusion and radiative levitation into account \citep{2020ApJ...895...51M}. The mode period shift due to atomic diffusion is illustrated graphically in Fig.\,5 of \citet{Aerts2021}. 
	
	We note that \citet{2016ApJ...823..130M} fitted the mode frequencies in their modelling of KIC\,7760680. Just as for the case of considering mode periods, their procedure makes the modelling result more subject to uncertainties in the input physics of the equilibrium models than our new approach in this work. This is particularly so because these authors did not include the covariance among the free parameters to estimate in their modelling. The results in our paper improve the asteroseismic modelling of KIC\,7760680 on these two fronts, also providing precision estimation that was not attempted in the first modelling efforts for this g-mode pulsator.
	
	\begin{figure}[!htp]
		\centering
		\includegraphics[width=\hsize]{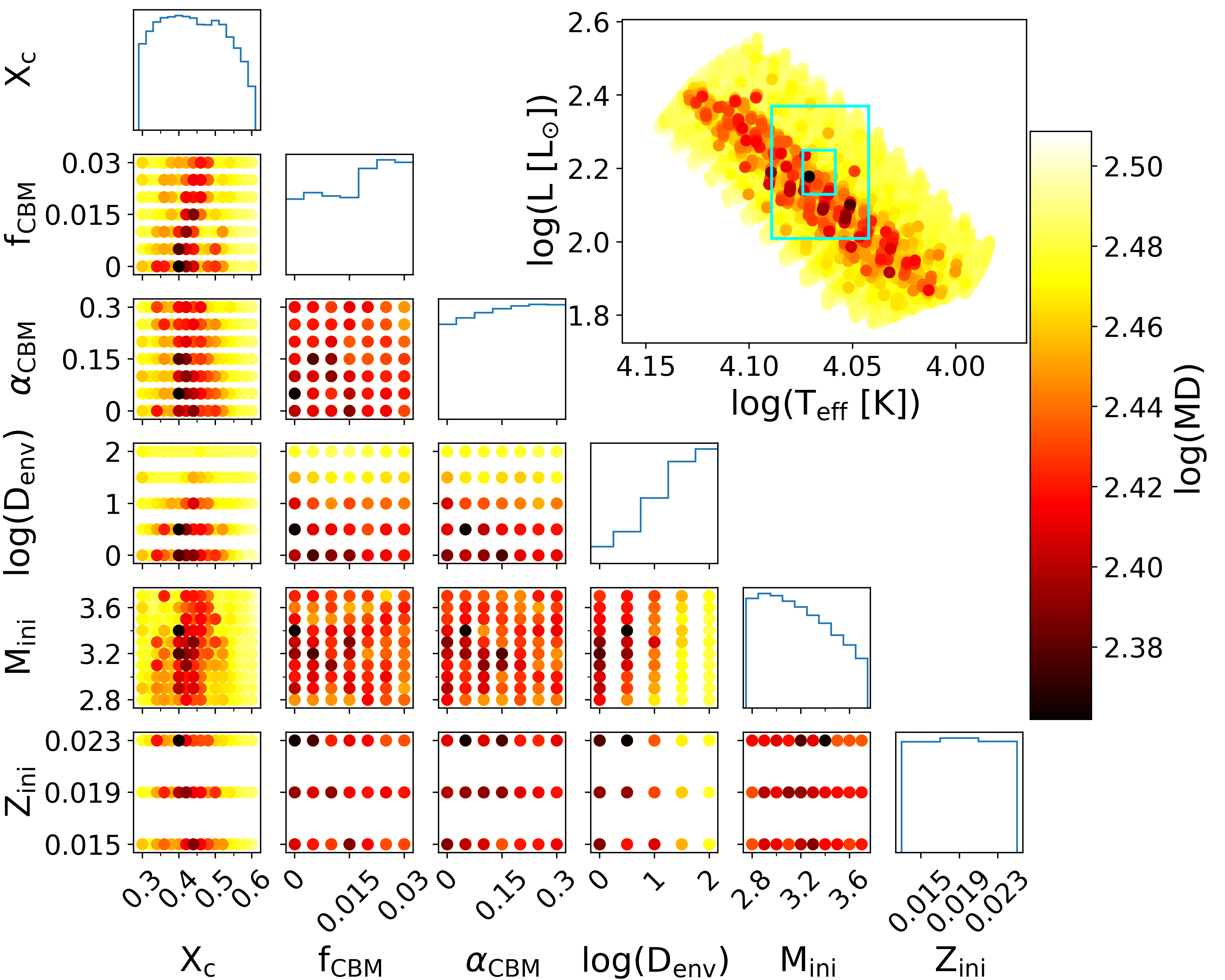}
		\caption{ Correlation plot (see \cref{fig:DO_chisq_longest_sequence_MD_P}) for the P\'eclet grid, using period spacings in a Mahalanobis distance merit function. The theoretical pulsation pattern was constructed according to the longest sequence method.}
		\label{fig:ECP_chisq_longest_sequence_MD_dP}
	\end{figure}
	
	\begin{figure}[!htp]
		\centering
		\includegraphics[width=\hsize]{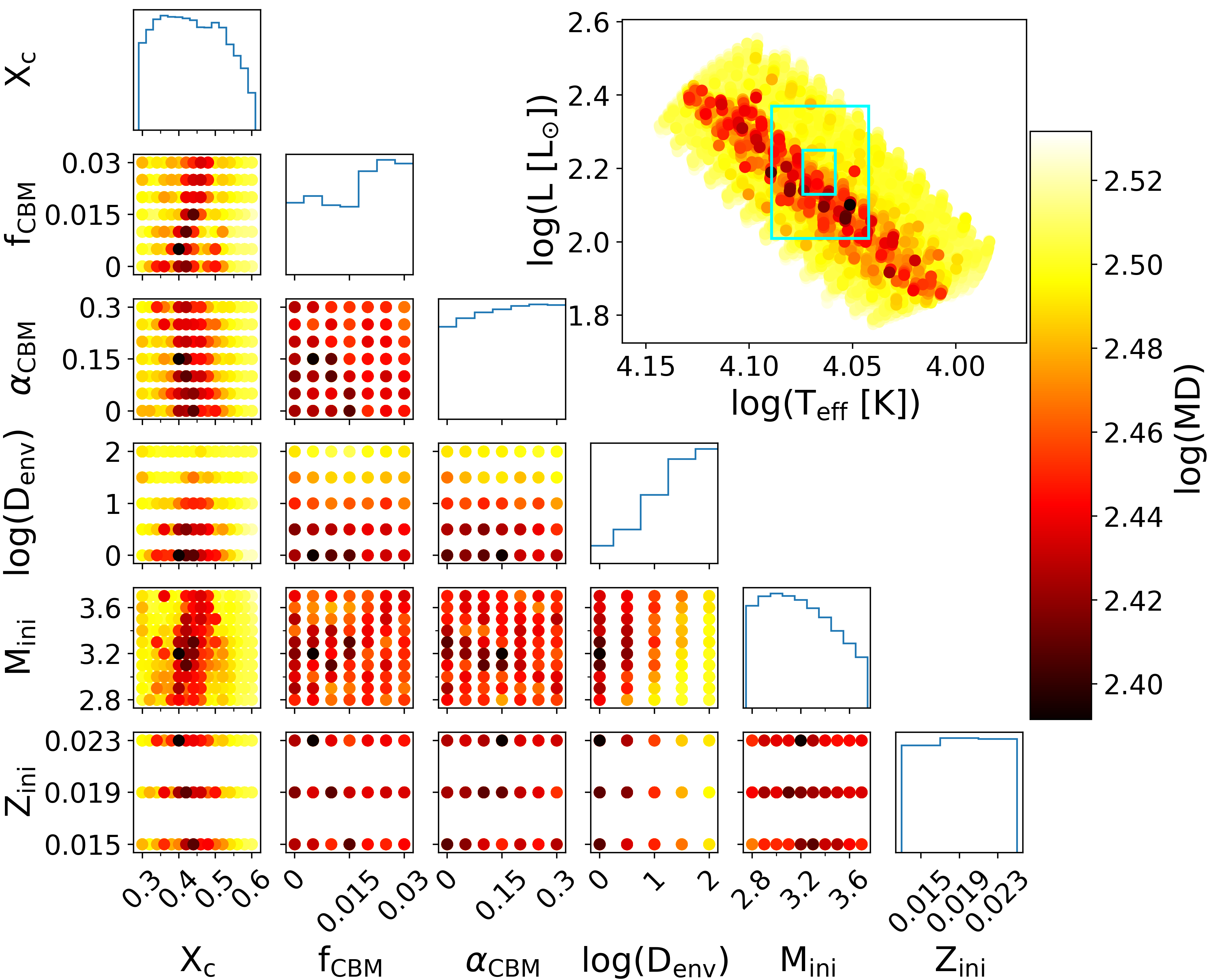}
		\caption{ Correlation plot (see \cref{fig:DO_chisq_longest_sequence_MD_P}) for the P\'eclet grid, using period spacings in a Mahalanobis distance merit function. The theoretical pulsation pattern was constructed according to the highest amplitude method.}
		\label{fig:ECP_highest_amplitude_MD_dP}
	\end{figure}
	
	\begin{figure}[!htp]
		\centering
		\includegraphics[width=\hsize]{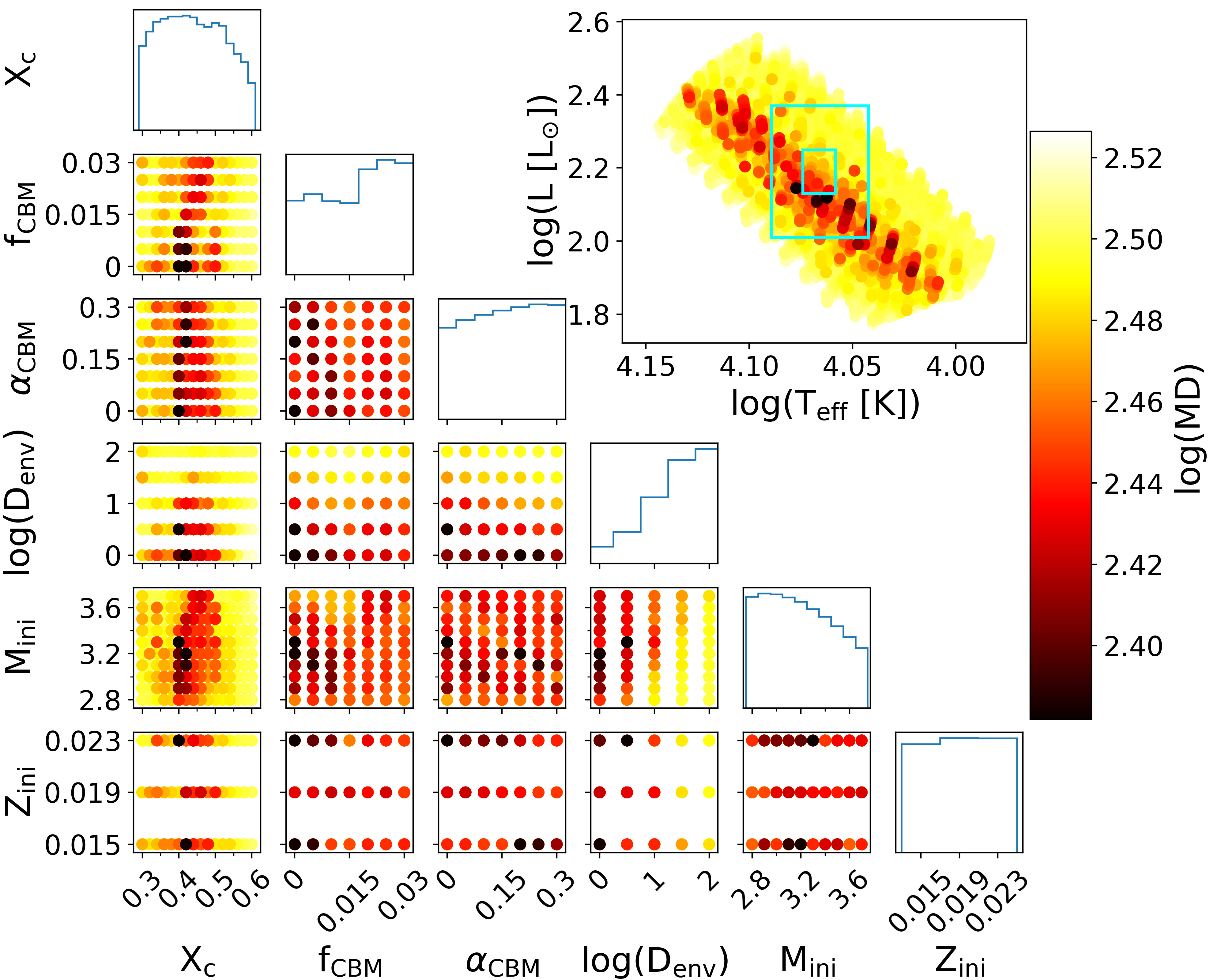}
		\caption{ Correlation plot (see \cref{fig:DO_chisq_longest_sequence_MD_P}) for the P\'eclet grid, using period spacings in a Mahalanobis distance merit function. The theoretical pulsation pattern was constructed according to the highest frequency method.}
		\label{fig:ECP_highest_frequency_MD_dP}
	\end{figure}
	
	%%%%%%%%%%%%%%%%%%%%%%%%%%%%%%%%%%%%%%%%%%%%%%%%%%%%%%%%%%%%%%%%%%%%%%%%%%%%%%%%%%%%%%%%%%%%%%
	\subsection{Theoretical mode period pattern construction}
	
	The Mahalanobis distance (MD) yields the same best model regardless of the method used for the theoretical pattern construction in the radiative grid. The MD and AICc values for this grid change slightly (\cref{tab:best_models_MD}) due to the changed correlation structure, as can be seen in the variance--covariance matrices in \cref{fig:Vmatrix_DO_chisq_longest_sequence_MD_dP,fig:Vmatrix_DO_highest_amplitude_MD_dP,fig:Vmatrix_DO_highest_frequency_MD_dP}, but they are relatively small differences. The highest amplitude method is slightly disfavoured to compose the pattern compared to the other two in terms of their MD values.
	
\begin{figure}[ht]
		\centering
		\includegraphics[width=\hsize]{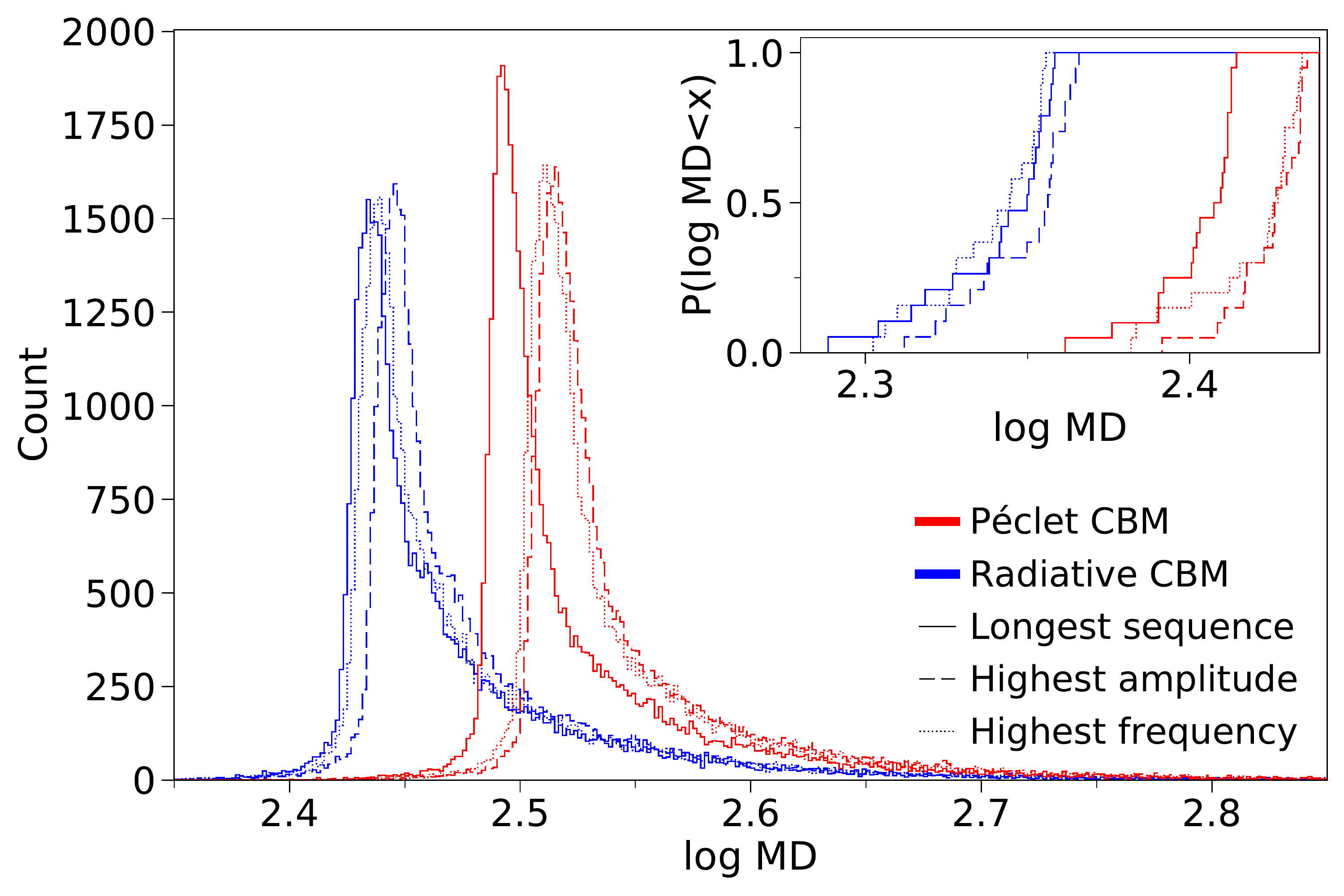}
		\caption{ Distribution of the Mahalanobis distances using $\Delta P$ as observables to fit. The inset shows the cumulative distribution function of the 20 best models in the grids.}
		\label{fig:pattern_cdf_MD_dP}
	\end{figure}

\begin{figure*}[ht]
		\centering 
		\begin{subfigure}{\hsize}
			\includegraphics[width=\hsize]{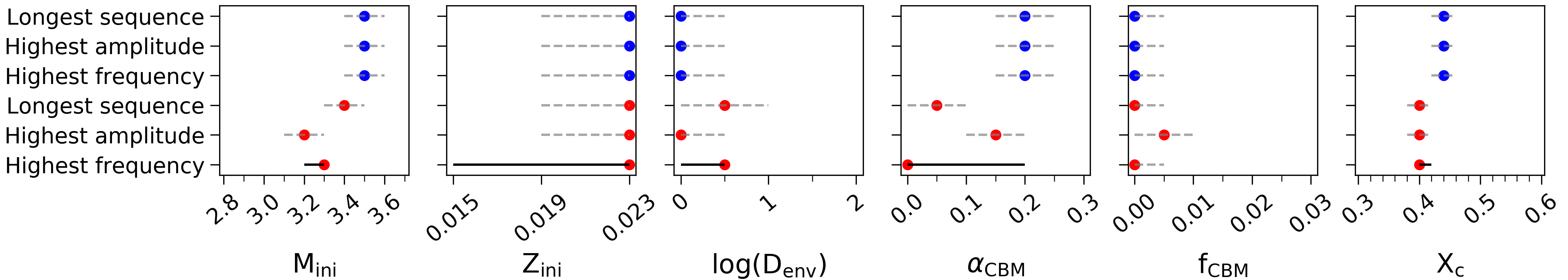}
		\end{subfigure}
		\begin{subfigure}{\hsize}
			\includegraphics[width=\hsize]{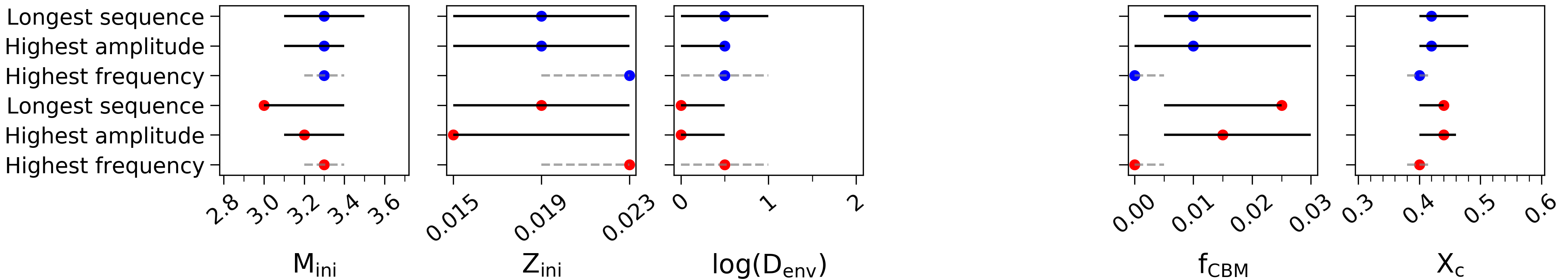}
		\end{subfigure}
		\begin{subfigure}{\hsize}
			\includegraphics[width=\hsize]{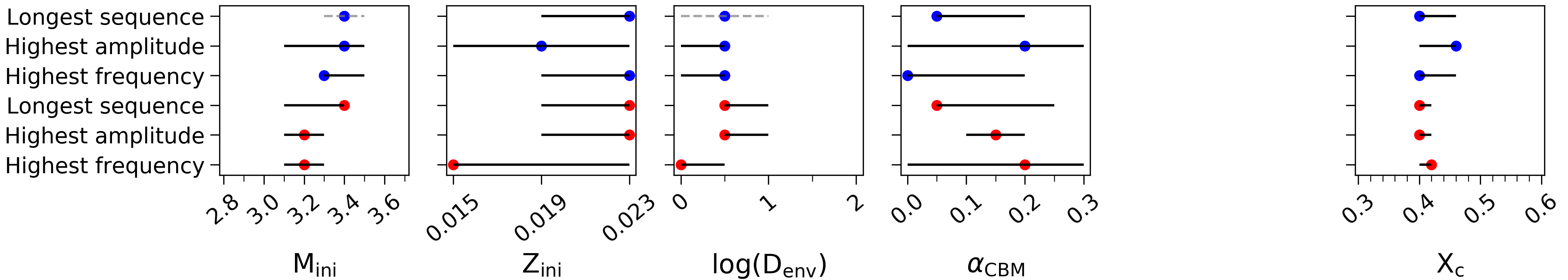}
		\end{subfigure}
		\begin{subfigure}{\hsize}
			\includegraphics[width=\hsize]{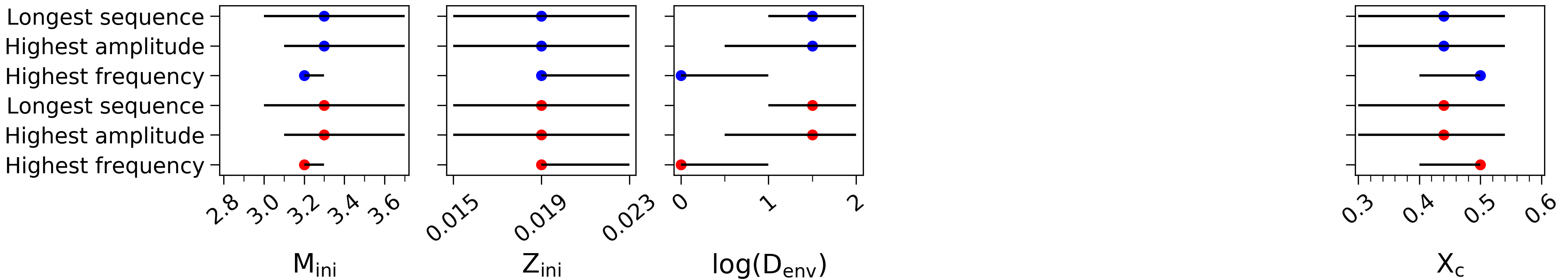}
		\end{subfigure}
		
		\caption{Parameters of the selected best model with their uncertainty ranges indicated by the black lines. If no uncertainty was found through Bayes' theorem, its upper limit is indicated by the dashed grey lines. The blue and red symbols indicate the results for the radiative and P\'eclet grid, respectively, while the labels on the y-axis indicate how the theoretical frequency pattern was constructed. The top row of figures represents the full grids, while the rows underneath represent the nested grids with fewer parameters.}
		\label{fig:uncertainties}
	\end{figure*}

	Although the global morphology of the correlation plots remains rather similar in the P\'eclet grid (\cref{fig:ECP_chisq_longest_sequence_MD_dP,fig:ECP_highest_amplitude_MD_dP,fig:ECP_highest_frequency_MD_dP}), the selected best model does change based on the way the theoretical pattern is constructed, as can be seen in the summary plot in \cref{fig:uncertainties} and in \cref{tab:best_models_MD}.  Looking at the differences in AICc values, the model returned using the longest sequence method is preferred over the other two methods, and the distributions of the MD values are in general shifted in favour of it (\cref{fig:pattern_cdf_MD_dP}). This method searches for a closely matching sequence of frequencies to start building the pattern from, as opposed to the other methods that match just one frequency and build the pattern from there onward. It thus comes as no surprise that this method yields better matching patterns overall.

	%%%%%%%%%%%%%%%%%%%%%%%%%%%%%%%%%%%%%%%%%%%%%%%%%%%%%%%%%%%%%%%%%%%%%%%%%%%%%%%%%%%%%%%%%%%%%
	
	\subsection{Radiative grid versus P\'eclet grid}
	
	When comparing the best models of the radiative and P\'eclet grid, the AICc cannot be coupled to a hypothesis test to determine how strongly one of these two astrophysical models is preferred because we are not dealing with evaluating nested regression models. We have an equal number of free parameters to fit for both of the grids. However, we can still rank the models with the same free parameters based on their AICc values.
	We find that the radiative grid has systematically lower AICc values than the P\'eclet grid, both for the best models (\cref{tab:best_models_MD}) and for the distribution of the MD values as a whole (\cref{fig:pattern_cdf_MD_dP}). Without being able to make firm statistical conclusions at the level of a significant difference, we   find that the radiative grid provides a better fit to the observations.
	
	In the next section we also consider nested regression models for both the radiative and P\'eclet grids. These models point out that, statistically speaking, the least complex grid without CBM is preferred (\cref{tab:best_models_MD_subgrid_noCBM}). From this point of view, the treatment of the temperature gradient in the CBM region becomes irrelevant since the presence of such a region altogether is statistically less favourable.
	
	\begin{figure*}[ht!]
		\centering 
		\includegraphics[width=\hsize]{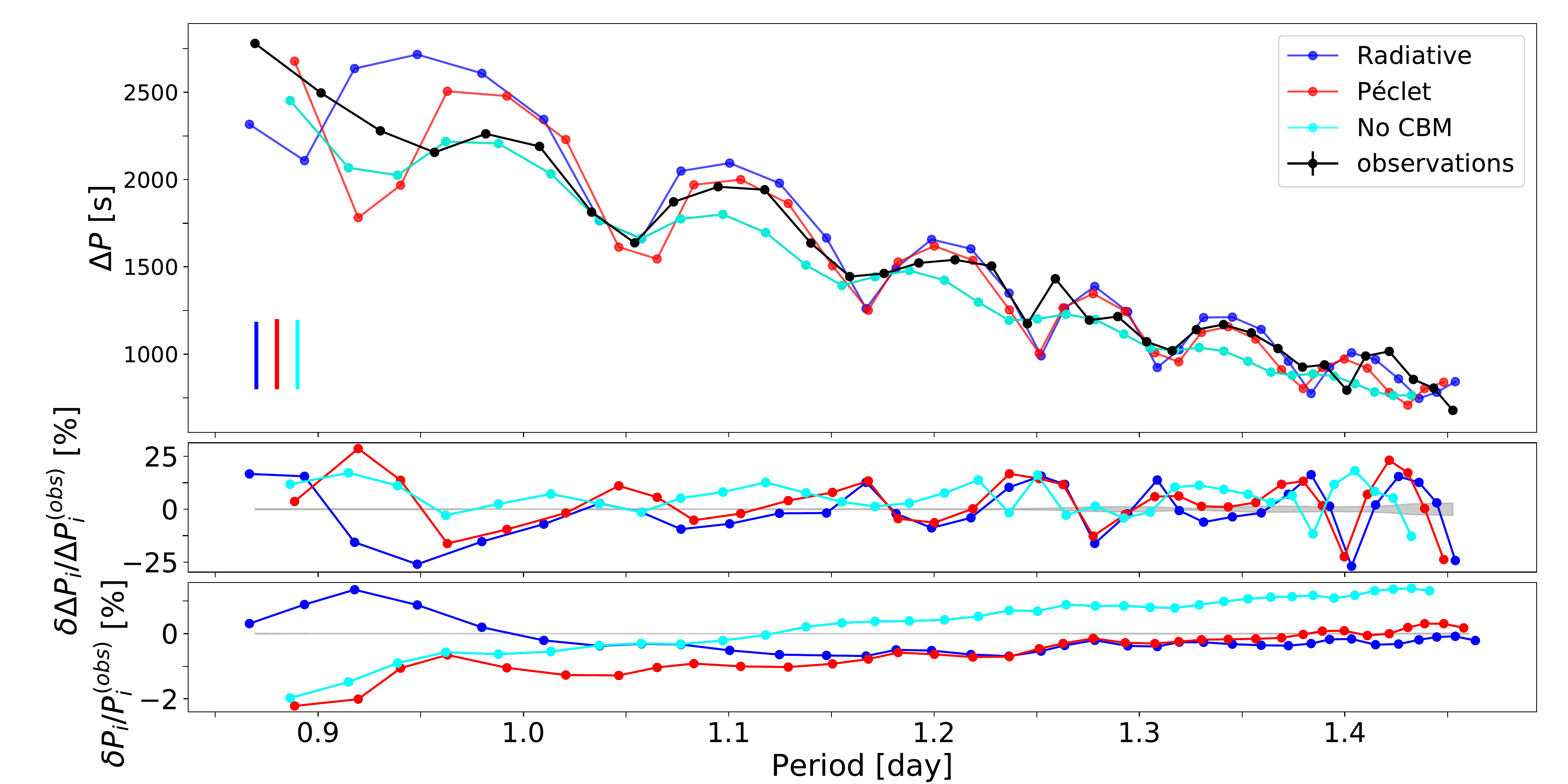}
		\caption{ Period spacing patterns of the observations, of the best models of the two grids, and of the nested grid where no CBM is included in the models, as extracted by the longest sequence method.
			The formal errors on the observations are smaller than the symbol sizes. The vertical bars in the bottom left corner show the maximum considered uncertainty for the theoretical predictions approximated by the variance--covariance matrix of that particular grid. 
			The middle and bottom panels show the relative difference in period spacing and period, respectively, between the observation and the model. The narrow grey areas indicate the formal 1$\sigma$ observational uncertainty from \citet{2015ApJ...803L..25P}, including the correction factor of four.}
		\label{fig:MD_period_spacing_dP}
	\end{figure*}
	
	\subsection{Nested statistical regression models}
	
	To verify if the better fit quality outweighs the penalties for increased model complexity, we compare the AICc values of the best regression models based on the full grid with six free parameters (\cref{tab:best_models_MD}) to those based on partial grids with five free parameters where either $\acbm$ (\cref{tab:best_models_MD_subgrid_fov}) or $\fcbm$ (\cref{tab:best_models_MD_subgrid_aov}) is fixed at 0. Additionally, we compare all of these regression model fits to the results from the partial grid composed of stellar models without any presence of CBM having just four free parameters (\cref{tab:best_models_MD_subgrid_noCBM}). 
	Since the discovery of their period spacing patterns \citep{Degroote2010}, slowly rotating B-type g-mode pulsators have been shown in the literature to have good solutions for a very small amount of CBM \citep{TaoWu2020}. This was our motivation for these tests on nested statistical models.
	
	On the other hand, mixing in the radiative envelope was found to be a necessary ingredient for these stars \citep[e.g.][]{2015A&A...580A..27M}. However, in none of these studies with asteroseismic modelling was a fixed zero CBM   considered in a way that takes proper account of a smaller number of degrees of freedom in the regression, nor were covariances due to theoretical uncertainties considered in the fitting problem. In particular, \citet{2016ApJ...823..130M} did not consider models without CBM in their modelling of KIC\,7760680.
	
	The added complexity in the regression models is more heavily   penalized statistically than rewarded by the better quality of the fit. We consistently find $\Delta \text{AICc}>10$ when comparing models with five free parameters to those with six, or when comparing four free parameters to five. This implies  a strong statistical preference for the less complex regression models in the case of this pulsator with exceptionally low mixing beyond its convective core.
	This capacity of the best models with and without CBM is  illustrated in 
	\cref{fig:MD_period_spacing_dP}, where we have indicated the maximum variance in the theoretical predictions according to the complete radiative and P\'eclet grids.
	
	We point out that a different conclusion would be reached if we were to rely on the $\chi^2$ merit function, as is usually done in the literature of asteroseismic modelling. \cref{app:chi2} offers all the results of the modelling 
	of the prograde dipole modes in  KIC\,7760680
	when ignoring any variance--covariance in the theoretical predictions (i.e. pretending that the theory adopted in the stellar models is fully correct) as was done by \citet{2016ApJ...823..130M}. Under this (incorrect) assumption, 
	the quality of the fit  increases by adding the complexity of a CBM layer with two free parameters to the models. 
	This is assessed quantitatively 
	by comparing the minima listed in 
	\cref{tab:best_models_chi2,tab:best_models_chi2_subgrid_fov,tab:best_models_chi2_subgrid_aov,tab:best_models_chi2_subgrid_noCBM} with 
	those in \cref{tab:best_models_MD,tab:best_models_MD_subgrid_fov,tab:best_models_MD_subgrid_aov,tab:best_models_MD_subgrid_noCBM}. It is visually shown in \cref{fig:MD_period_spacing_dP} if one ignores the theoretical uncertainties whose maxima are indicated by the coloured lines in the bottom left of the upper panel. 
	A firm conclusion irrespective of the choice of merit function, even in the case where   any theoretical uncertainties are ignored, is that the radiative grid delivers the better regression solution.
	
	We conclude that if we allow for theoretical uncertainties 
	in the regression problem at the level of the variance--covariance covered by the model grids, the prograde dipole modes of KIC\,7760680 do not provide statistical evidence of the occurrence of CBM in this star, which has an exceptionally low mixing beyond the convective core. For the models without CBM, an increase in the uncertainties on the inferred model parameters occurs compared to the situation where more complex models with CBM are considered. This can be seen by the increased uncertainty of the solutions for the case of the decreasing number of free parameters in \cref{fig:uncertainties}.
	
	%%%%%%%%%%%%%%%%%%%%%%%%%%%%%%%%%%%%%%%%%%%%%%%%%%%%%%%%%%%%%%%%%%%%%%%%%%%%%%%%%%%%%%%%%%%%%%
	\subsection{Bootstrapping observed frequencies}
	In this paper we have been relying on the observational input from
	\citet{2015ApJ...803L..25P}, but we   wondered how the frequency resolution of the data may have impacted the selection of modes during the extraction and identification of period spacing patterns via iterative pre-whitening (see Sect.\,2). This is particularly important for modes of rather low amplitude as choices in the method of iterative pre-whitening can impact the resultant period spacing pattern. To assess if and how our asteroseismic modelling might depend on the robustness of an extracted period spacing pattern, we considered the statistical resampling method of bootstrapping where we replaced one of the modes in the observed period spacing pattern by another value, relying on the
	frequency resolution of the light curve,  R=$1/ \Delta T$=0.00068d$^{-1}$. 
	
	We perturbed different observed frequencies within the pattern one at a time by increasing or decreasing its value by R. We performed this experiment for a mode at high period (at 1.45261d), a trapped mode (at 1.24544d), a non-trapped mode (at 1.117507d), and for a mode at low period (at 0.90147d) within the extracted pattern from \citet{2015ApJ...803L..25P}. For each of these cases we
	then repeated the whole analysis for these patterns based on the one perturbed mode frequency instead of the one from \citet{2015ApJ...803L..25P}. This gives us an idea of how sensitive the asteroseismic modelling solutions are to an individual frequency being different in the pattern. Such a sensitivity analysis is standard in modern statistics and is of particular relevance in the case of a trapped gravito-inertial mode leading to a dip in the period spacing pattern.
	
\begin{figure*}[ht]
		\centering 
		\includegraphics[width=\hsize]{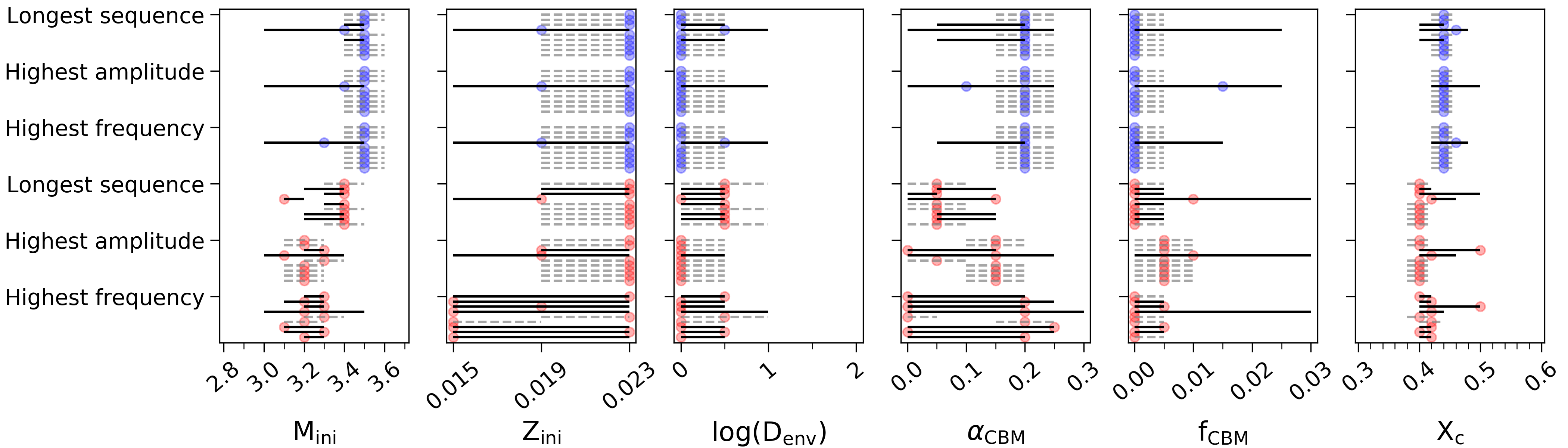}
		\caption{Same as \cref{fig:uncertainties}, but the different lines per extraction method are for each of the different perturbed patterns from the bootstrapping;   the upper lines are  for the unperturbed pattern.}
		\label{fig:uncertainties_bootstrap}
	\end{figure*}

	The results of remodelling all these bootstrapped patterns 
	are shown in \cref{fig:uncertainties_bootstrap}.
	The best solutions sometimes deviate from the one found for the unperturbed pattern, but for the cases where this occurs the uncertainty range almost always includes the unperturbed solution. Our modelling results hence reveal to be mostly robust against the replacement of one mode in the pattern according to the frequency resolution. We recall that this resolution is a gross overestimation of the actual frequency uncertainties, but it must be kept in mind in the context of a pre-whitening procedure \citep{Papics2014}.
	
	%%%%%%%%%%%%%%%%%%%%%%%%%%%%%%%%%%%%%%%%%%%%%%%%%%%%%%%%%%%%%%%%%%%%%%%%%%%%%%%%%%%%%%%%%%%%%%
	\section{Discussion and conclusion}
	
	In this work we investigated the relative importance of various aspects when performing forward asteroseismic modelling of gravito-inertial modes. It is important to not only consider which physical prescriptions or free parameters are used in the computational setup, but to also take different methods for the observational input of the regression problem into account.
	We highlighted the impact of choices in the procedures via the case study of KIC\,7760680, which is the B-type star with the largest number of identified prograde dipole gravito-intertial modes and the best assessed near-core rotation rate among all the published main-sequence gravito-inertial mode pulsators to date \citep{2015ApJ...803L..25P,2016ApJ...823..130M}.
	
	We focused our study on the question of whether it is possible to infer the most likely temperature gradient in the near-core boundary layer from identified high-order gravito-inertial modes. This was done by considering two model grids based on two different prescriptions for this temperature gradient, following the potential capacity of dipole g-mode asteroseismology published in the twin papers by \citet{2018A&A...614A.128P} and \citet{2019A&A...628A..76M}.
	While doing so, we considered two different merit functions to solve the regression problem, following \citet{2018ApJS..237...15A}. One of these is $\chi^2$ while the other is the Mahalanobis distance taking into account the variance--covariance properties of the theoretical predictions, including the entire correlation structure of the regression problem at hand. Moreover, we assessed the quality of different sets of observables to perform the regression based on the comparison between observed and theoretically predicted pulsation patterns.
	
	In terms of a comparison between the two merit functions, we found that using the Mahalanobis distance not only changes the preferred model parameters as compared to $\chi^2$, but it constrains the optimal solution better. Irrespective of the merit function, we found a clear preference for the model grid based on the radiative temperature gradient in the small core-boundary layer for 
	KIC\,7760680 in models that include this layer, instead of a gradually changing temperature gradient between the adiabatic and radiative gradient based on the P\'eclet number. We emphasise that this preference for a radiative temperature gradient in the small core-boundary layer has been obtained for just this one rotating B-type pulsator, and it should not be considered a general result for all B stars given the immense diversity in observed period spacing patterns and the large range of CBM and envelope mixing levels for 26 such stars found by \citet{2021NatAs.tmp...80P}, as summarized in \citet[][Table\,1]{Aerts2021}.  The summary of all available g-mode asteroseismology of B stars has so far revealed that KIC\,7760680 has exceptionally low envelope mixing among those stars rotating at a considerable fraction of the critical rotation rate. We also found that models without any convective penetration or overshooting are statistically preferred for this rotating star, as shown by our evaluation of nested regression models.
	
	For the regression problem based on the Mahalanobis distance,  the use of different sets of quantities to be compared with the respective observables can heavily influence the condition number of the used variance--covariance matrix. In particular, 
	the way the theoretical pulsation quantities are used deserves attention when adopting the Mahalanobis distance as merit function since it impacts the structure of the variance--covariance matrix to be inverted in the regression. 
	The effects for the regression based on a $\chi^2$ merit function are smaller since in that case we ignore the covariance among the theoretical prediction or even the theoretical uncertainty as a whole. We should find a compromise between the best numerical stability and the use of maximum observable information in the definition of the regression problem. The application of singular value decomposition or principle component analysis as adopted for some of the modelling efforts of low-mass stars  \citep[e.g.][]{Angelou2017} or intermediate-mass pulsators \citep{Mombarg2019}
	might be a useful additional or complementary approach to the one we adopted here for g-mode asteroseismology of B stars.
	
	\begin{table}
		\caption{Stellar parameters of the best asteroseismic model 
			of KIC\texorpdfstring{\,}{TEXT}7760680
			obtained in this work for the cases without and with CBM.} 
		\label{tab:ourbestsolutions}
		\centering
		\begin{tabular}{l c c}     
			\hline               
			Parameter & Models w/o CBM & Models with CBM \\  
			\hline                        
			$M_{\rm ini}$ [\msol] & 3.3 & 3.5 \\   
			$Z_{\rm ini}$ & 0.019 & 0.023 \\
			$\acbm$  & (...) &  0.2 \\
			$\fcbm$  & (...) &0.0 \\
			log($\D[env]$) & 1.5 & 0.0\\ 
			$X_c$ & 0.44 & 0.44 \\ 
			\hline
		\end{tabular}
	\end{table}
	The best model parameters we found for the cases with and without CBM are provided in \cref{tab:best_models_MD,tab:best_models_MD_subgrid_noCBM} and are listed in \cref{tab:ourbestsolutions}. Overall, the model parameters obtained by \citet{2016ApJ...823..130M} listed in \cref{tab:Moravveji2016} are outside of the uncertainties on the model with CBM, but within those of the statistically preferred model without CBM. The well-known mass--metallicity and mass--CBM degeneracies for pulsators in this mass range \citep[][their Fig.\,5]{2015A&A...580A..27M} are   the basis of the uncertainty regions of the estimated parameters, whose projection in one dimension are shown graphically in \cref{fig:uncertainties}. 
	The solved regression problem based on the Mahalanobis distance merit function properly takes into account these correlations and places the solution based on the $\chi^2$ as obtained by 
	\citet{2016ApJ...823..130M}  within the uncertainty regions of the estimated parameters. This is remarkably consistent given that we used a non-flat $D_{\rm mix}(r)$ profile and performed the fitting from the $\Delta$P values instead of the mode  frequencies as used by \citet{2016ApJ...823..130M}. Our less refined grid step implied a major reduction in the analysis time of the problem and allowed us to treat the primary goal set for this paper,
	which was  to investigate if we can distinguish between a different thermal and chemical structure in the CBM region while having fixed the shape of the mixing profile as diffusive core overshooting in the core-boundary layer, and as being due to wave mixing in the envelope. With this choice of core overshooting, which is  often taken in the literature, we
	found that a radiative temperature gradient in the CBM region gives a better fit to the data than the models with a temperature gradient based on the P\'eclet number. However, the presence of such a CBM region altogether turned out to be statistically disfavoured by the AICc for this particular star rotating at about a quarter of its critical rate.
	
	Finally, it is noteworthy that the best model with CBM listed in \cref{tab:ourbestsolutions} falls within the $3\sigma$ error box of the spectroscopic $T_{\rm eff}$ and astrometrically derived 
	$\log(L/L_\odot)$ as shown in \cref{fig:DO_chisq_longest_sequence_MD_dP}, while these observables were not used in the asteroseismic modelling process. This  a posteriori evaluation of the asteroseismic solution is   reassuring as there was no need to reject models based on these non-asteroseismic observables.
	
	Overall, we conclude that choices in the modelling methodology should be made with great care and consideration. The optimal modelling setup might vary from star to star because some pulsators reveal signatures of trapped modes, while others do not \citep{Papics2017}. A fully automated modelling strategy may therefore be premature for ensemble modelling of gravito-inertial B-type pulsators covering the entire instability strip of such stars \citep[see][]{2020MNRAS.495.2738P}, particularly as our current case study was done for a star with very high-precision observables and an exceptionally low level of mixing in the core-boundary layer and stellar envelope compared to most other gravito-inertial pulsators of this kind among the B-type stars
	\citep{2021NatAs.tmp...80P}.

	\begin{acknowledgements}
		The authors are grateful to the 
		\texttt{MESA} and \texttt{GYRE} developers teams led by Bill Paxton and Rich Townsend for their efforts and for releasing their software publicly; this study would  not have been possible without their codes.
		The authors thank Geert Molenberghs for his helpful advice on the statistical aspects of the study, Joey Mombarg for the fruitful discussions, and Cole Johnston for his insights in theoretical pulsation pattern construction. We thank the referee for the comments and suggestions towards improving the manuscript.
		The research leading to these results has received
		funding from the Research Foundation Flanders (FWO) by means of a PhD scholarship to MM under project No. 11F7120N and a senior postdoctoral fellowship to DMB with grant agreement No. 1286521N, from the European Research Council (ERC) under the European Union’s Horizon 2020 research and innovation programme (grant agreement no. 670519: MAMSIE to CA), and from the KU\,Leuven Research Council (grant C16/18/005: PARADISE).
		
	\end{acknowledgements}
	
	\bibliographystyle{aa}

	%%%%%%%%%%%%%%%%%%%%%%%%%%%%%%%%%%%%%%%%%%%%%%%%%%%%%%%%%%%%%%%%%%%%%%%%%%%%%%%%%%%%%%%%%%%%%%
	%%%%%%%%%%%%%%%%%%%%%%%%%%%%%%%%%%%%%%%%%%%%%%%%%%%%%%%%%%%%%%%%%%%%%%%%%%%%%%%%%%%%%%%%%%%%%%
	\begin{appendix}
		
		\section{MESA and GYRE Inlists} \label{appendix:inlist} 
		
		The example MESA and GYRE inlists used for this work are
		available from the MESA Inlists section of the MESA Marketplace: 
		\url{cococubed.asu.edu/mesa_market/inlists.html.}
		
		%%%%%%%%%%%%%%%%%%%%%%%%%%%%%%%%%%%%%%%%%%%%%%%%%%%%%%%%%%%%%%%%%%%%%%%%%%%%%%%%%%%%%%%%%%%%%%
		\section{Reduced \texorpdfstring{$\chi^2$}{TEXT} as merit function} \label{app:chi2}
		
		\subsection{Model selection criterion}
		
		The most commonly used merit function to assess the goodness of fit in forward asteroseismic modelling of SPB pulsators is a $\chi^2$, which is based on a Euclidian distance.
		Denoting a set of uncorrelated observables as $\Y{Obs}$ with errors according to a normal distribution with variance $\sigma^2_{\text{Y}^{\text{Obs}}}$, and the set of corresponding values predicted by a theoretical model as $\Y{Theo}$, we write the reduced $\chi^2$ score as
		
		\begin{align} \label{eq:red-chi2}
			\chi^2_{\text{red}} = \frac{1}{N-k} \sum^N_i \lb \frac{\text{Y}^{\text{Theo}}_i - \text{Y}^{\text{Obs}}_i}{\sigma_{\text{Y}^{\text{Obs}}_i}} \rb ^2,
		\end{align}
		where $N$ is the number of observables and $k$ is the number of free parameters.
		In general, our MD formulation used in the main part of the paper can be connected to the $\chi^2_{\text{red}}$ merit function by ignoring uncertainties on the theoretical predictions of the observables and by ignoring correlations among the observables. Ignoring theoretical uncertainties comes down to ignoring the covariance structures shown in Appendix\,D. 
		We provide the  $\chi^2_{\text{red}}$ results here to assess the differences in the solutions to the regression problem by making these simplifications, following the earlier assessment of theoretical uncertainties in \citet{2018ApJS..237...15A}. The modelling procedures applied to the g-mode frequencies of rotating SPB stars in the literature so far consider the mode frequencies as independent observables \citep{2016ApJ...823..130M,2018MNRAS.478.2243S}, such that using a $\chi^2_{\text{red}}$ is meaningful. Nevertheless, we show the results obtained by using this simplified merit function for two  cases, of mode periods and of mode period spacings, where the latter in principle demands using a formalism suitable for correlated data such as the MD. This allows the reader to assess the impact induced by adopting a simplified 
		$\chi^2_{\text{red}}$ for applications to correlated data, while also ignoring theoretical uncertainties.

		Using $\chi^2_{\text{red}}$ as the merit function, the AICc reduces to  
		\begin{equation}
		\text{AICc} = \chi^2_{\text{red}} + \frac{2kN}{N-k-1}.
		\end{equation}

		%%%%%%%%%%%%%%%%%%%%%%%%%%%%%%%%%%%%%%%%%%%%%%%%%%%%%%%%%%%%%%%%%%%%%%%%%%%%%%%%%%%%%%%%%%%%%%
		\subsection{Observables to fit}
		When considering $\chi^2$ as the merit function, there is no variance--covariance matrix whose numerical stability needs to be taken into account through condition numbers. We simply compare the use of different sets of observables and look at the correlation plots and modelling results to assess their use.
		
		\begin{figure}[ht]
			\centering
			\includegraphics[width=\hsize]{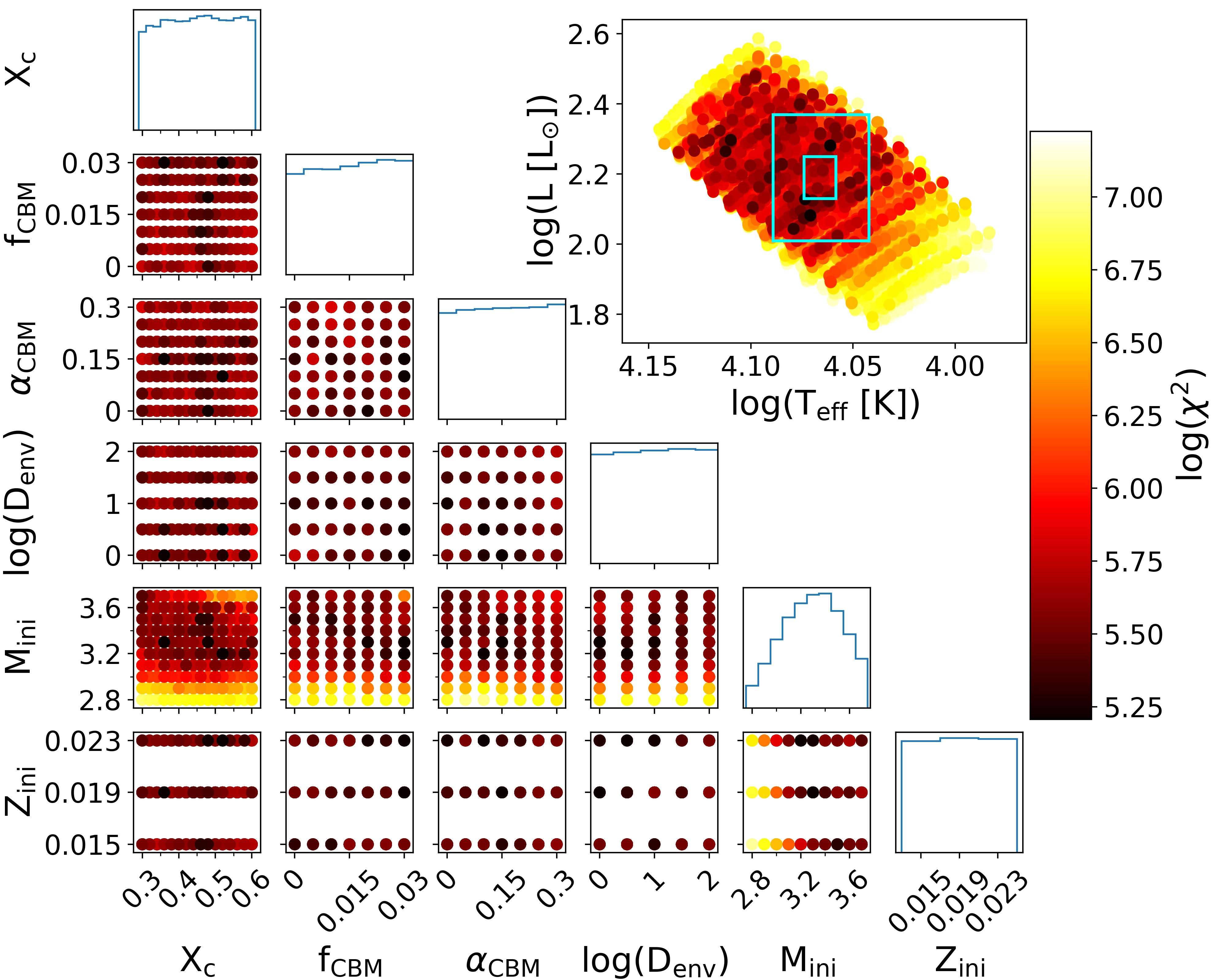}
			\caption{ Correlation plot for the radiative grid, using periods in a $\chi^2$ merit function. The theoretical pulsation pattern was constructed according to the longest sequence method. The 50\% best models are shown, colour-coded according to the log of their merit function value (at right). The figures on the diagonal show binned parameter distributions, and the larger panel at  the top right shows an Hertzsprung–Russell diagram with the 1 and 3$\sigma$ spectroscopic $T_{\rm eff}$ and astrometric luminosity error boxes.}
			\label{fig:DO_chisq_longest_sequence_CS_P}
			
		\end{figure}
		
\begin{figure}[!ht]
			\centering
			\includegraphics[width=\hsize]{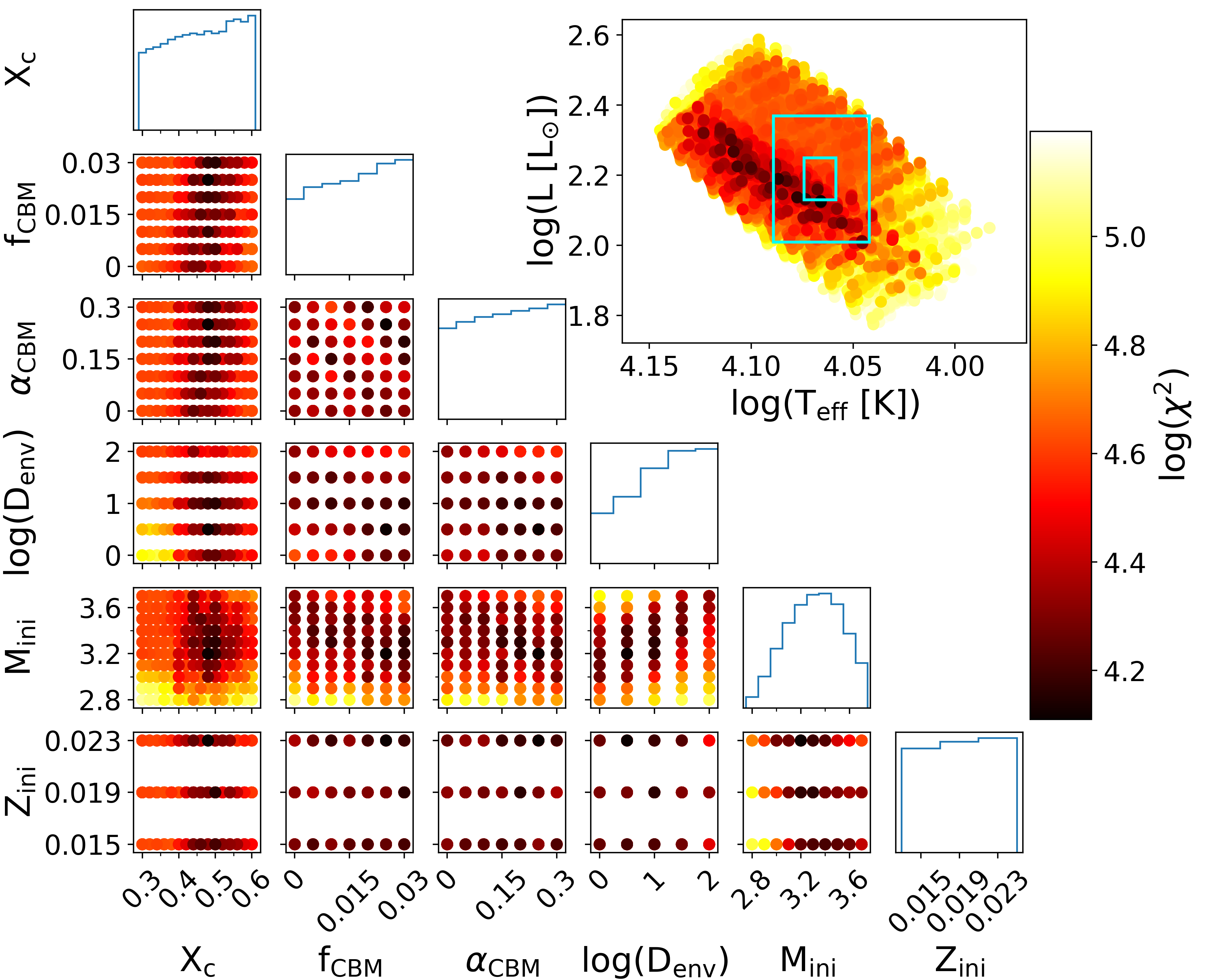}
			\caption{ Correlation plot (see \cref{fig:DO_chisq_longest_sequence_CS_P}) for the radiative grid, using period spacings in a $\chi^2$ merit function. The theoretical pulsation pattern was constructed according to the longest sequence method.}
			\label{fig:DO_chisq_longest_sequence_CS_dP}
		\end{figure}

\begin{figure}[!t]
			\centering
			\includegraphics[width=\hsize]{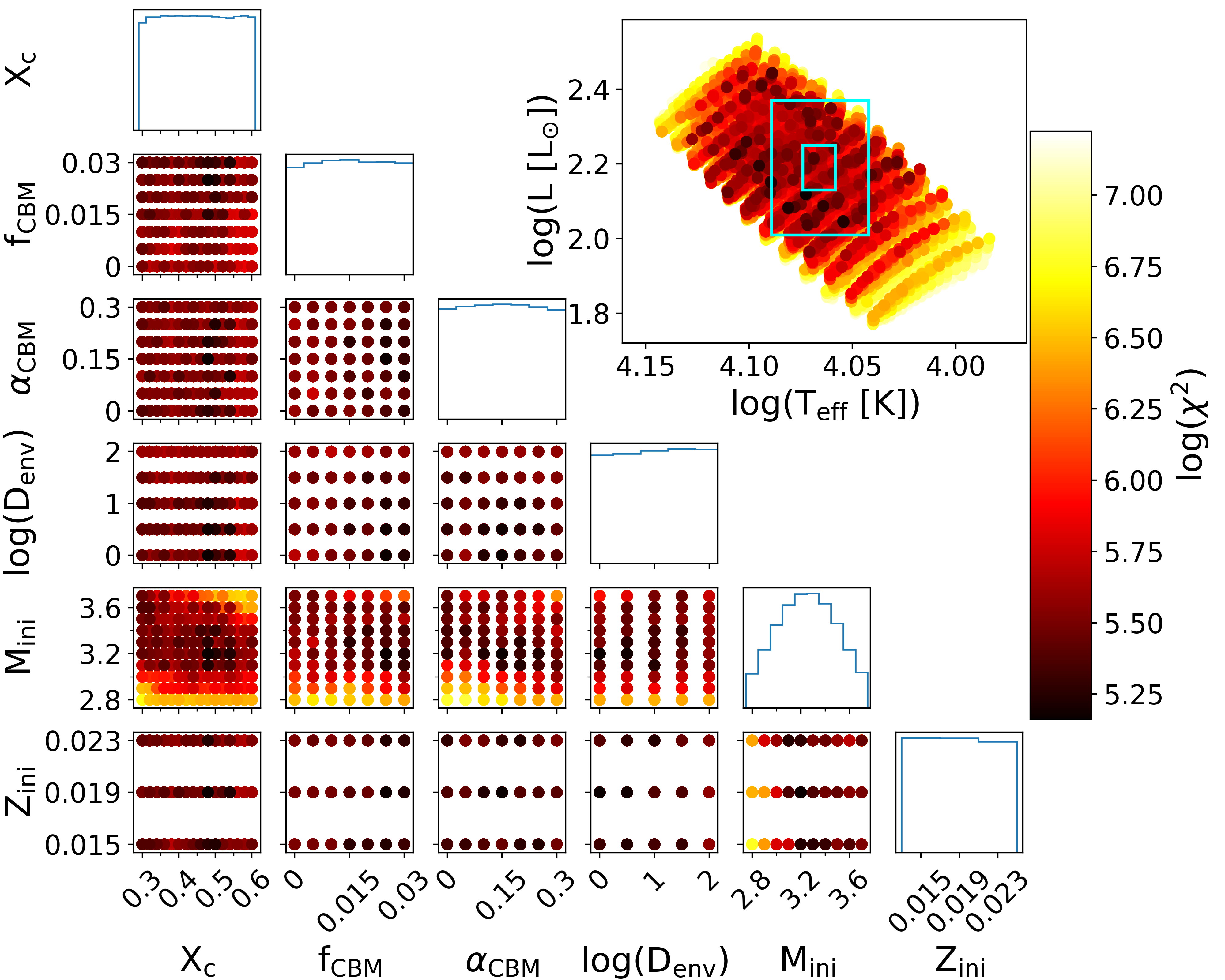}
			\caption{ Correlation plot (see \cref{fig:DO_chisq_longest_sequence_CS_P}) for the P\'eclet grid, using periods in a $\chi^2$ merit function. The theoretical pulsation pattern was constructed according to the longest sequence method.}
			\label{fig:ECP_chisq_longest_sequence_CS_P}
		\end{figure}

\begin{figure}[!ht]
			\centering
			\includegraphics[width=\hsize]{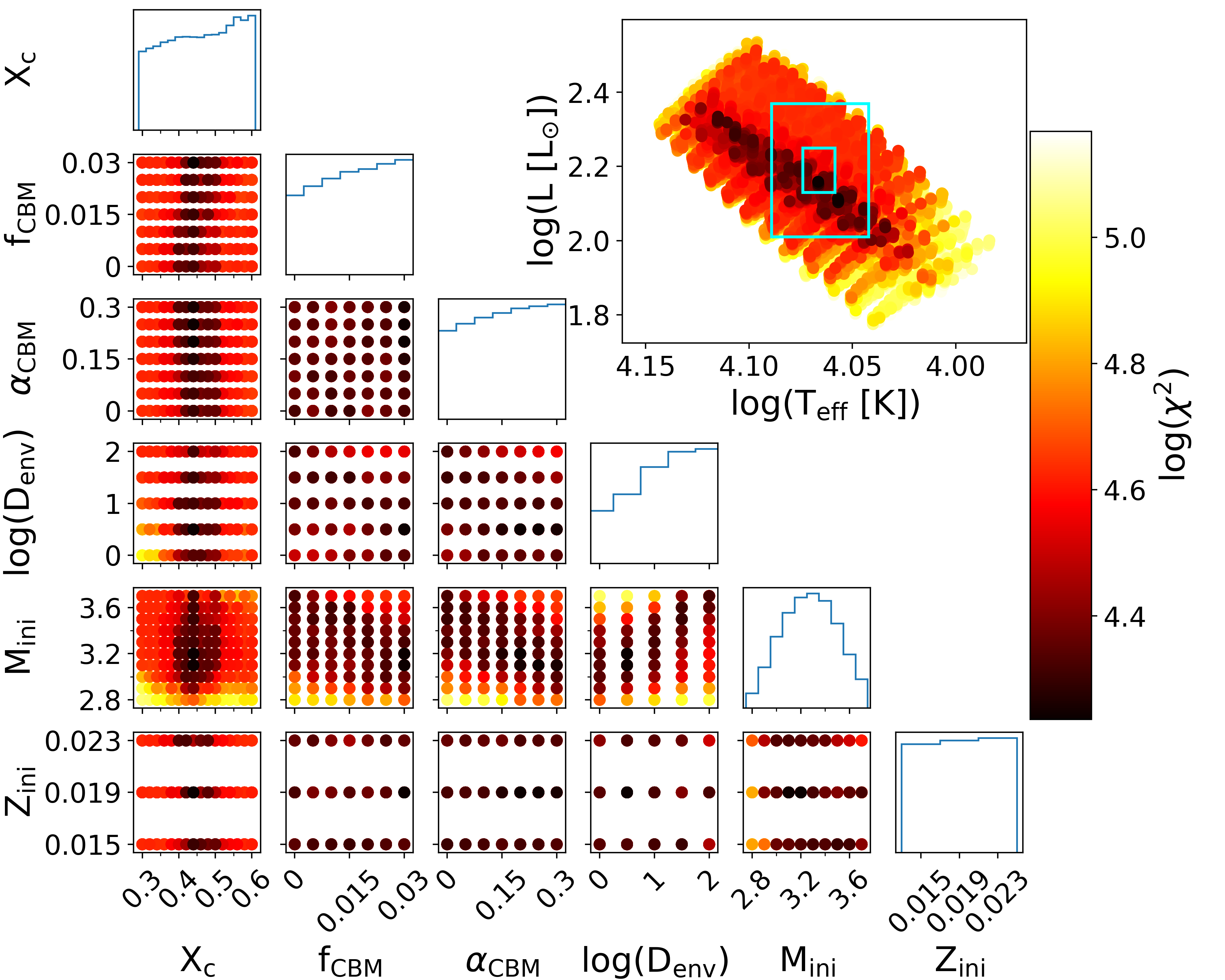}
			\caption{ Correlation plot (see \cref{fig:DO_chisq_longest_sequence_CS_P}) for the P\'eclet grid, using period spacings in a $\chi^2$ merit function. The theoretical pulsation pattern was constructed according to the longest sequence method.}
			\label{fig:ECP_chisq_longest_sequence_CS_dP}
		\end{figure}

		The correlation plots show a better constrained region with optimal solutions when selecting the period spacings as observables instead of the periods themselves. This can be seen comparing \cref{fig:DO_chisq_longest_sequence_CS_P,fig:DO_chisq_longest_sequence_CS_dP} for the radiative grid, or \cref{fig:ECP_chisq_longest_sequence_CS_P,fig:ECP_chisq_longest_sequence_CS_dP} for the P\'eclet grid.

		Apart from the correlation structure, the actual $\chi^2$ values themselves are also lower when fitting period spacings.
		This is no surprise since the mode periods have much smaller relative errors, of order 0.003\% or even smaller in the case of KIC\,7760680 \citep{2015ApJ...803L..25P}, than the propagated errors on $\Delta P$, which are about 10 to 20 times larger.
		
		%%%%%%%%%%%%%%%%%%%%%%%%%%%%%%%%%%%%%%%%%%%%%%%%%%%%%%%%%%%%%%%%%%%%%%%%%%%%%%%%%%%%%%%%%%%%%%
		\subsection{Theoretical mode period pattern construction}
		
\begin{figure}[ht]
			\centering
			\includegraphics[width=\hsize]{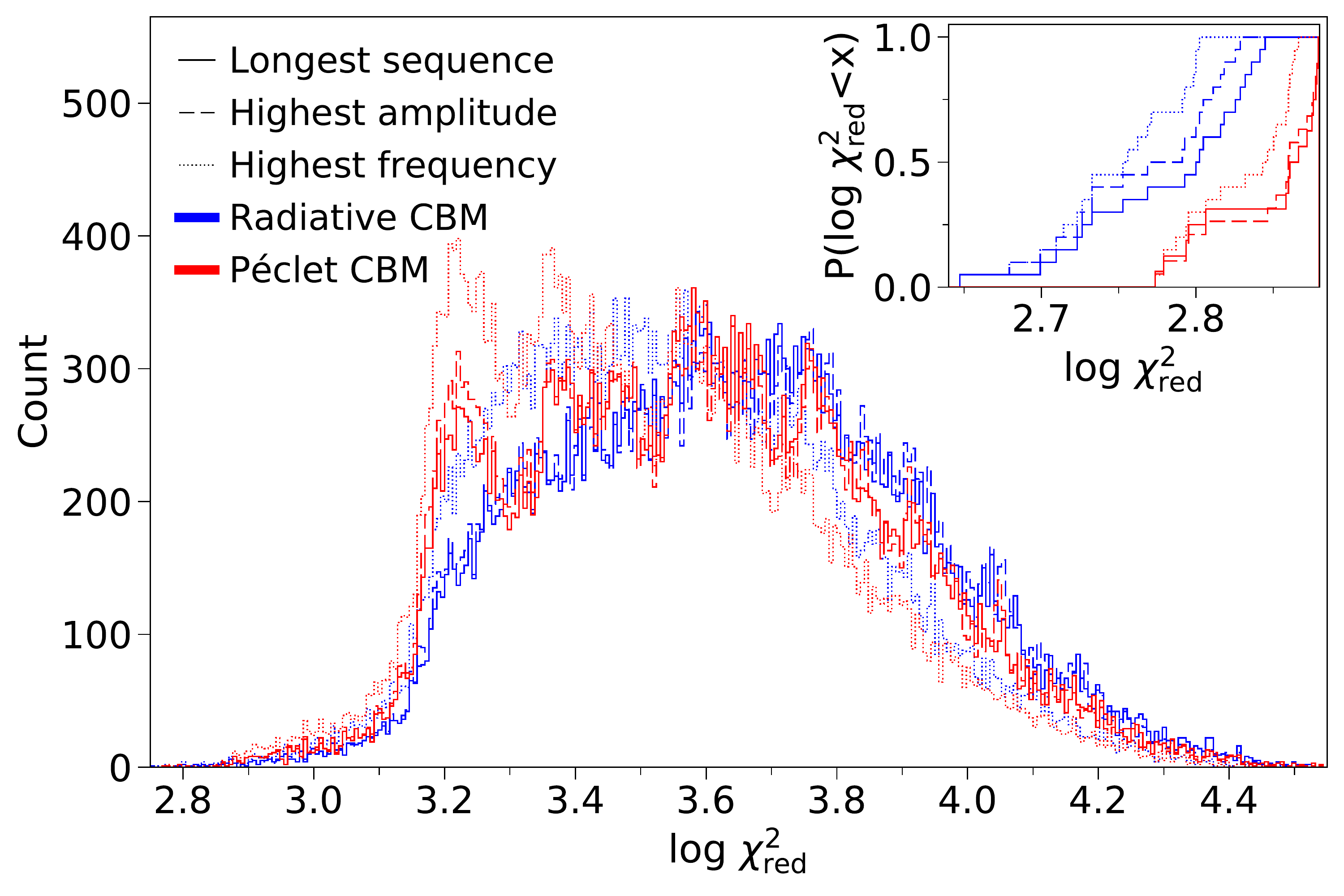}
			\caption{ Distribution of the reduced $\chi^2$, using $\Delta P$ as observables to fit. The inset shows the cumulative distribution function of the 20 best models in the grids.}
			\label{fig:pattern_cdf_CS_dP}
		\end{figure}

\begin{figure}[ht]
			\centering
			\includegraphics[width=\hsize]{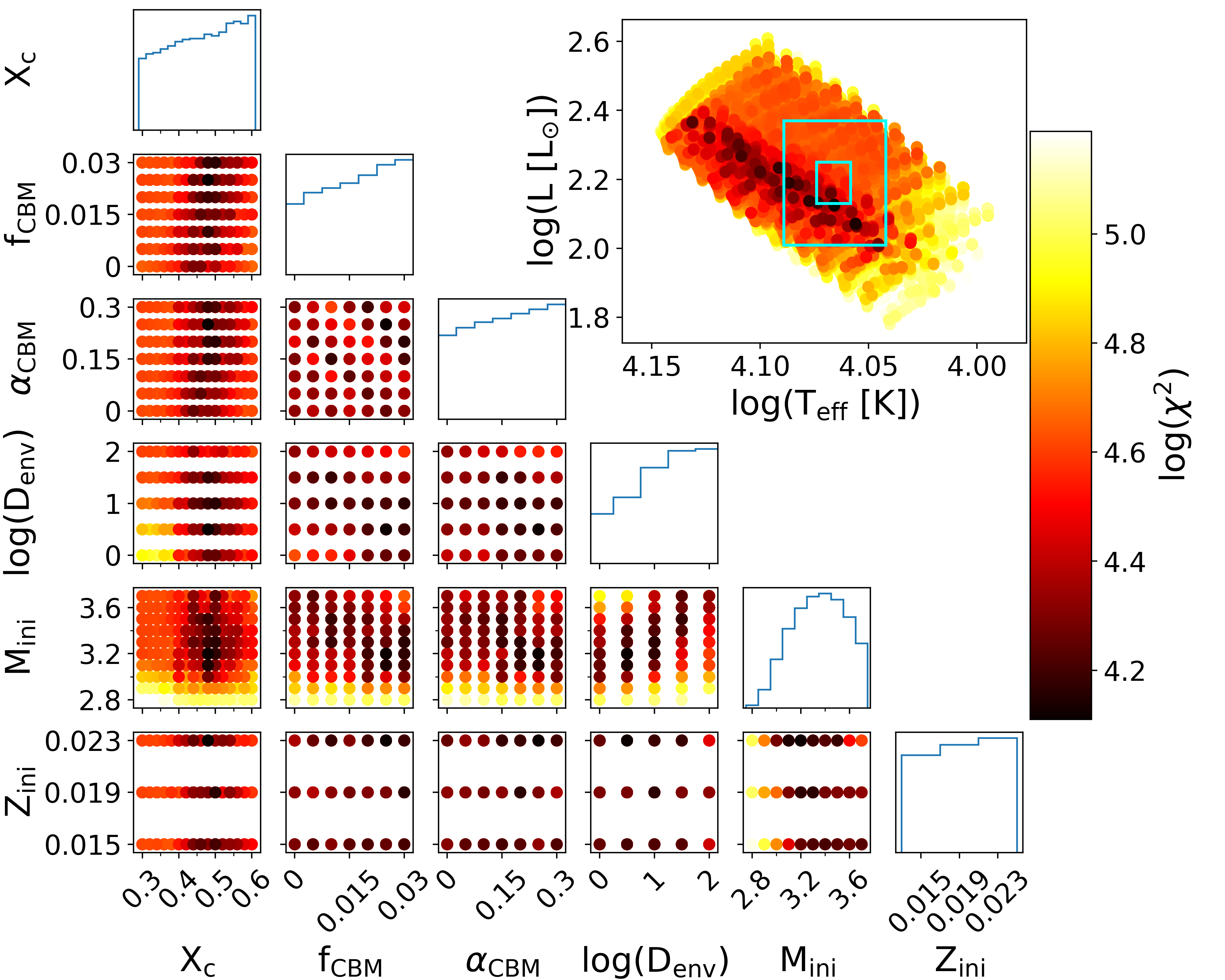}
			\caption{ Correlation plot (see \cref{fig:DO_chisq_longest_sequence_CS_P}) for the radiative grid, using period spacings in a $\chi^2$ merit function. The theoretical pulsation pattern was constructed according to the highest amplitude method.}
			\label{fig:DO_highest_amplitude_CS_dP}
		\end{figure}

\begin{figure}[ht]
			\centering
			\includegraphics[width=\hsize]{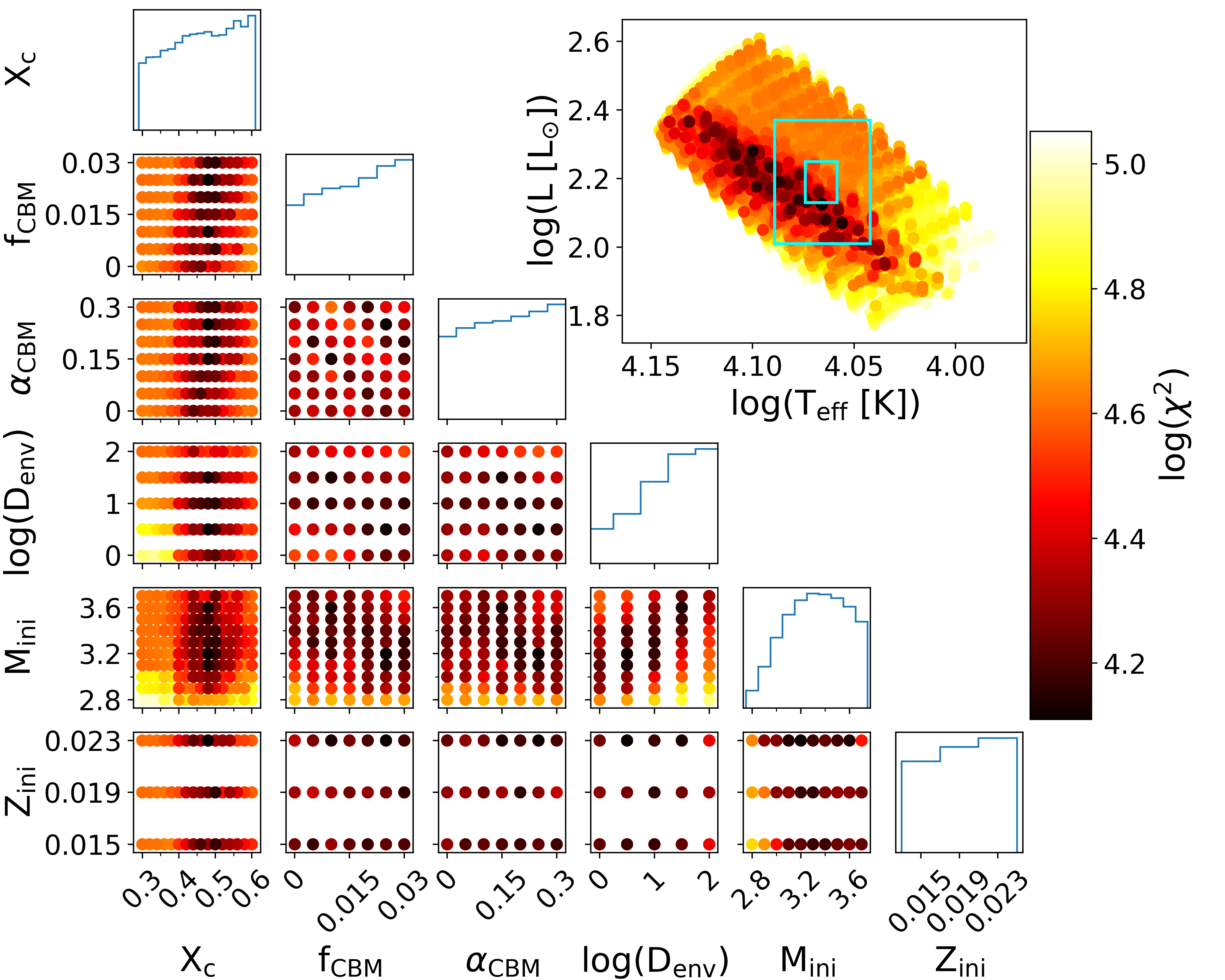}
			\caption{ Correlation plot (see \cref{fig:DO_chisq_longest_sequence_CS_P}) for the radiative grid, using period spacings in a $\chi^2$ merit function. The theoretical pulsation pattern was constructed according to the highest frequency method.}
			\label{fig:DO_highest_frequency_CS_dP}
		\end{figure}

		Although the different methods used to construct the theoretical frequency pattern can influence the distribution of the $\chi^2$ values (\cref{fig:pattern_cdf_CS_dP}), the best model in the grid and its $\chi^2_{\text{red}}$ value remain identical when considering the full grids of models (\cref{tab:best_models_chi2}). The optimal solution   slightly changes for some of the nested statistical models (\cref{tab:best_models_chi2_subgrid_noCBM,tab:best_models_chi2_subgrid_fov,tab:best_models_chi2_subgrid_aov}), but this change in most cases only amounts to one grid step-size for one parameter.
		Additionally, the differences in the correlation plots are minimal, and the same global morphology applies (\cref{fig:DO_chisq_longest_sequence_CS_dP,fig:DO_highest_amplitude_CS_dP,fig:DO_highest_frequency_CS_dP}).

		%%%%%%%%%%%%%%%%%%%%%%%%%%%%%%%%%%%%%%%%%%%%%%%%%%%%%%%%%%%%%%%%%%%%%%%%%%%%%%%%%%%%%%%%%%%%%%
		\subsection{Nested statistical regression models}
		Investigating the best regression models with varying complexity in the CBM of the models reveals a different trend than for the Mahalanobis distance.
		Comparing the best models of the full grids with six free parameters (\cref{tab:best_models_chi2}) to those of the partial grids with five (\cref{tab:best_models_chi2_subgrid_fov,tab:best_models_chi2_subgrid_aov}) or four free parameters (\cref{tab:best_models_chi2_subgrid_noCBM}), reveals that the more complex models are generally preferred over the simpler ones. 
		In contrast to when using the Mahalanobis distance, the increased fit quality plays a dominant role over the penalty for the increased model complexity. The main reason for this is that no uncertainty is propagated for the theoretical predictions when using a $\chi^2$ merit function.
		
		When comparing the different partial grids we find a clear preference for the models with an exponentially decaying CBM coefficient. This is in agreement with the results of \citet{2016ApJ...823..130M}. Using a $\chi^2$ merit function the authors found that an exponentially decaying CBM prescription outperformed a step-like mixing coefficient in the CBM region. We thus recover the previous result under the restricted assumption that the theoretical model predictions are error-free.

		\subsection{Radiative grid versus P\'eclet grid}
		
\begin{figure}[ht]
			\centering
			\includegraphics[width=\hsize]{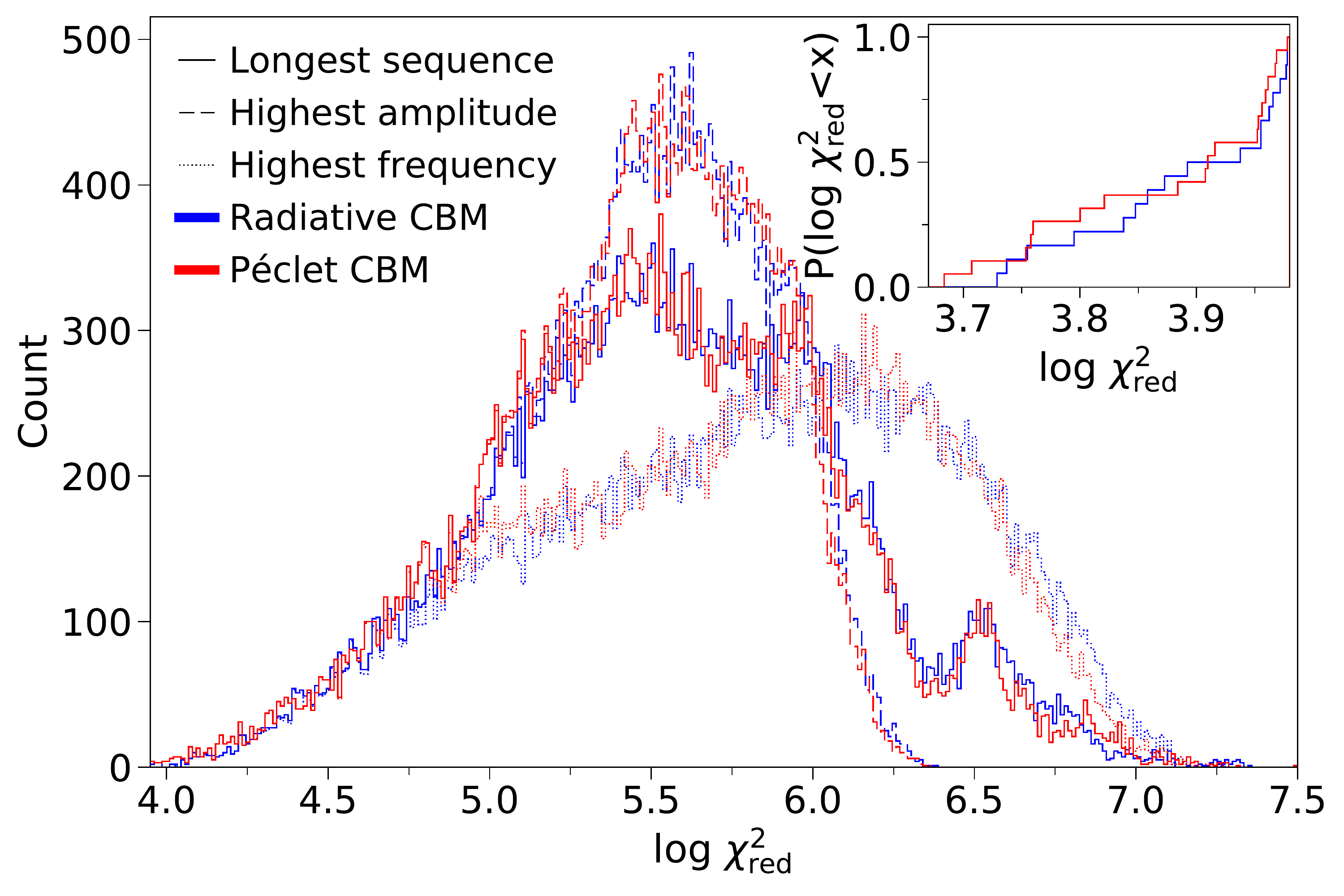}
			\caption{ Distribution of the reduced $\chi^2$, using periods as observables to fit. The inset shows the cumulative distribution function of the 20 best models in the grids.}
			\label{fig:pattern_cdf_CS_P}
		\end{figure}

		According to the $\chi^2_{\rm red}$ values the radiative grid gives a better fit to the observations than the P\'eclet grid. Although there is a clear difference between the best models of the two grids (\cref{tab:best_models_chi2}), the distribution of their $\chi^2_{\text{red}}$ values does not differ too much (\cref{fig:pattern_cdf_CS_dP}). However, a different conclusion arises when using periods as the observables without considering the period spacings. Specifically,  the $\chi^2_{\text{red}}$ distributions are again similar between the two grids, but the preferred best model stems from the P\'eclet grid (\cref{fig:pattern_cdf_CS_P,tab:best_models_chi2}).
		
		We note that the best solutions according to the Mahalanobis distance are not among the best solutions delivered by the $\chi^2_{\text{red}}$ metric (\cref{tab:best_models_MD}). On the other hand, the best models according to the $\chi^2_{\text{red}}$ still score among the better models following the Mahalanobis distance (listed in \cref{tab:best_models_chi2}), although they are rated significantly worse than the best ones for that metric.
		
		Even though the conclusions of the analysis of this star with the $\chi^2_{\text{red}}$ merit function would not have been influenced by the way we construct the theoretical pulsation pattern, the choice of observables (mode periods versus period spacings) changes the answer to the question of which temperature gradient is preferred when only looking at the best point estimator, without taking into account any precision estimation.
		
		\newpage
		
		%%%%%%%%%%%%%%%%%%%%%%%%%%%%%%%%%%%%%%%%%%%%%%%%%%%%%%%%%%%%%%%%%%%%%%%%%%%%%%%%%%%%%%%%
		\section{Best models}
		Tables listing the parameters of the best fit models.
		
		\begin{table*}
			\caption{Best fit models according to Mahalanobis distance.} 
			\label{tab:best_models_MD}
			\centering
			\setlength\tabcolsep{4.0pt} % default value: 6pt
			\renewcommand{\arraystretch}{1.12}
			\begin{tabular}{l l l l l l l l l r r r}     
				\hline               
				Grid & Observables & Pattern construction & $M_{\rm ini}$ [\msol] & $Z_{\rm ini}$ & $\acbm$ & $\fcbm$ & log($\D[env]$) & $X_c$ & $\chi^2_{\text{red}}$ & MD & AICc(MD) \vspace{1pt} \\
				\hline
				Radiative & $\Delta $P & longest sequence & 3.5 & 0.023 & 0.2 & 0.0 & 0.0 & 0.44 & 3479 & 182 & -294.0 \\ 
				&  & highest amplitude & 3.5 & 0.023 & 0.2 & 0.0 & 0.0 & 0.44 & 3479 & 186 & -290.6 \\ 
				&  & highest frequency & 3.5 & 0.023 & 0.2 & 0.0 & 0.0 & 0.44 & 3479 & 184 & -293.6 \\ 
				
				& Period & longest sequence & 3.5 & 0.023 & 0.2 & 0.0 & 0.0 & 0.44 & 362028 & 168 & -315.8 \\ 
				&  & highest amplitude & 3.5 & 0.023 & 0.2 & 0.0 & 0.0 & 0.44 & 362028 & 173 & -313.4 \\ 
				&  & highest frequency & 3.5 & 0.023 & 0.2 & 0.0 & 0.0 & 0.44 & 362028 & 170 & -315.2 \\ 
				\hline
				P\'eclet & $\Delta $P & longest sequence & 3.4 & 0.023 & 0.05 & 0.0 & 0.5 & 0.40 & 2398 & 230 & -255.7 \\ 
				&  & highest amplitude & 3.2 & 0.023 & 0.15 & 0.005 & 0.0 & 0.40 & 1616 & 246 & -241.1 \\ 
				&  & highest frequency & 3.3 & 0.023 & 0.0 & 0.0 & 0.5 & 0.40 & 2450 & 241 & -247.3 \\ 
				
				& Period & longest sequence & 3.4 & 0.023 & 0.05 & 0.0 & 0.5 & 0.40 & 744142 & 218 & -276.5 \\ 
				&  & highest amplitude & 3.2 & 0.023 & 0.15 & 0.005 & 0.0 & 0.40 & 101511 & 226 & -270.6 \\ 
				&  & highest frequency & 3.2 & 0.023 & 0.15 & 0.005 & 0.0 & 0.40 & 101511 & 233 & -263.9 \\ 
				\hline
			\end{tabular}
		\end{table*}
		
		%%%%%
		\begin{table*}
			\caption{Best fit models according to Mahalanobis distance where  $\acbm$=0 was fixed.} 
			\label{tab:best_models_MD_subgrid_fov}
			\centering
			\setlength\tabcolsep{4.0pt} % default value: 6pt
			\renewcommand{\arraystretch}{1.12}
			\begin{tabular}{l l l l l l l l l r r r}     
				\hline               
				Grid & Observables & Pattern construction & $M_{\rm ini}$ [\msol] & $Z_{\rm ini}$ & $\acbm$ & $\fcbm$ & log($\D[env]$) & $X_c$ & $\chi^2_{\text{red}}$ & MD & AICc(MD) \vspace{1pt} \\  
				\hline    
				Radiative & $\Delta $P &  longest sequence & 3.3 & 0.019 & (...) & 0.01 & 0.5 & 0.42 & 1335 & 171 & -307.5 \\ 
				&  & highest amplitude & 3.3 & 0.019 & (...) & 0.01 & 0.5 & 0.42 & 1335 & 171 & -308.1 \\ 
				&  & highest frequency & 3.3 & 0.023 & (...) & 0.0 & 0.5 & 0.40 & 2369 & 150 & -329.1 \\ 
				
				& Period & longest sequence & 3.4 & 0.023 & (...) & 0.02 & 1.0 & 0.46 & 464512 & 152 & -334.9 \\ 
				&  & highest amplitude & 3.5 & 0.023 & (...) & 0.02 & 1.0 & 0.46 & 209399 & 155 & -334.1 \\ 
				&  & highest frequency & 3.3 & 0.023 & (...) & 0.0 & 0.5 & 0.40 & 818181 & 147 & -340.9 \\ 
				\hline
				P\'eclet & $\Delta $P & longest sequence & 3.0 & 0.019 & (...) & 0.025 & 0.0 & 0.44 & 1714 & 170 & -310.6 \\ 
				&  & highest amplitude & 3.2 & 0.015 & (...) & 0.015 & 0.0 & 0.44 & 1934 & 178 & -303.7 \\ 
				&  & highest frequency & 3.3 & 0.023 & (...) & 0.0 & 0.5 & 0.40 & 2369 & 164 & -317.9 \\ 
				
				& Period & longest sequence & 3.3 & 0.023 & (...) & 0.03 & 0.5 & 0.46 & 450592 & 163 & -326.5 \\ 
				&  & highest amplitude & 3.3 & 0.023 & (...) & 0.03 & 0.5 & 0.46 & 450592 & 168 & -323.8 \\ 
				&  & highest frequency & 3.4 & 0.023 & (...) & 0.03 & 0.0 & 0.46 & 265614 & 171 & -319.6 \\ 
				\hline
			\end{tabular}
		\end{table*}
		
		%%%%%%%%%%%
		\begin{table*}
			\caption{Best fit models according to Mahalanobis distance where $\fcbm$=0 was fixed.} 
			\label{tab:best_models_MD_subgrid_aov}
			\centering
			\setlength\tabcolsep{4.0pt} % default value: 6pt
			\renewcommand{\arraystretch}{1.12}
			\begin{tabular}{l l l l l l l l l r r r}     
				\hline               
				Grid & Observables & Pattern construction & $M_{\rm ini}$ [\msol] & $Z_{\rm ini}$ & $\acbm$ & $\fcbm$ & log($\D[env]$) & $X_c$ & $\chi^2_{\text{red}}$ & MD & AICc(MD) \vspace{1pt} \\  
				\hline    
				Radiative & $\Delta $P & longest sequence & 3.4 & 0.023 & 0.05 & (...) & 0.5 & 0.40 & 2227 & 112 & -355.7 \\ 
				&  & highest amplitude & 3.4 & 0.019 & 0.2 & (...) & 0.5 & 0.46 & 1783 & 121 & -347.6 \\ 
				&  & highest frequency & 3.3 & 0.023 & 0.0 & (...) & 0.5 & 0.40 & 2369 & 116 & -351.7 \\ 
				
				& Period & longest sequence & 3.4 & 0.023 & 0.05 & (...) & 0.5 & 0.40 & 758935 & 105 & -370.5 \\ 
				&  & highest amplitude & 3.4 & 0.019 & 0.2 & (...) & 0.5 & 0.46 & 92393 & 108 & -369.7 \\ 
				&  & highest frequency & 3.4 & 0.019 & 0.2 & (...) & 0.5 & 0.46 & 92393 & 110 & -367.0 \\ 
				\hline
				P\'eclet & $\Delta $P & longest sequence & 3.4 & 0.023 & 0.05 & (...) & 0.5 & 0.40 & 2318 & 130 & -344.1 \\ 
				&  & highest amplitude & 3.2 & 0.023 & 0.15 & (...) & 0.5 & 0.40 & 1299 & 138 & -337.2 \\ 
				&  & highest frequency & 3.2 & 0.015 & 0.2 & (...) & 0.0 & 0.42 & 2630 & 132 & -343.3 \\ 
				
				& Period & longest sequence & 3.2 & 0.023 & 0.25 & (...) & 0.0 & 0.40 & 363903 & 130 & -352.4 \\ 
				&  & highest amplitude & 3.2 & 0.023 & 0.25 & (...) & 0.0 & 0.40 & 363903 & 134 & -351.1 \\ 
				&  & highest frequency & 3.2 & 0.023 & 0.25 & (...) & 0.0 & 0.40 & 363903 & 140 & -345.2 \\ 
				\hline
			\end{tabular}
		\end{table*}
		%%%%%%%%%%%
		\begin{table*}
			\caption{Best fit models according to Mahalanobis distance in the subgrid without CBM.} 
			\label{tab:best_models_MD_subgrid_noCBM}
			\centering
			\setlength\tabcolsep{4.0pt} % default value: 6pt
			\renewcommand{\arraystretch}{1.12}
			\begin{tabular}{l l l l l l l l l r r r}     
				\hline               
				Grid & Observables & Pattern construction & $M_{\rm ini}$ [\msol] & $Z_{\rm ini}$ & $\acbm$ & $\fcbm$ & log($\D[env]$) & $X_c$ & $\chi^2_{\text{red}}$ & MD & AICc(MD) \vspace{1pt} \\  
				\hline    
				Radiative & $\Delta $P &  longest sequence & 3.3 & 0.019 & (...) & (...) & 1.5 & 0.44 & 1816 & 84 & -380.2 \\ 
				&  & highest amplitude & 3.3 & 0.019 & (...) & (...) & 1.5 & 0.44 & 1816 & 92 & -373.5 \\ 
				&  & highest frequency & 3.2 & 0.019 & (...) & (...) & 0.0 & 0.50 & 8601 & 83 & -381.8 \\ 
				
				& Period & longest sequence & 3.4 & 0.019 & (...) & (...) & 1.5 & 0.44 & 965418 & 97 & -376.5 \\ 
				&  & highest amplitude & 3.1 & 0.019 & (...) & (...) & 0.5 & 0.40 & 377038 & 99 & -377.6 \\ 
				&  & highest frequency & 3.3 & 0.023 & (...) & (...) & 0.5 & 0.40 & 792613 & 107 & -367.9 \\
				\hline
			\end{tabular}
		\end{table*}

		%%%%%%%%%%%%%%%%%%%%%%%%%%%%%%%%%%%%%%%%%%%%%%%%%%%%%%%%%%%%%%%%%%%%%%%%%%%%%%%%%%%%%%%%
		\begin{table*}
			\caption{Best fit models according to $\chi^2_{\text{red}}$.} 
			\label{tab:best_models_chi2}
			\centering
			\setlength\tabcolsep{4.0pt} % default value: 6pt
			\renewcommand{\arraystretch}{1.12}
			\begin{tabular}{l l l l l l l l l r r r}     
				\hline               
				Grid & Observables & Pattern construction & $M_{\rm ini}$ [\msol] & $Z_{\rm ini}$ & $\acbm$ & $\fcbm$ & log($\D[env]$) & $X_c$ & $\chi^2_{\text{red}}$ & MD & AICc($\chi^2_{\text{red}}$) \vspace{1pt} \\
				\hline    
				Radiative &$\Delta $P & longest sequence & 3.2 & 0.023 & 0.25 & 0.025 & 0.5 & 0.48 & 444 & 244 & 459.3 \\ 
				&  & highest amplitude & 3.2 & 0.023 & 0.25 & 0.025 & 0.5 & 0.48 & 444 & 247 & 459.3 \\ 
				&  & highest frequency & 3.2 & 0.023 & 0.25 & 0.025 & 0.5 & 0.48 & 444 & 245 & 459.3 \\
				
				& Period & longest sequence & 3.2 & 0.023 & 0.1 & 0.03 & 0.5 & 0.52 & 5369 & 258 & 5384.2 \\ 
				&  & highest amplitude & 3.2 & 0.023 & 0.1 & 0.03 & 0.5 & 0.52 & 5369 & 265 & 5384.2 \\ 
				&  & highest frequency & 3.2 & 0.023 & 0.1 & 0.03 & 0.5 & 0.52 & 5369 & 262 & 5384.2 \\ 
				\hline
				P\'eclet & $\Delta $P & longest sequence & 3.2 & 0.019 & 0.2 & 0.03 & 0.5 & 0.44 & 594 & 297 & 609.1 \\ 
				&  & highest amplitude & 3.2 & 0.019 & 0.2 & 0.03 & 0.5 & 0.44 & 594 & 308 & 609.1 \\ 
				&  & highest frequency & 3.2 & 0.019 & 0.2 & 0.03 & 0.5 & 0.44 & 594 & 308 & 609.1 \\ 
				
				& Period & longest sequence & 3.2 & 0.019 & 0.15 & 0.025 & 0.0 & 0.48 & 4836 & 315 & 4850.5 \\ 
				&  & highest amplitude & 3.2 & 0.019 & 0.15 & 0.025 & 0.0 & 0.48 & 4836 & 330 & 4850.5 \\ 
				&  & highest frequency & 3.2 & 0.019 & 0.15 & 0.025 & 0.0 & 0.48 & 4836 & 328 & 4850.5 \\ 
				\hline
			\end{tabular}
		\end{table*}
		
		%%%%%%%%%%%
		\begin{table*}
			\caption{Best fit models according to $\chi^2_{\text{red}}$ where $\acbm$=0 was fixed.} 
			\label{tab:best_models_chi2_subgrid_fov}
			\centering
			\setlength\tabcolsep{4.0pt} % default value: 6pt
			\renewcommand{\arraystretch}{1.12}
			\begin{tabular}{l l l l l l l l l r r r}     
				\hline               
				Grid & Observables & Pattern construction & $M_{\rm ini}$ [\msol] & $Z_{\rm ini}$ & $\acbm$ & $\fcbm$ & log($\D[env]$) & $X_c$ & $\chi^2_{\text{red}}$ & MD & AICc($\chi^2_{\text{red}}$) \vspace{1pt} \\
				\hline    
				Radiative &$\Delta $P & longest sequence & 3.3 & 0.023 & (...) & 0.025 & 1.0 & 0.44 & 600 & 187 & 612.4 \\ 
				&  & highest amplitude & 3.3 & 0.023 & (...) & 0.025 & 1.0 & 0.44 & 600 & 189 & 612.4 \\ 
				&  & highest frequency & 3.4 & 0.023 & (...) & 0.025 & 1.0 & 0.44 & 572 & 190 & 583.8 \\ 
				
				& Period & longest sequence & 3.3 & 0.023 & (...) & 0.02 & 1.0 & 0.48 & 5511 & 196 & 5522.9 \\ 
				&  & highest amplitude & 3.3 & 0.023 & (...) & 0.02 & 1.0 & 0.48 & 5511 & 203 & 5522.9 \\ 
				&  & highest frequency & 3.3 & 0.023 & (...) & 0.02 & 1.0 & 0.48 & 5511 & 200 & 5522.9 \\ 
				\hline
				P\'eclet & $\Delta $P & longest sequence & 3.3 & 0.019 & (...) & 0.03 & 1.0 & 0.44 & 701 & 196 & 712.7 \\ 
				&  & highest amplitude & 3.3 & 0.019 & (...) & 0.03 & 1.0 & 0.44 & 701 & 203 & 712.7 \\ 
				&  & highest frequency & 3.3 & 0.019 & (...) & 0.03 & 1.0 & 0.44 & 701 & 202 & 712.7 \\ 
				
				& Period & longest sequence & 3.2 & 0.023 & (...) & 0.03 & 0.5 & 0.48 & 6122 & 212 & 6133.7 \\ 
				&  & highest amplitude & 3.2 & 0.023 & (...) & 0.03 & 0.5 & 0.48 & 6122 & 220 & 6133.7 \\ 
				&  & highest frequency & 3.2 & 0.023 & (...) & 0.03 & 0.5 & 0.48 & 6122 & 225 & 6133.7 \\ 
				\hline
			\end{tabular}
		\end{table*}
		
		%%%%%%%%%%%
		\begin{table*}
			\caption{Best fit models according to $\chi^2_{\text{red}}$ where $\fcbm$=0 was fixed.} 
			\label{tab:best_models_chi2_subgrid_aov}
			\centering
			\setlength\tabcolsep{4.0pt} % default value: 6pt
			\renewcommand{\arraystretch}{1.12}
			\begin{tabular}{l l l l l l l l l r r r}     
				\hline               
				Grid & Observables & Pattern construction & $M_{\rm ini}$ [\msol] & $Z_{\rm ini}$ & $\acbm$ & $\fcbm$ & log($\D[env]$) & $X_c$ & $\chi^2_{\text{red}}$ & MD & AICc($\chi^2_{\text{red}}$) \vspace{1pt} \\
				\hline    
				Radiative &$\Delta $P & longest sequence & 3.4 & 0.023 & 0.0 & (...) & 1.5 & 0.42 & 779 & 133 & 790.9 \\ 
				&  & highest amplitude & 3.4 & 0.023 & 0.0 & (...) & 1.5 & 0.42 & 779 & 136 & 790.9 \\ 
				&  & highest frequency & 3.3 & 0.023 & 0.0 & (...) & 1.5 & 0.42 & 762 & 140 & 773.9 \\ 
				
				& Period & longest sequence & 3.2 & 0.019 & 0.3 & (...) & 0.5 & 0.50 & 10435 & 144 & 10447.4 \\ 
				&  & highest amplitude & 3.2 & 0.019 & 0.3 & (...) & 0.5 & 0.50 & 10435 & 148 & 10447.4 \\ 
				&  & highest frequency & 3.5 & 0.023 & 0.15 & (...) & 1.5 & 0.34 & 11681 & 140 & 11693.2 \\ 
				\hline
				P\'eclet & $\Delta $P & longest sequence & 3.2 & 0.019 & 0.25 & (...) & 1.0 & 0.42 & 751 & 143 & 763.3 \\ 
				&  & highest amplitude & 3.2 & 0.019 & 0.25 & (...) & 1.0 & 0.42 & 751 & 153 & 763.3 \\ 
				&  & highest frequency & 3.2 & 0.023 & 0.3 & (...) & 1.0 & 0.40 & 721 & 156 & 733.4 \\ 
				
				& Period & longest sequence & 3.5 & 0.019 & 0.3 & (...) & 0.5 & 0.30 & 10182 & 195 & 10193.8 \\ 
				&  & highest amplitude & 3.5 & 0.019 & 0.3 & (...) & 0.5 & 0.30 & 10182 & 201 & 10193.8 \\ 
				&  & highest frequency & 3.5 & 0.019 & 0.3 & (...) & 0.5 & 0.30 & 10182 & 204 & 10193.8 \\ 
				\hline
			\end{tabular}
		\end{table*}

		%%%%%%%%%%%
		
		\begin{table*}
			\caption{Best fit models according to $\chi^2_{\text{red}}$ in the subgrid without CBM.} 
			\label{tab:best_models_chi2_subgrid_noCBM}
			\centering
			\setlength\tabcolsep{4.0pt} % default value: 6pt
			\renewcommand{\arraystretch}{1.12}
			\begin{tabular}{l l l l l l l l l r r r}     
				\hline               
				Grid & Observables & Pattern construction & $M_{\rm ini}$ [\msol] & $Z_{\rm ini}$ & $\acbm$ & $\fcbm$ & log($\D[env]$) & $X_c$ & $\chi^2_{\text{red}}$ & MD & AICc($\chi^2_{\text{red}}$) \vspace{1pt} \\
				\hline    
				Radiative &$\Delta $P & longest sequence & 3.4 & 0.023 & (...) & (...) & 1.5 & 0.42 & 754 & 88 & 763.1 \\ 
				&  & highest amplitude & 3.4 & 0.023 & (...) & (...) & 1.5 & 0.42 & 754 & 99 & 763.1 \\ 
				&  & highest frequency & 3.3 & 0.023 & (...) & (...) & 1.5 & 0.42 & 737 & 96 & 746.6 \\ 
				
				& Period & longest sequence & 3.6 & 0.023 & (...) & (...) & 1.5 & 0.38 & 15639 & 127 & 15648.4 \\ 
				&  & highest amplitude & 3.6 & 0.023 & (...) & (...) & 1.5 & 0.38 & 15639 & 126 & 15648.4 \\ 
				&  & highest frequency & 3.6 & 0.023 & (...) & (...) & 1.5 & 0.38 & 15639 & 129 & 15648.4 \\ 
				\hline
			\end{tabular}
		\end{table*}
		
		%%%%%%%%%%%%%%%%%%%%%%%%%%%%%%%%%%%%%%%%%%%%%%%%%%%%%%%%%%%%%%%%%%%%%%%%%%%%%%%%%%%%%%%%%%%%%%
		% 
		% \newpage
		
		\section{Variance--covariance matrices} 
		Figures showing the variance--covariance matrices for the two grids, the different pattern construction methods, and the different sets of asteroseismic observables. Mode periods are labelled according to decreasing period value.
		\newpage
		%%% Periods
		\begin{figure}[!htp]
			\centering
			\includegraphics[width=0.95\hsize]{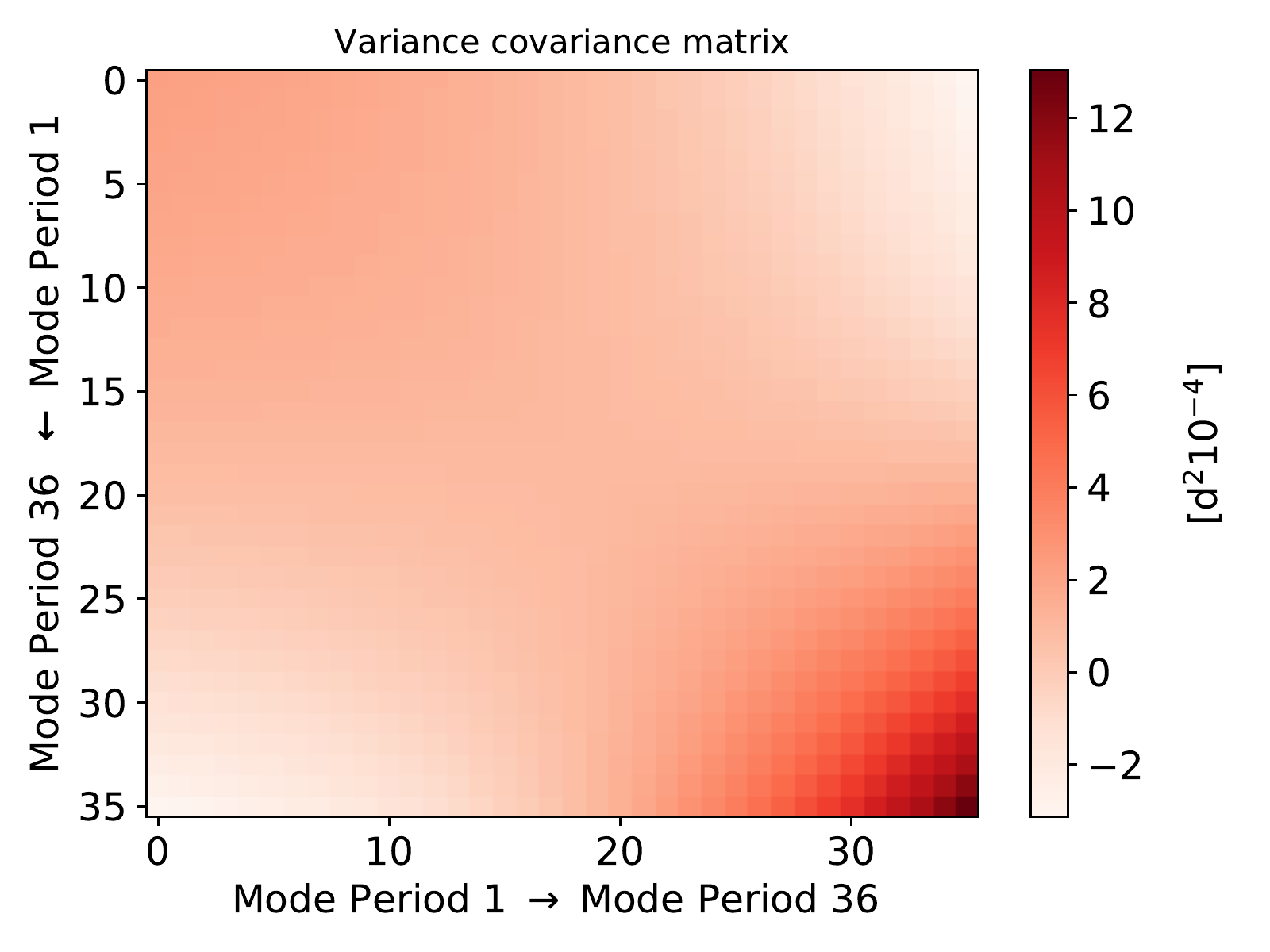}
			\caption{ Variance--covariance matrix for the periods of the radiative grid, selected using the longest sequence method.}
			\label{fig:Vmatrix_DO_chisq_longest_sequence_MD_P}
		\end{figure}
		\vspace{-0.43cm}
		\begin{figure}[!htp]
			\centering
			\includegraphics[width=0.95\hsize]{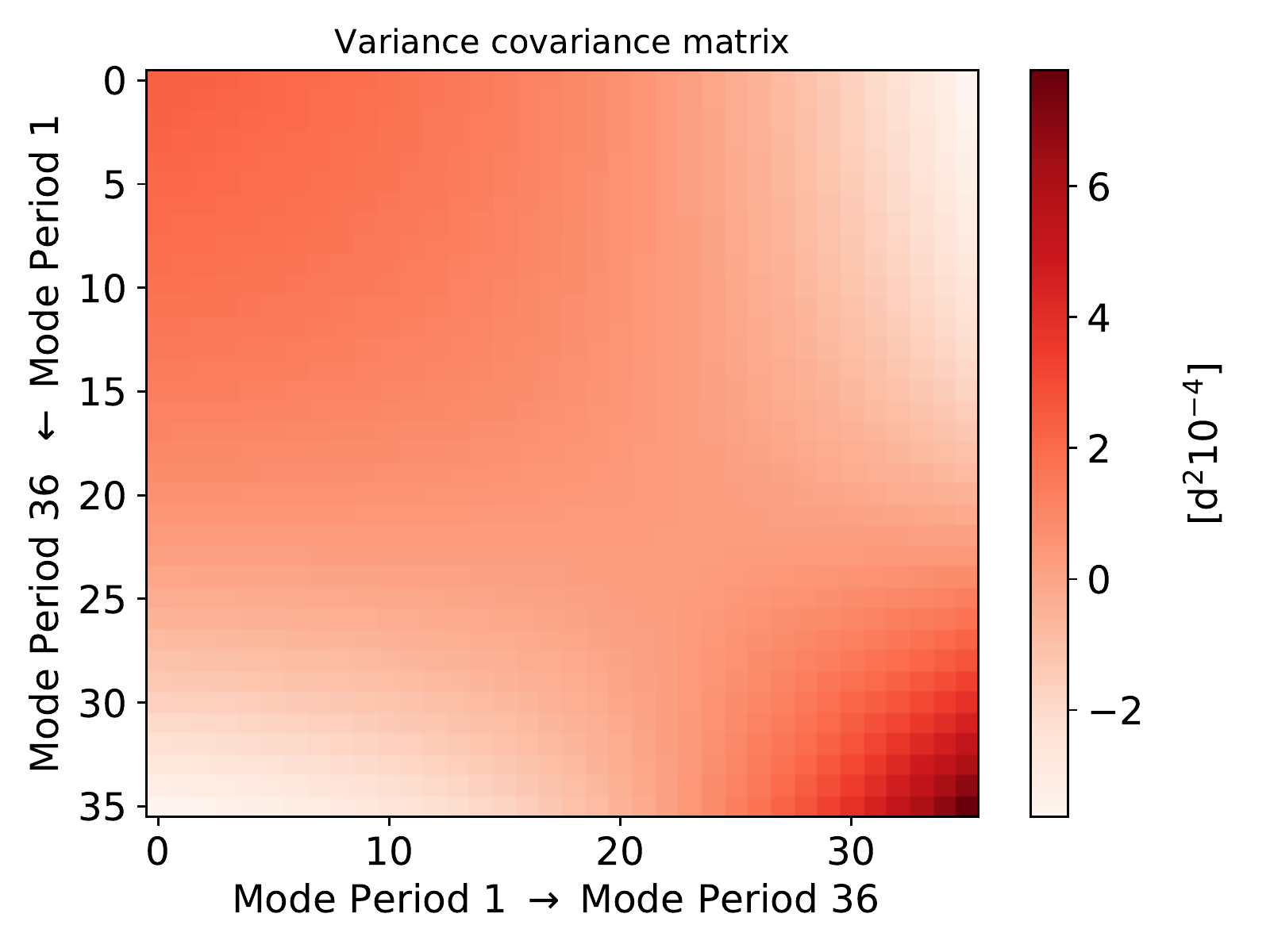}
			\caption{ Variance--covariance matrix for the periods of the radiative grid, selected using the highest amplitude method.}
			\label{fig:Vmatrix_DO_highest_amplitude_MD_P}
		\end{figure}
		\vspace{-0.43cm}
		\begin{figure}[!htp]
			\centering
			\includegraphics[width=0.95\hsize]{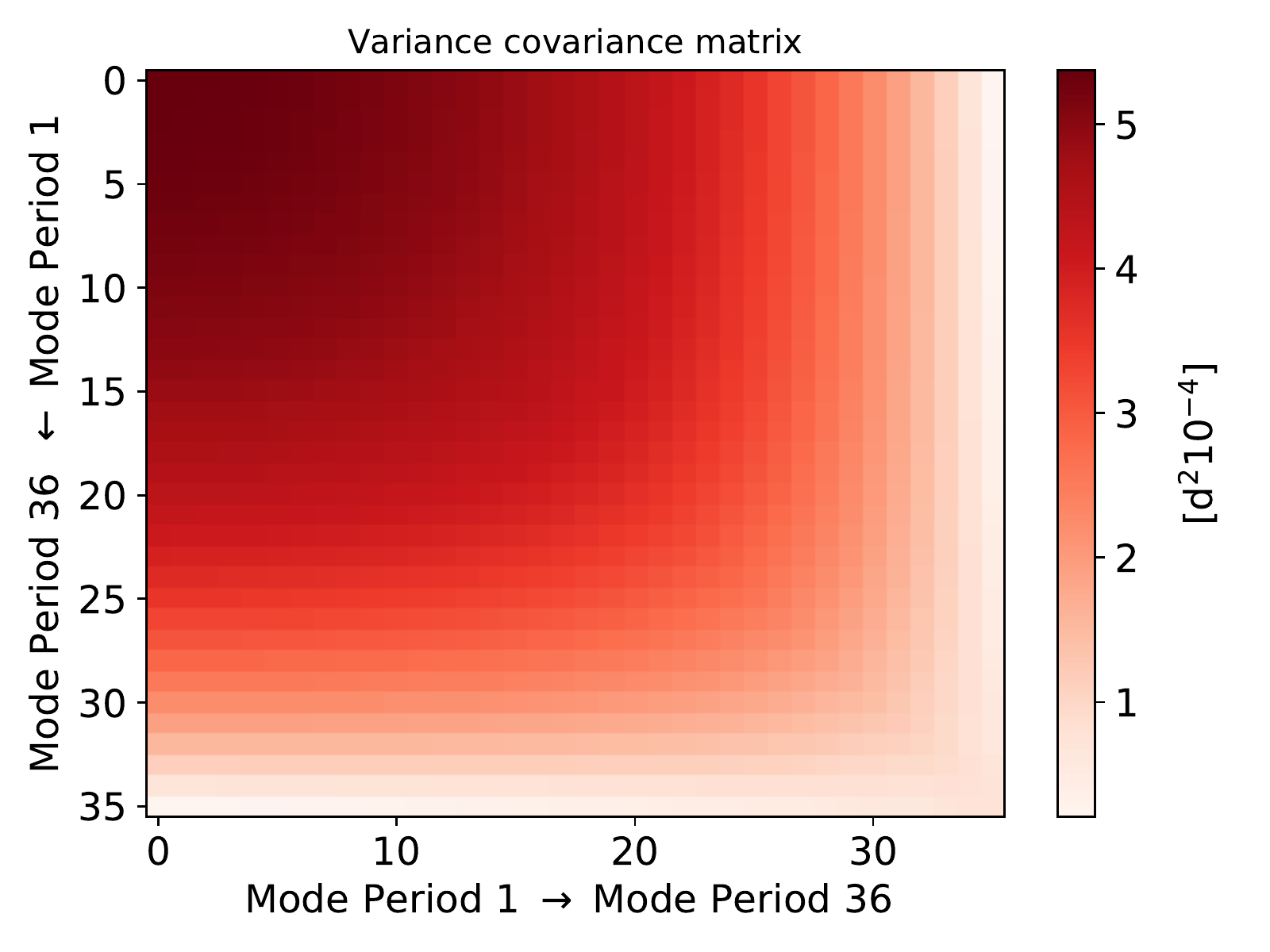}
			\caption{ Variance--covariance matrix for the periods of the radiative grid, selected using the highest frequency method.}
			\label{fig:Vmatrix_DO_highest_frequency_MD_P}
		\end{figure}
		
		\begin{figure}[!htp]
			\centering
			\includegraphics[width=0.95\hsize]{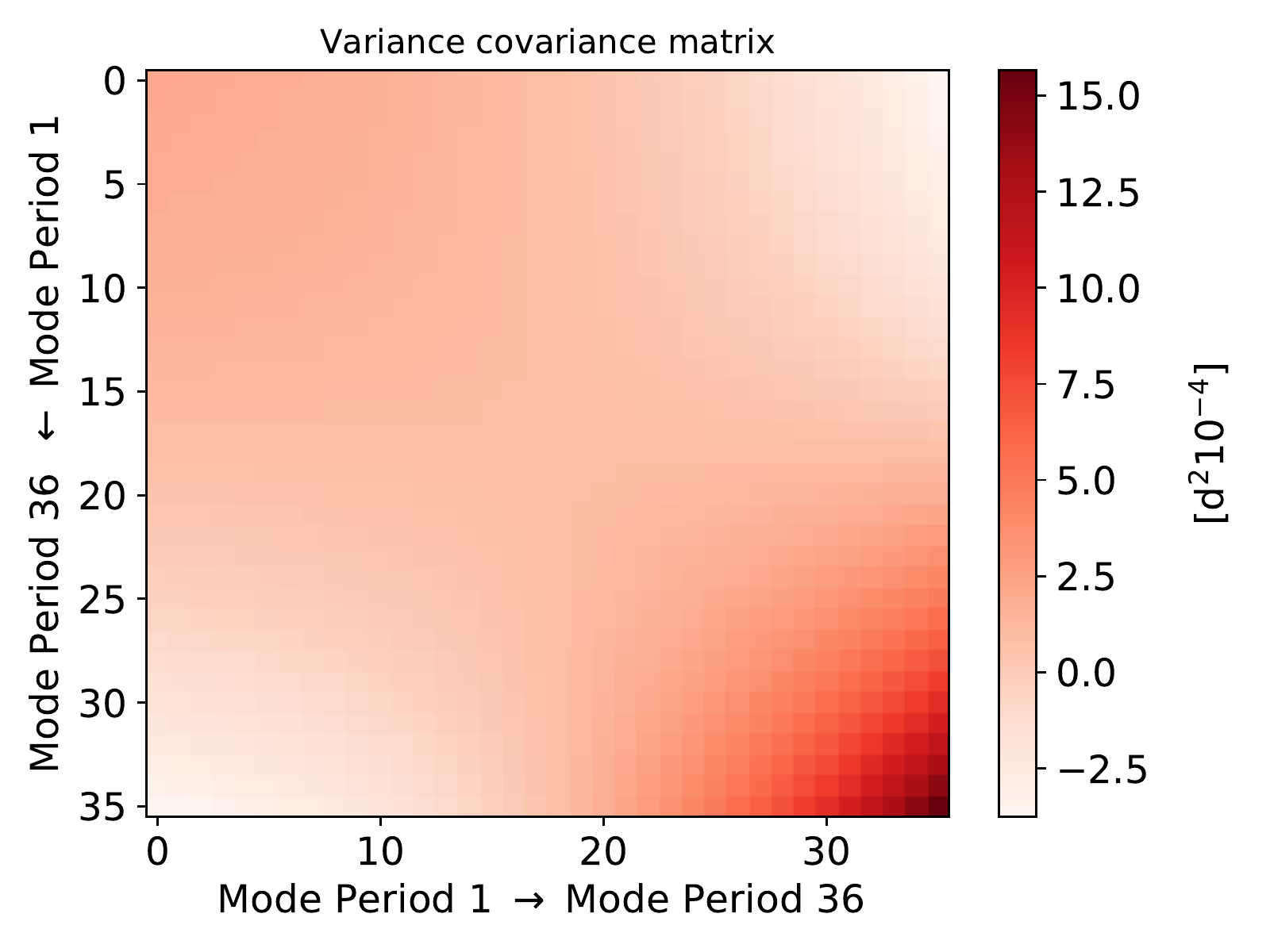}
			\caption{ Variance--covariance matrix for the periods of the P\'eclet grid, selected using the longest sequence method.}
			\label{fig:Vmatrix_ECP_chisq_longest_sequence_MD_P}
		\end{figure}
		\begin{figure}[!htp]
			\centering
			\includegraphics[width=0.95\hsize]{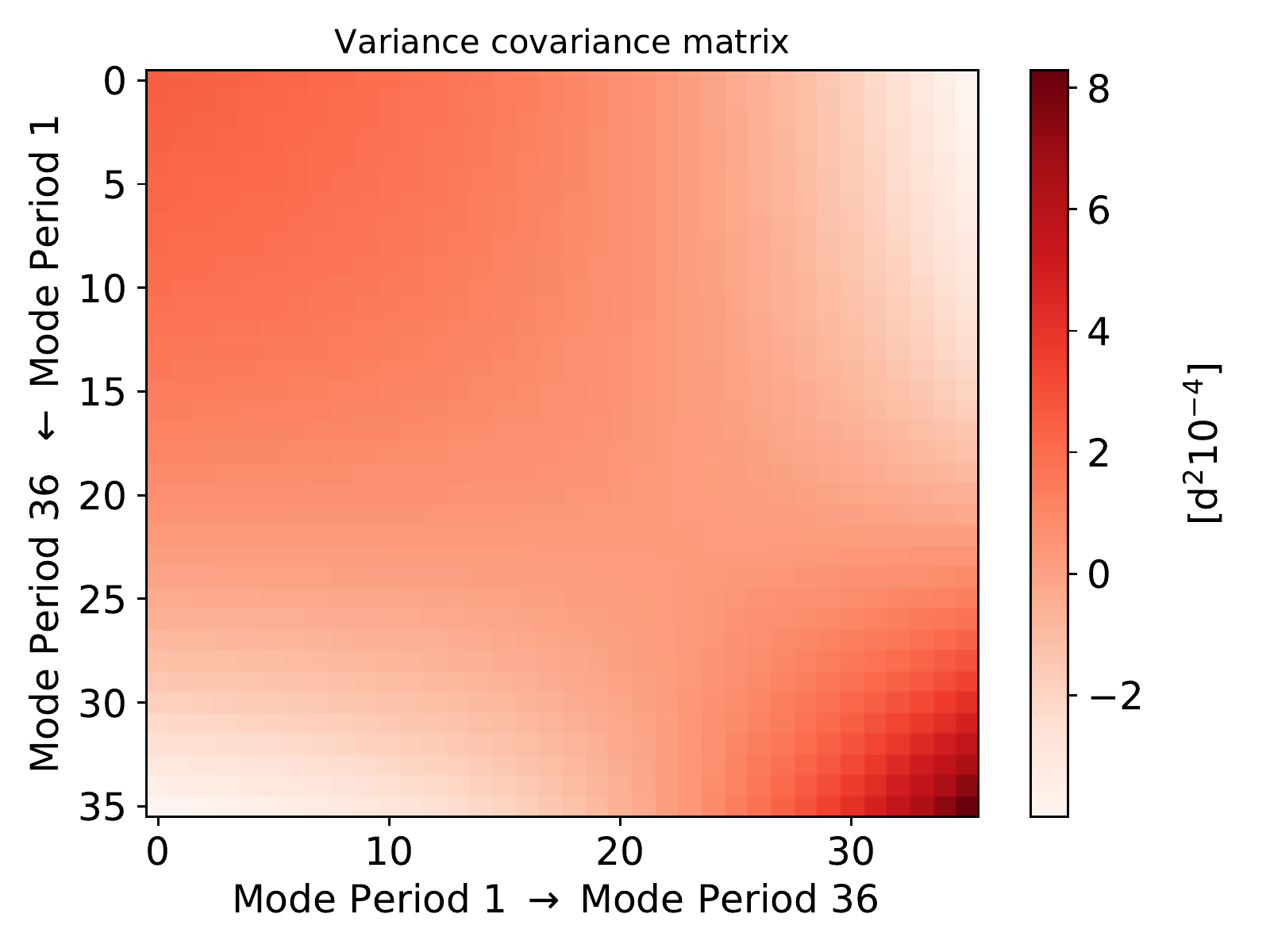}
			\caption{ Variance--covariance matrix for the periods of the P\'eclet grid, selected using the highest amplitude method.}
			\label{fig:Vmatrix_ECP_highest_amplitude_MD_P}
		\end{figure}
		\begin{figure}[!htp]
			\centering
			\includegraphics[width=0.95\hsize]{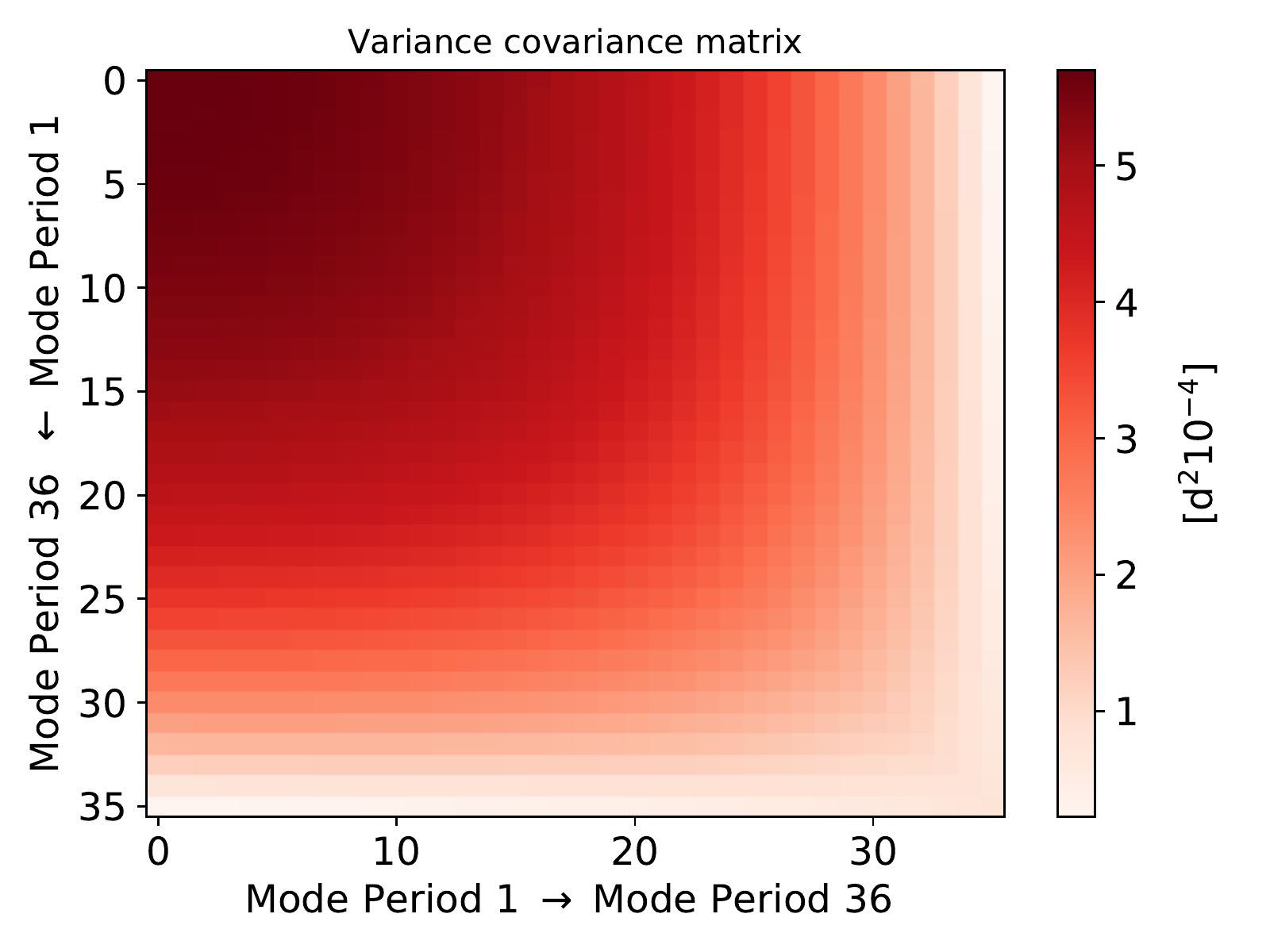}
			\caption{ Variance--covariance matrix for the periods of the P\'eclet grid, selected using the highest frequency method.}
			\label{fig:Vmatrix_ECP_highest_frequency_MD_P}
		\end{figure}
		
		%%% Period spacings
		\begin{figure}[!htp]
			\centering
			\includegraphics[width=\hsize]{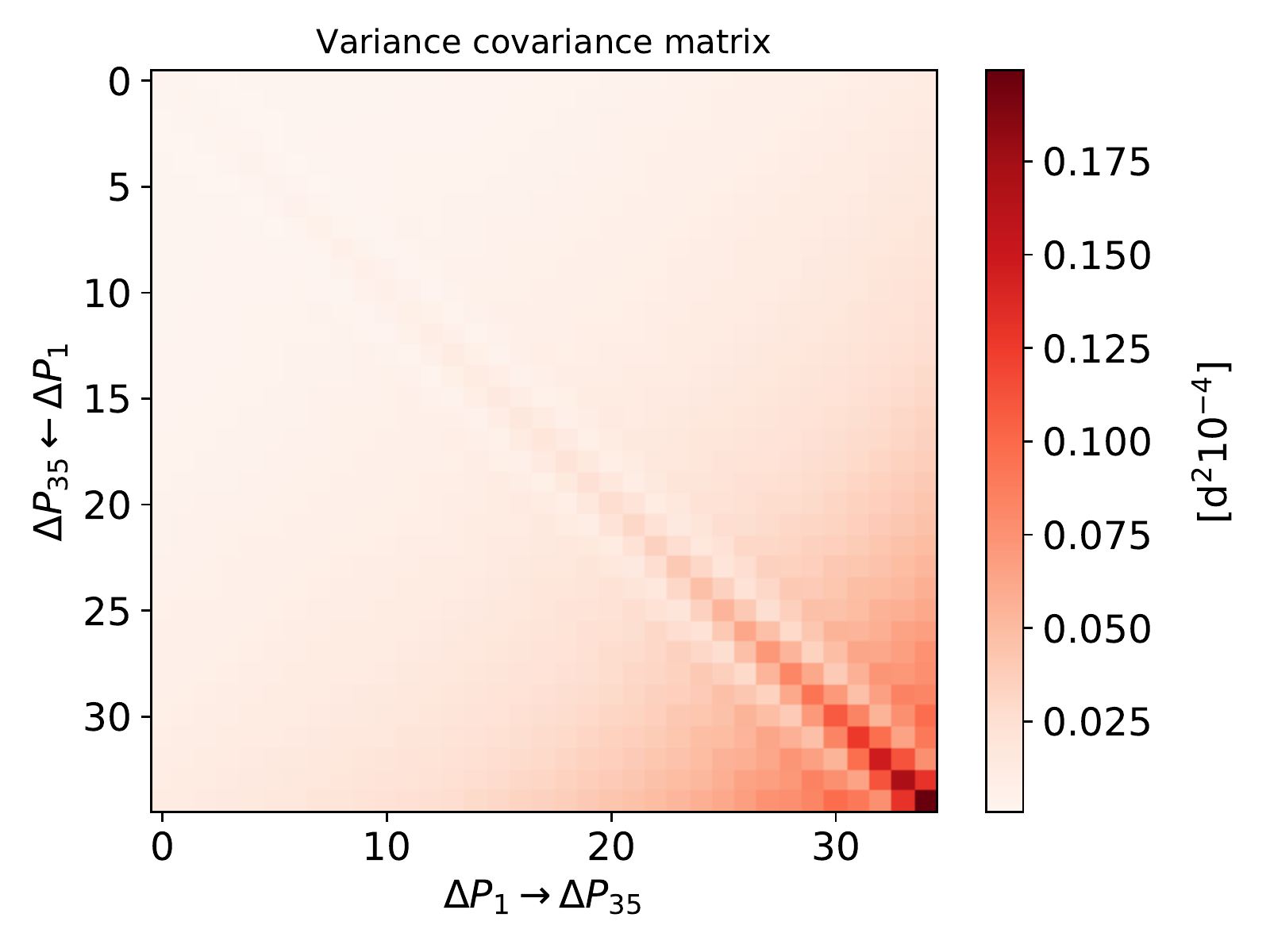}
			\caption{ Variance--covariance matrix for the period spacings of the radiative grid, selected using the longest sequence method.}
			\label{fig:Vmatrix_DO_chisq_longest_sequence_MD_dP}
		\end{figure}
		\begin{figure}[!htp]
			\centering
			\includegraphics[width=\hsize]{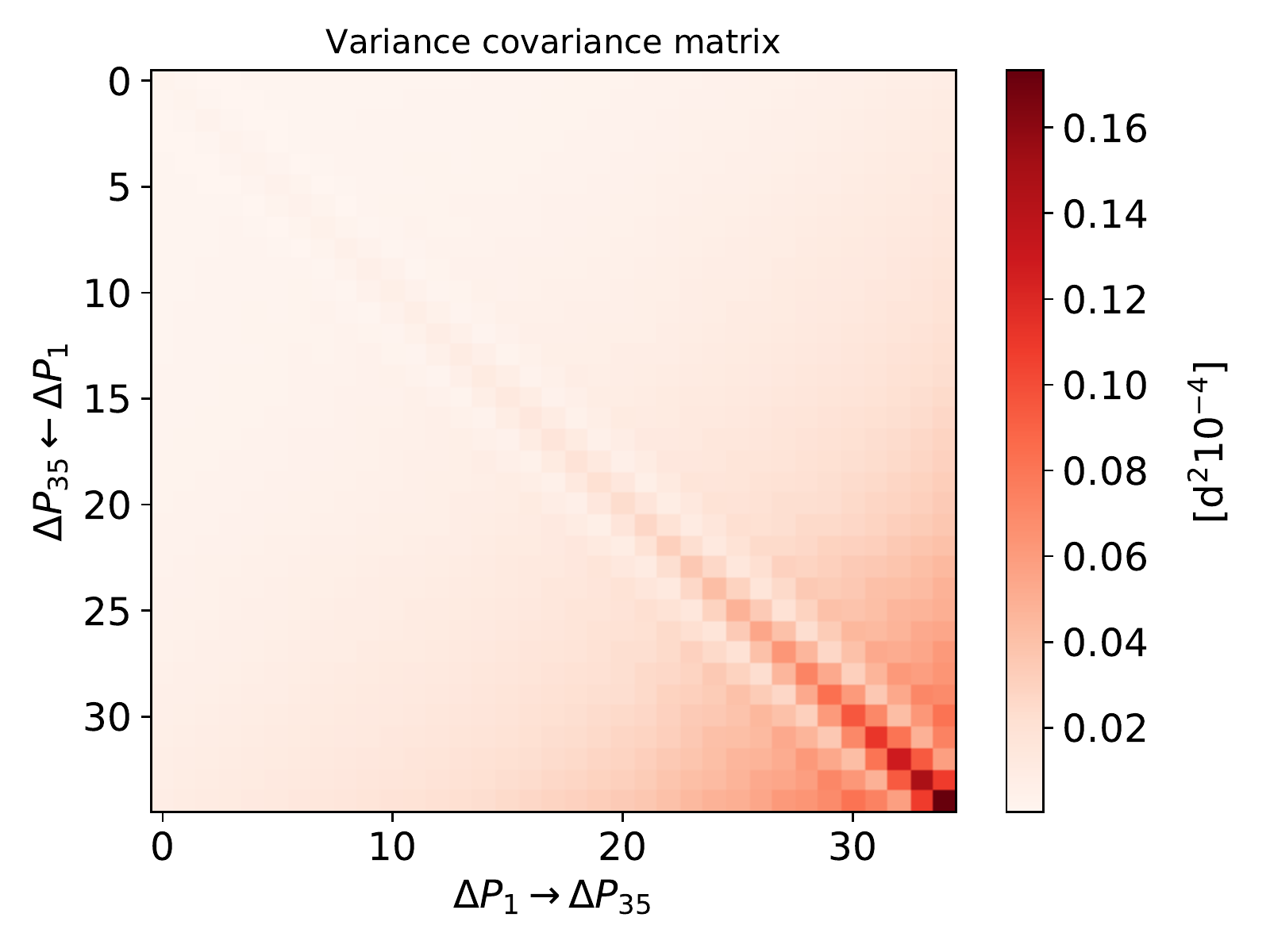}
			\caption{ Variance--covariance matrix for the period spacings of the radiative grid, selected using the highest amplitude method.}
			\label{fig:Vmatrix_DO_highest_amplitude_MD_dP}
		\end{figure}
		\begin{figure}[!htp]
			\centering
			\includegraphics[width=\hsize]{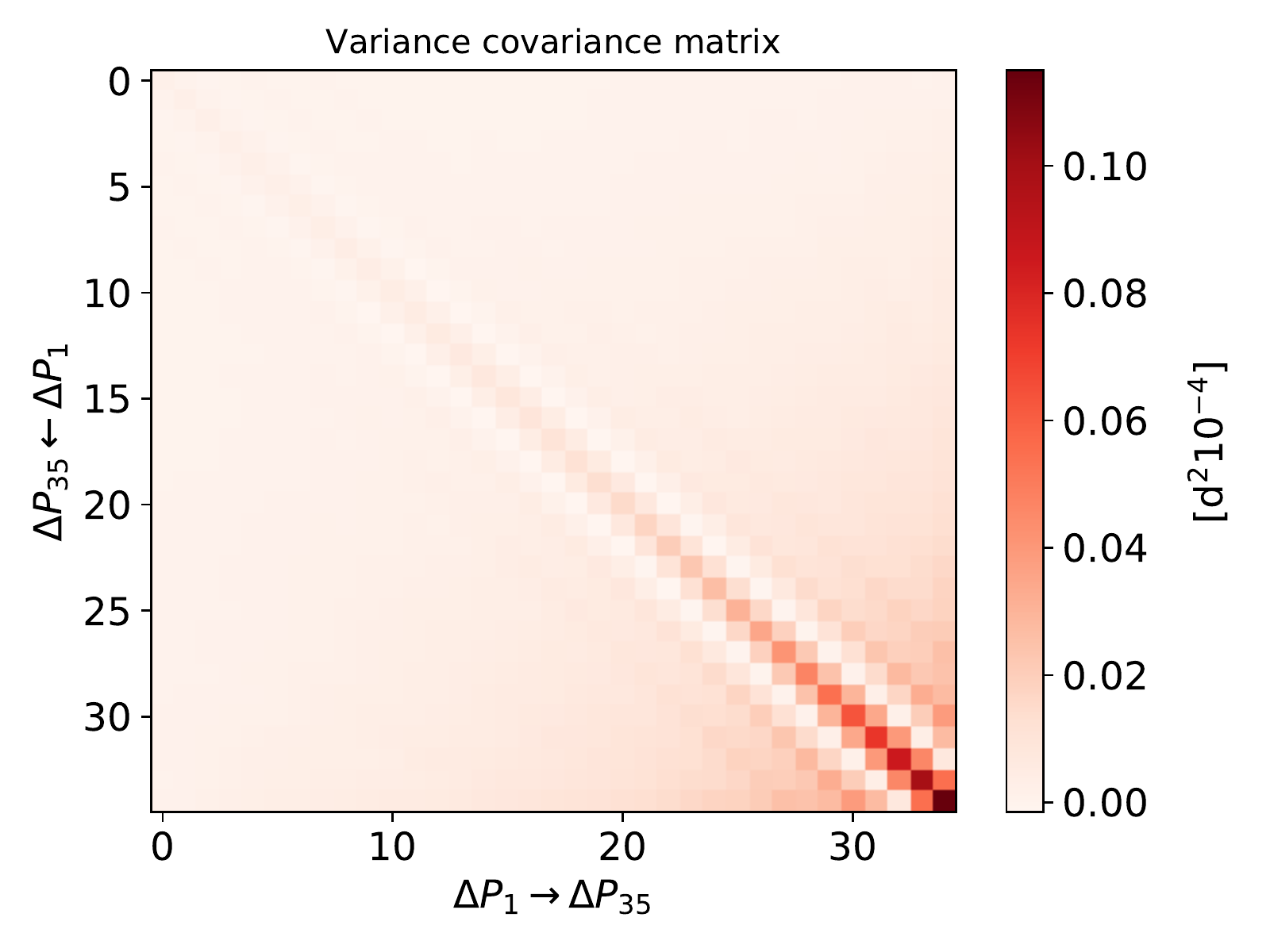}
			\caption{ Variance--covariance matrix for the period spacings of the radiative grid, selected using the highest frequency method.}
			\label{fig:Vmatrix_DO_highest_frequency_MD_dP}
		\end{figure}

		\begin{figure}[!htp]
			\centering
			\includegraphics[width=\hsize]{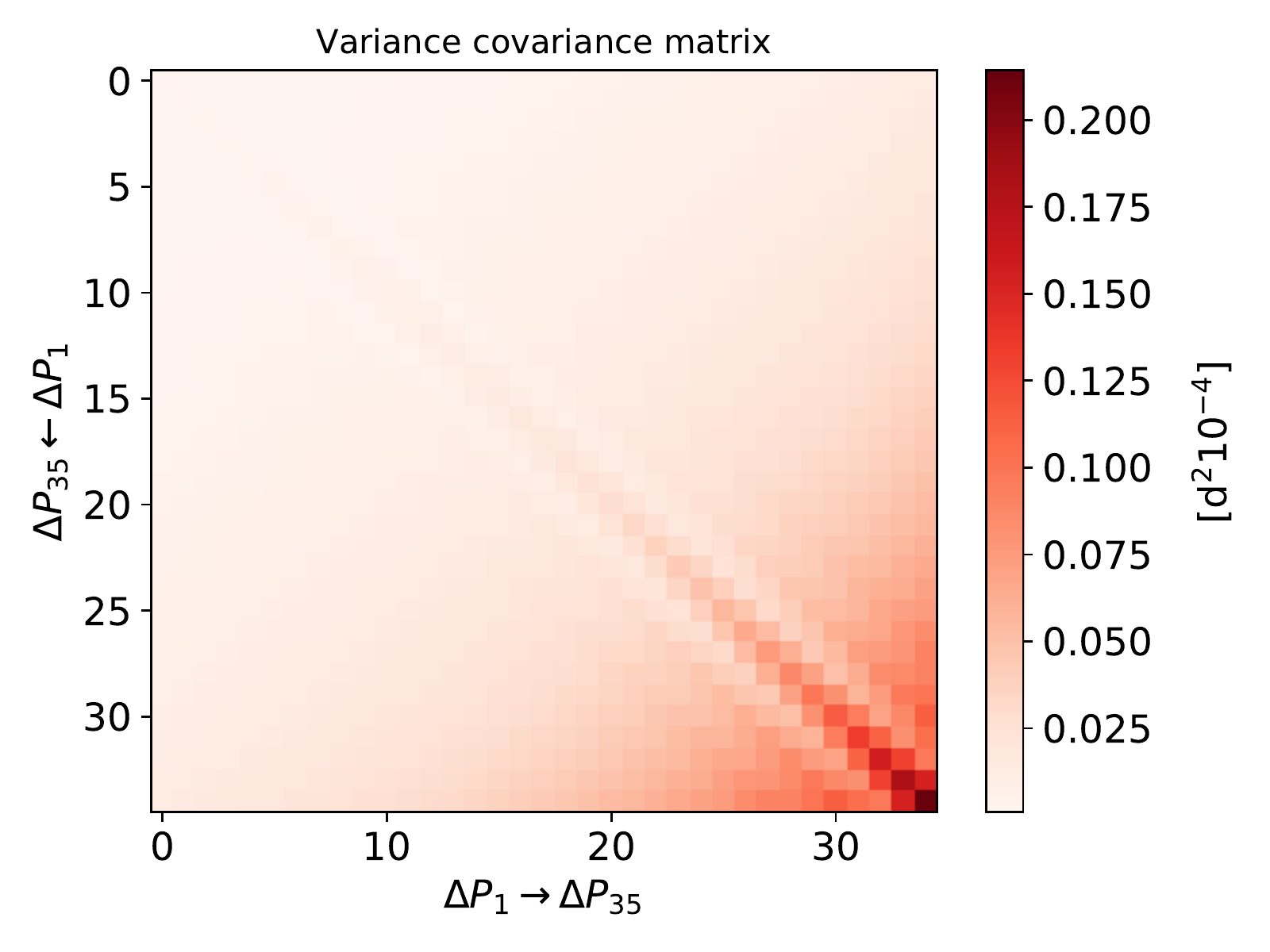}
			\caption{ Variance--covariance matrix for the period spacings of the P\'eclet grid, selected using the longest sequence method.}
			\label{fig:Vmatrix_ECP_chisq_longest_sequence_MD_dP}
		\end{figure}
		\begin{figure}[!htp]
			\centering
			\includegraphics[width=\hsize]{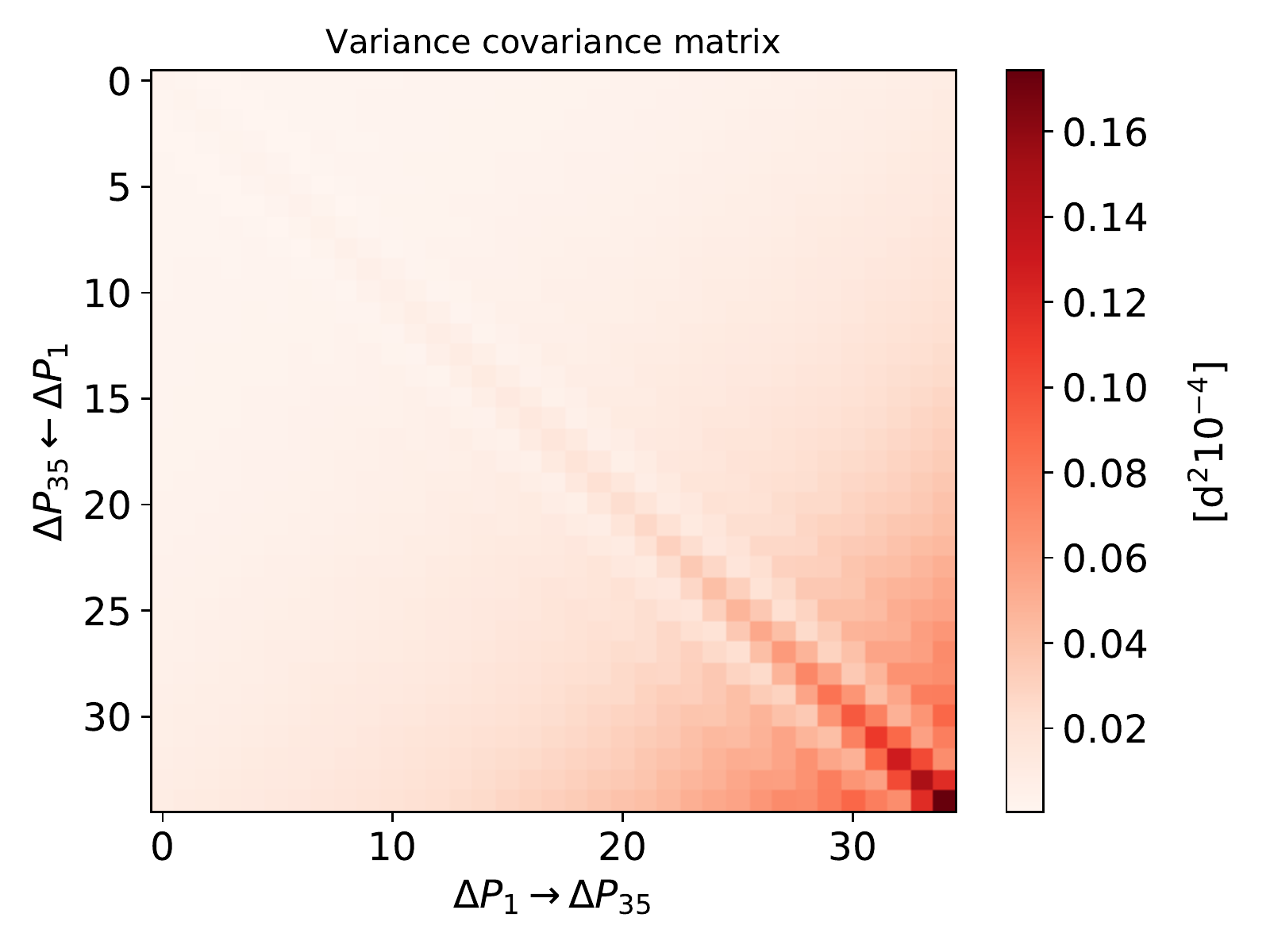}
			\caption{ Variance--covariance matrix for the period spacings of the P\'eclet grid, selected using the highest amplitude method.}
			\label{fig:Vmatrix_ECP_highest_amplitude_MD_dP}
		\end{figure}
		\begin{figure}[!hbp]
			\centering
			\includegraphics[width=\hsize]{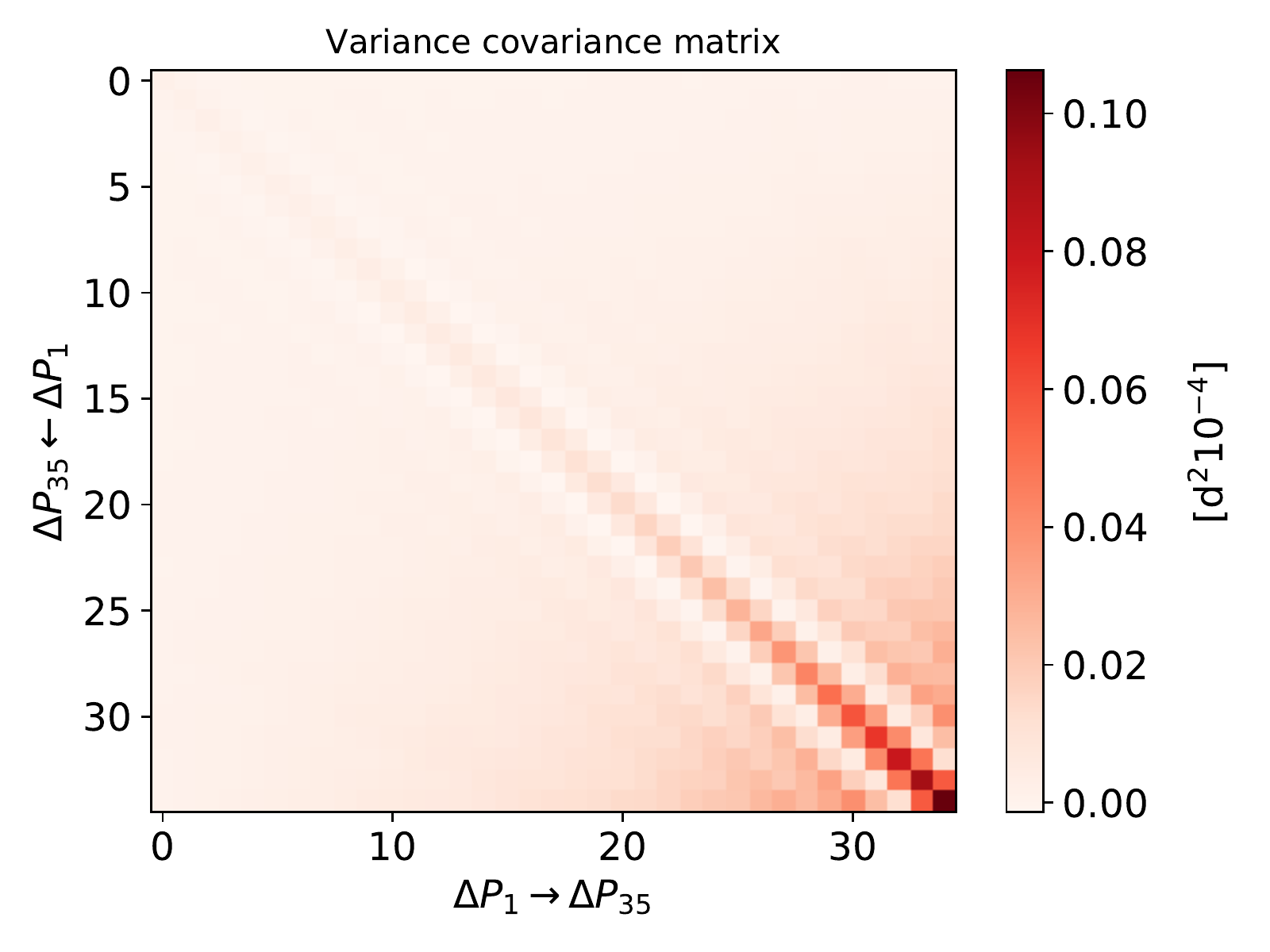}
			\caption{ Variance--covariance matrix for the period spacings of the P\'eclet grid, selected using the highest frequency method.}
			\label{fig:Vmatrix_ECP_highest_frequency_MD_dP}
		\end{figure}
		
	\end{appendix}
\end{document}